\documentclass[a4wide]{article}

\usepackage[longnamesfirst]{natbib}
\usepackage{ifthen}
\usepackage{calc}
\usepackage{fancyhdr}
\usepackage{xspace}
\usepackage[dvips]{graphicx}
\usepackage{mathrsfs}
\usepackage{amsmath}
\usepackage{amsthm}
\usepackage{amssymb}
\usepackage{float}
\usepackage{bm}
\usepackage{xcolor}
\usepackage{colortbl}
\usepackage{tabularx}
\usepackage{subfigure}
\usepackage{rotating}
\usepackage{caption}
\usepackage[absolute]{textpos}
\usepackage{marvosym}
\usepackage[backref,pagebackref=true,colorlinks=true,breaklinks=true,citecolor=blue,linkcolor=blue,bookmarks=true]{hyperref}

\usepackage{algorithmic,algorithm,listings}
\usepackage{graphicx}

\graphicspath{{Figs/}}
\DeclareMathOperator*{\argmax}{argmax}
\floatname{algorithm}{Code}

\renewcommand{\cite}{\citet}

\renewcommand{\ast}{\star}              
\renewcommand{\geq}{\geqslant}          
\renewcommand{\leq}{\leqslant}          

\newcommand{\bea}{\begin{eqnarray}}      
\newcommand{\eea}{\end{eqnarray}}

\newcommand{\ben}{\begin{enumerate}[parsep=15pt]}  
\newcommand{\een}{\end{enumerate}}

\newcommand{\Ito}{It\={o}\xspace}
\newcommand{\DD}{\triangle}

\begin{document}

\title{{A Stochastic Processes Toolkit for Risk Management}\thanks{Updated version accepted for publication in the Journal of Risk Management for Financial Institutions. We are grateful to Richard Hrvatin and Lars Jebjerg for reviewing the manuscript and for helpful comments. Kyriakos Chourdakis furhter helped with comments and formatting issues.}}
\author{Damiano Brigo\thanks{Fitch Solutions, and Department of Mathematics, Imperial College, London. {\tt damiano.brigo@fitchratings.com}}, Antonio Dalessandro\thanks{Fitch Solutions, and Department of Mathematics, Imperial College, London. {\tt antonio.dalessandro@ic.ac.uk}}, Matthias Neugebauer\thanks{Fitch Ratings. {\tt matthias.neugebauer@fitchratings.com}}, Fares Triki\thanks{Fitch Solutions, and Paris School of Economics, Paris. {\tt fares.triki@fitchratings.com}}}
\date{15 November 2007}

\maketitle

\begin{abstract}
In risk management it is desirable to grasp the essential statistical features of a time series representing a risk factor. This tutorial aims to introduce a number of different stochastic processes that can help in grasping the essential features of risk factors describing different asset classes or behaviors. This paper does not aim at being exhaustive, but gives examples and a feeling for practically implementable models allowing for stylised features in the data. The reader may also use these models as building blocks to build more complex models, although for a number of risk management applications the models developed here suffice for the first step in the quantitative analysis. The broad qualitative features addressed here are { fat tails} and { mean reversion}. We give some orientation on the initial choice of a suitable stochastic process and then explain how the process parameters can be estimated based on historical data. Once the process has been calibrated, typically through maximum likelihood estimation, one may simulate the risk factor and build future scenarios for the risky portfolio. On the terminal simulated distribution of the portfolio one may then single out several risk measures, although here we focus on the stochastic processes estimation preceding the simulation of the risk factors Finally, this first survey report focuses on single time series. Correlation or more generally dependence across risk factors, leading to multivariate processes modeling, will be addressed in future work.
\end{abstract}

\medskip

{\bf JEL Classification code:} G32, C13, C15, C16.

\medskip

{\bf AMS Classification code:} 60H10, 60J60, 60J75, 65C05, 65c20, 91B70

\medskip

\noindent {\bf Key words:} Risk Management, Stochastic Processes, Maximum Likelihood Estimation, Fat Tails, Mean Reversion, Monte Carlo Simulation

\vspace{0.2cm}
\begin{center}
\mbox{
\begin{minipage}{14 cm}
\tableofcontents
\end{minipage}
}
\end{center}

\newpage

\pagestyle{myheadings} \markboth{}{{\footnotesize  D. Brigo, A. Dalessandro, M. Neugebauer, F. Triki: A stochastic processes toolkit for Risk Management}}

\section{Introduction}
In risk management and in the rating practice it is desirable to grasp the essential statistical features of a time series representing a risk factor to begin a detailed technical analysis of the product or the entity under scrutiny. The quantitative analysis is not the final tool, since it has to be integrated with judgemental analysis and possible stresses, but represents a good starting point for the process leading to the final decision.

This tutorial aims to introduce a number of different stochastic processes that, according to the economic and financial nature of the specific risk factor, can help in grasping the essential features of the risk factor itself. For example, a family of stochastic processes that can be suited to capture foreign exchange behaviour might not be suited to model hedge funds or a commodity. Hence there is need to have at one's disposal a sufficiently large set of practically implementable stochastic processes that may be used to address the quantitative analysis at hand to begin a risk management or rating decision process.

This report does not aim at being exhaustive, since this would be a formidable task beyond the scope of a survey paper. Instead, this survey gives examples and a feeling for practically implementable models allowing for stylised features in the data. The reader may also use these models as building blocks to build more complex models, although for a number of risk management applications the models developed here suffice for the first step in the quantitative analysis.

The broad qualitative features addressed here are {\bf fat tails} and {\bf mean reversion}. This report begins with the geometric Brownian motion (GBM) as a fundamental example of an important stochastic process featuring neither mean reversion nor fat tails, and then it goes through generalisations featuring any of the two properties or both of them. According to the specific situation, the different processes are formulated either in (log-) returns space or in levels space, as is more convenient. Levels and returns can be easily converted into each other, so this is indeed a matter of mere convenience.

This tutorial will address the following processes:

\begin{itemize}

\item \textbf{Basic process:} Arithmetic Brownian Motion (ABM, returns) or GBM (levels).

\item \textbf{Fat tails processes:} GBM with lognormal jumps (levels), ABM with normal jumps (returns), GARCH (returns), Variance Gamma (returns).

\item \textbf{Mean Reverting processes:} Vasicek, CIR (levels if interest rates or spreads, or returns in general), exponential Vasicek (levels).

\item \textbf{Mean Reverting processes with Fat Tails:} Vasicek with jumps (levels if interest rates or spreads, or returns in general), exponential Vasicek with jumps (levels).

\end{itemize}

Different financial time series are better described by different processes above. In general, when first presented with a historical time series of data for a risk factor, one should decide what kind of general properties the series features. The financial nature of the time series can give some hints at whether mean reversion and fat tails are expected to be present or not. Interest rate processes, for example, are often taken to be mean reverting, whereas foreign exchange series are supposed to be often fat tailed and credit spreads feature both characteristics. In general one should:

\medskip

\noindent {\bf Check for mean reversion or stationarity}\\
Some of the tests mentioned in the report concern mean reversion. The presence of an autoregressive (AR) feature can be tested in the data, typically on the returns of a series or on the series itself if this is an interest rate or spread series. In linear processes with normal shocks, this amounts to checking stationarity. If this is present, this can be a symptom for mean reversion and a mean reverting process can be adopted as a first assumption for the data. If the AR test is rejected this means that there is no mean reversion, at least under normal shocks and linear models, although the situation can be more complex for nonlinear processes with possibly fat tails, where the more general notions of stationarity and ergodicity may apply. These are not addressed in this report.

If the tests find AR features then the process is considered as mean reverting and one can compute autocorrelation and partial autocorrelation functions to see the lag in the regression. Or one can go directly to the continuous time model and estimate it on the data through maximum likelihood. In this case, the main model to try is the Vasicek model.

If the tests do not find AR features then the simplest form of mean reversion for linear processes with Gaussian shocks is to be rejected. One has to be careful though that the process could still be mean reverting in a more general sense. In a way, there could be some sort of mean reversion even under non-Gaussian shocks, and example of such a case are the jump-extended Vasicek or exponential Vasicek models, where mean reversion is mixed with fat tails, as shown below in the next point.

\noindent {\bf Check for fat tails}
If the AR test fails, either the series is not mean reverting, or it is but with fatter tails that the Gaussian distribution and possibly nonlinear behaviour. To test for fat tails a first graphical tool is the QQ-plot, comparing the tails in the data with the Gaussian distribution. The QQ-plot gives immediate graphical evidence on possible presence of fat tails. Further evidence can be collected by considering the third and fourth sample moments, or better the skewness and excess kurtosis, to see how the data differ from the Gaussian distribution. More rigorous tests on normality should also be run following this preliminary analysis\footnote{such as, for example, the Jarque Bera, the Shapiro-Wilk and the Anderson-Darling tests, that are not addressed in this report}.
If fat tails are not found and the AR test failed, this means that one is likely dealing with a process lacking mean reversion but with Gaussian shocks, that could be modeled as an arithmetic Brownian motion. If fat tails are found (and the AR test has been rejected), one may start calibration with models featuring fat tails and no mean reversion (GARCH, NGARCH, Variance Gamma, arithmetic Brownian motion with jumps) or fat tails and mean reversion (Vasicek with jumps, exponential Vasicek with jumps). Comparing the goodness of fit of the models may suggest which alternative is preferable. Goodness of fit is determined again graphically through QQ-plot of the model implied distribution against the empirical distribution, although more rigorous approaches can be applied~\footnote{Examples are the Kolmogorov Smirnov test, likelihood ratio methods and the Akaike information criteria, as well as methods based on the Kullback Leibler information or relative entropy, the Hellinger Distance and other divergences}. The predictive power can also be tested, following the calibration and the goodness of fit tests. \footnote{The Diebold Mariano statistics can be mentioned as an example for AR processes. These approaches are not pursued here.}.

The above classification may be summarized in the table, where typically the referenced variable is the return process or the process itself in case of interest rates or spreads:

\begin{table}[h!]
\begin{center}
\begin{tabular}{|c|c|c|c|}
\hline    &  Normal tails & Fat tails \\
\hline           NO mean reversion   & ABM   & ABM+Jumps, \\
            &  &  (N)GARCH, VG  \\
\hline  Mean Reversion    & Vasicek & Exponential Vasicek\\
     &   &  CIR, Vasicek with Jumps \\  \hline
\end{tabular}
\end{center}
\end{table}

Once the process has been chosen and calibrated to historical data, typically through regression or maximum likelihood estimation, one may use the process to simulate the risk factor over a given time horizon and build future scenarios for the portfolio under examination. On the terminal simulated distribution of the portfolio one may then single out several risk measures. This report does not focus on the risk measures themselves but on the stochastic processes estimation preceding the simulation of the risk factors. In case more than one model is suited to the task, one can analyze risks with different models and compare the outputs as a way to reduce model risk.

Finally, this first survey report focuses on single time series. Correlation or more generally dependence across risk factors, leading to multivariate processes modeling, will be addressed in future work.

\medskip

\noindent {\bf Prerequisites}

The following discussion assumes that the reader is familiar with basic probability theory, including probability distributions (density and cumulative density functions, moment generating functions), random variables and expectations. It is also assumed that the reader has some practical knowledge of stochastic calculus and of \Ito's formula, and basic coding skills. For each section, the MATLAB$^{\circledR}$ code is provided to implement the specific models. Examples illustrating the possible use of the models on actual financial time series are also presented.

\section{Modelling with Basic Stochastic Processes: GBM}

This section provides an introduction to modelling through stochastic processes. All the concepts will be introduced using the fundamental process for financial modelling, the geometric Brownian motion with constant drift and constant volatility. The GBM is ubiquitous in finance, being the process underlying the Black and Scholes formula for pricing European options.

\subsection{The Geometric Brownian Motion}

The geometric Brownian motion (GBM) describes the random behaviour of the asset price level $S(t)$ over time. The GBM is specified as follows:

\begin{equation}\label{eq:gbm1}
	dS(t) =\mu S(t) dt +\sigma S(t) dW(t)
\end{equation}

Here $W$ is a standard Brownian motion, a special diffusion process\footnote{Also referred to as a Wiener Process.} that is characterised by independent identically distributed (iid) increments that are normally (or Gaussian) distributed with zero mean and a standard deviation equal to the square root of the time step. Independence in the increments implies that the model is a Markov Process, which is a particular type of process for which only the current asset price is relevant for computing probabilities of events involving future asset prices. In other terms, to compute probabilities involving future asset prices, knowledge of the whole past does not add anything to knowledge of the present.

The $d$ terms indicate that the equation is given in its continuous time version\footnote{A continuous-time stochastic process is one where the value of the price can change at any point in time. The theory is very complex and actually this differential notation is just short-hand for an integral equation. A discrete-time stochastic process on the other hand is one where the price of the financial asset can only change at certain fixed times. In practice, discrete behavior is usually observed, but continuous time processes prove to be useful to analyse the properties of the model, besides being paramount in valuation where the assumption of continuous trading is at the basis of the Black and Scholes theory and its extensions.}. The property of independent identically distributed increments is very important and will be exploited in the calibration as well as in the simulation of the random behaviour of asset prices. Using some basic stochastic calculus the equation can be rewritten as follows:

\begin{equation}\label{eq:logS}
	d \log S(t)=\left(\mu-\frac{1}{2}\sigma^{2}\right) dt +\sigma dW(t)
\end{equation}

where log denotes the standard natural logarithm. The process followed by the log is called an Arithmetic Brownian Motion.
The increments in the logarithm of the asset value are normally distributed. This equation is straightforward to integrate between any two instants, $t$ and $u$, leading to:

\begin{equation}\label{eq:logSdistrib}
 \log S(u) - \log S(t)=\left(\mu-\frac{1}{2}\sigma^{2}\right) (u-t) +\sigma (W(u)- W(t)) \sim N\left(\left(\mu-\frac{1}{2}\sigma^{2}\right) (u-t),\sigma^2(u-t) \right).
\end{equation}

Through exponentiation and taking $u=T$ and $t=0$ the solution for the GBM equation is obtained:

\begin{equation}
	S(T)=S(0)\exp\left(\left[\mu-\frac{1}{2}\sigma^{2} \right]T+\sigma W(T) \right)
\end{equation}

This equation shows that the asset price in the GBM follows a log-normal distribution, while the logarithmic returns $\log(S_{t+\Delta t}/S_{t})$ are normally distributed.\\

The moments of $S(T)$ are easily found using the above solution, the fact that $W(T)$ is Gaussian with mean zero and variance $T$, and finally the moment generating function of a standard normal random variable $Z$ given by:

\begin{equation}
	\mathbb{E}\left[e^{aZ} \right]=e^{\frac{1}{2}a^{2}}.
\end{equation}

Hence the first two moments (mean and variance) of $S(T)$ are:

\begin{equation}
	\mathbb{E}\left[S(T)\right]=S(0)e^{\mu T} \qquad\qquad \mbox{Var}\left[S(T)\right]=e^{2\mu T}S^{2}(0)\left(e^{\sigma^{2}T}-1 \right)
\end{equation}
To simulate this process, the continuous equation between discrete instants  $t_{0}<t_{1}<\ldots<t_{n}$ needs to be solved as follows:

\begin{equation}\label{GBMdiscreeteSol}
	S(t_{i+1})=S(t_{i})\exp\left(\left[\mu-\frac{1}{2}\sigma^{2} \right]\left(t_{i+1}-t_{i} \right)+\sigma\sqrt{t_{i+1}-t_{i}} Z_{i+1} \right)
\end{equation}

where $Z_{1},Z_{2},\ldots Z_{n}$ are independent random draws from the standard normal distribution. The simulation is exact and does not introduce any discretisation error, due to the fact that the equation can be solved exactly. The following charts plot a number of simulated sample paths using the above equation and the mean plus/minus one standard deviation and the first and 99th percentiles over time. The mean of the asset price grows exponentially with time.

\begin{figure}[h]\label{GBMmeanstdev}
\begin{center}
\includegraphics[width=0.45\textwidth]{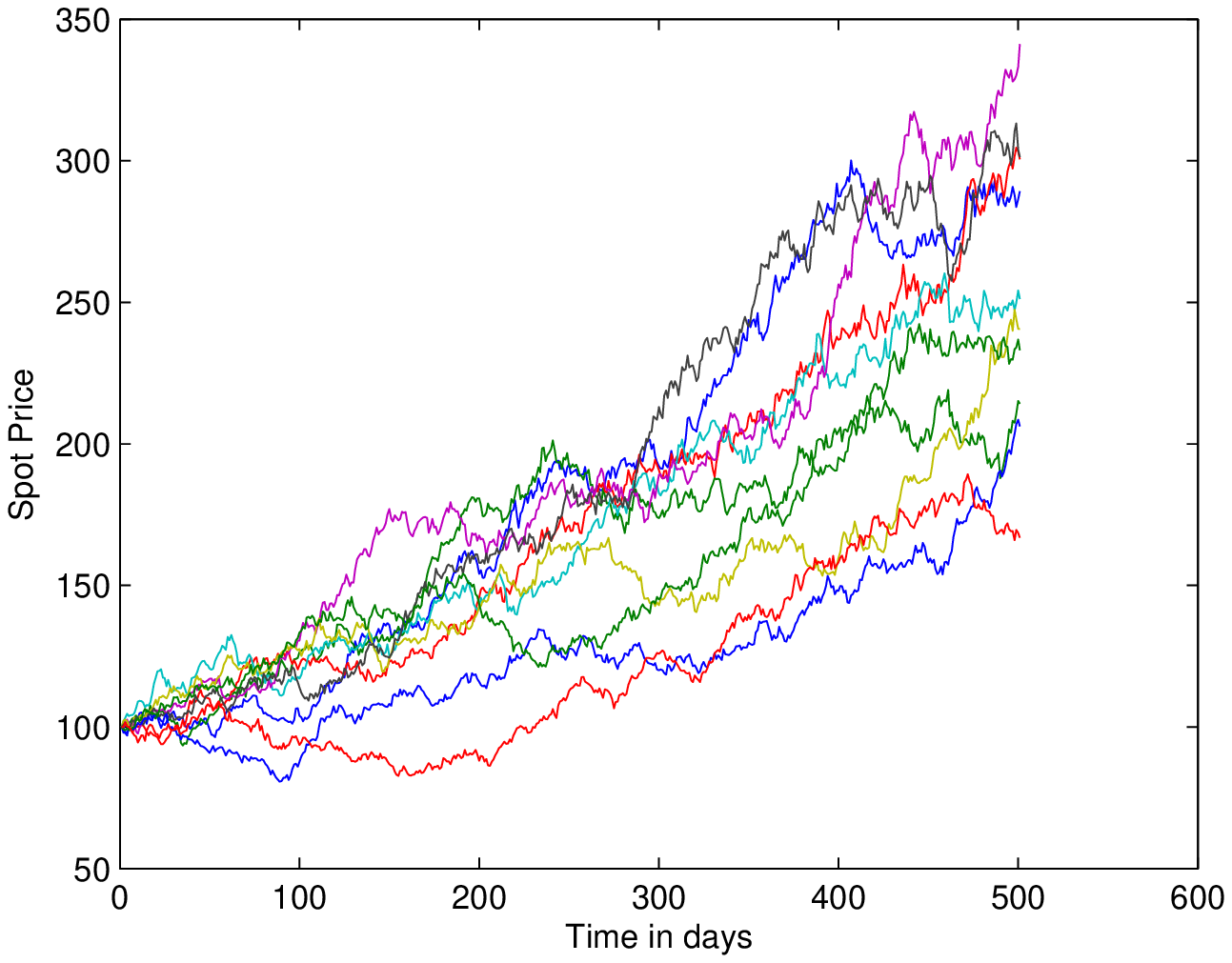}
\includegraphics[width=0.45\textwidth]{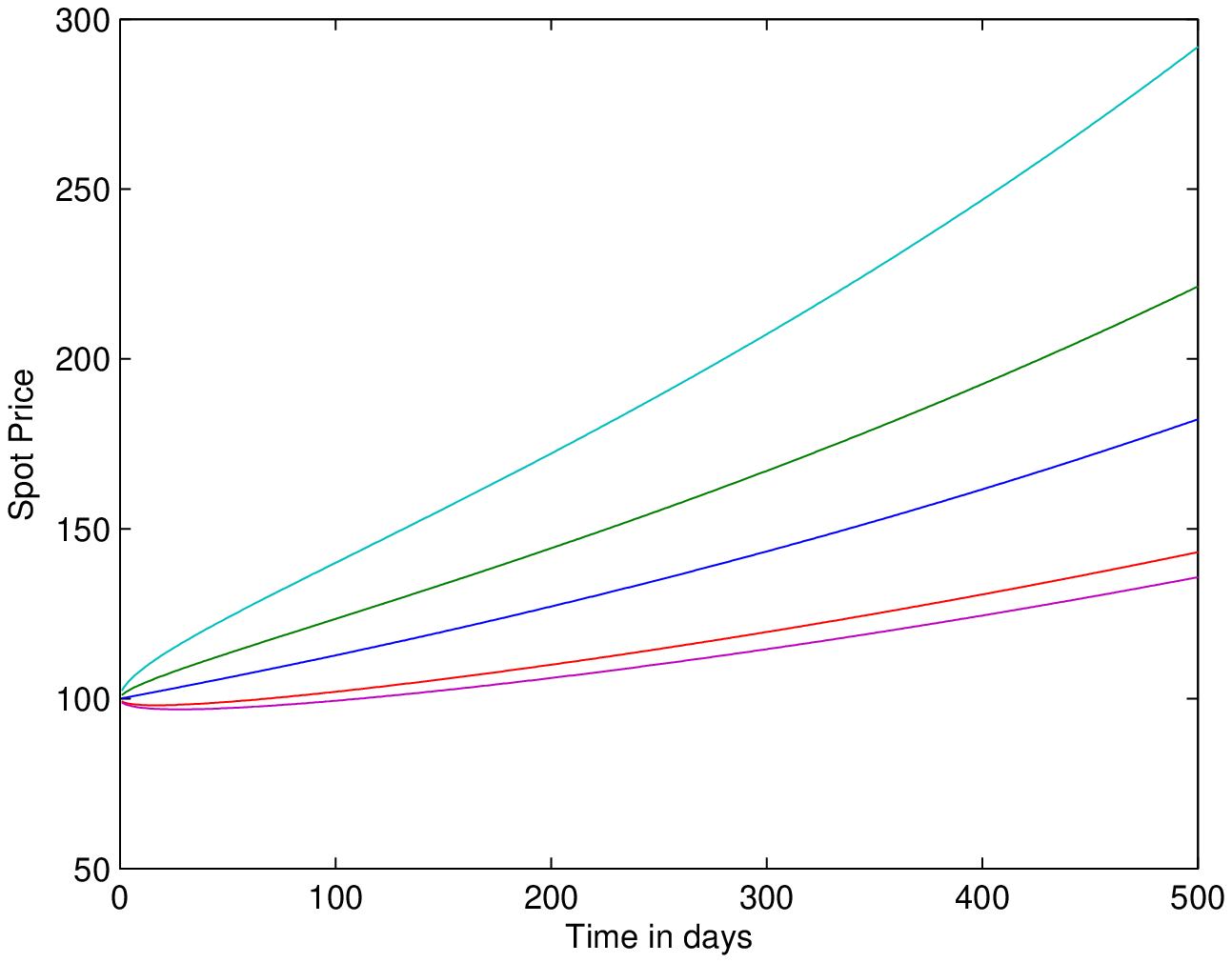}
\caption{GBM Sample Paths and Distribution Statistics }
\end{center}
\end{figure}

The following Matlab function simulates sample paths of the GBM using equation \eqref{GBMdiscreeteSol}, which was vectorised in Matlab.

\begin{algorithm}[!h]
\begin{lstlisting}
function S=GBM_simulation(N_Sim,T,dt,mu,sigma,S0)

mean=(mu-0.5*sigma^2)*dt;
S=S0*ones(N_Sim,T+1);

BM=sigma*sqrt(dt)*normrnd(0,1,N_Sim,T);
S(:,2:end)=S0*exp(cumsum(mean+BM,2));

end
\end{lstlisting}
\caption{$MATLAB^{\circledR}$ Code to simulate GBM Sample Paths.}\label{GBMsim}
\end{algorithm}

\subsection{Parameter Estimation}
\begin{figure}[h]
\begin{center}
\includegraphics[width=0.45\textwidth]{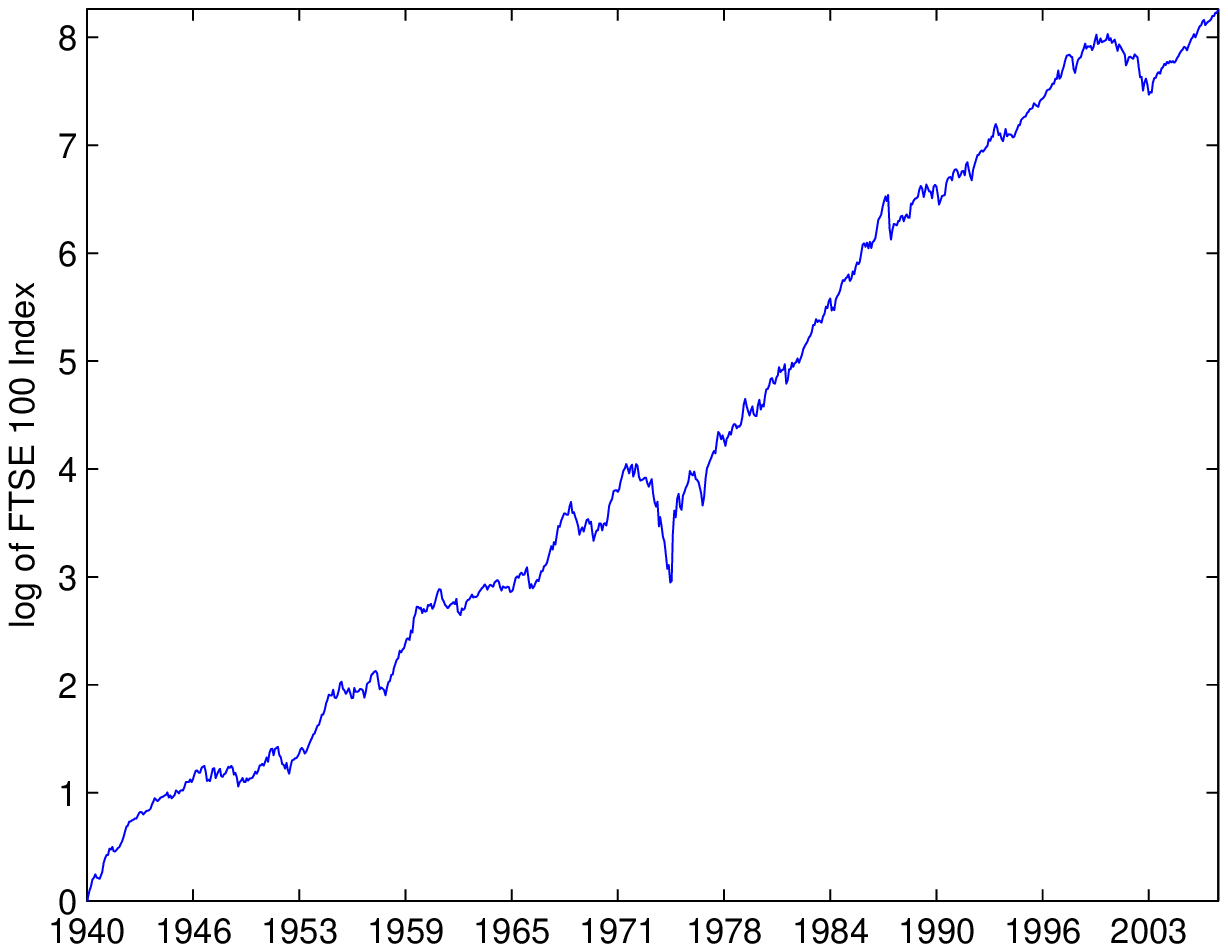}
\includegraphics[width=0.45\textwidth]{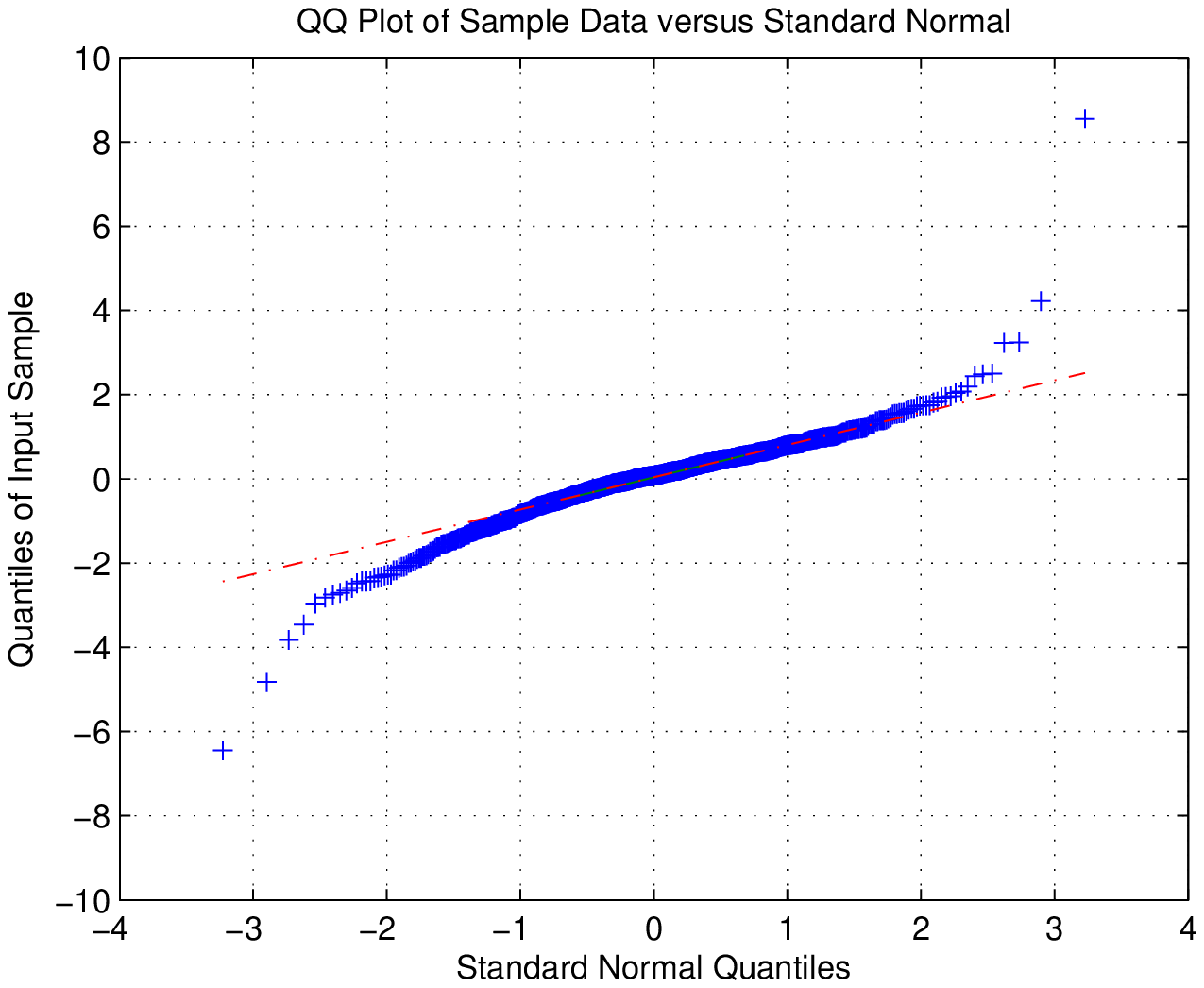}
\caption{Historical FTSE100 Index and QQ Plot FTSE 100}\label{ftse100}
\end{center}
\end{figure}

This section describes how the GBM can be used as an attempt to model the random behaviour of the FTSE100 stock index, the log of which is plotted in the left chart of Figure~\ref{ftse100}.\\

The second chart, on the right of Figure~\ref{ftse100}, plots the quantiles of the log return distribution against the quantiles of the standard normal distribution. This QQ-plot allows one  to compare distributions and to check the assumption of normality. The quantiles of the historical distribution are plotted on the Y-axis and the quantiles of the chosen modeling distribution on the X-axis. If the comparison distribution provides a good fit to the historical returns, then the QQ-plot approximates a straight line. In the case of the FTSE100 log returns, the QQ-plot show that the historical quantiles in the tail of the distribution are significantly larger compared to the normal distribution. This is in line with the general observation about the presence of fat tails in the return distribution of financial asset prices. Therefore the GBM at best provides only a rough approximation for the FTSE100.  The next sections will outline some extensions of the GBM that provide a better fit to the historical distribution.

Another important property of the GBM is the independence of returns. Therefore, to ensure the GBM is appropriate for modelling the asset price in a financial time series one has to check that the returns of the observed data are truly independent. The assumption of independence in statistical terms means that there is no autocorrelation present in historical returns. A common test for autocorrelation (typically on returns of a given asset) in a time series sample $x_1,x_2,\ldots,x_n$ realised from a random process $X(t_i)$ is to plot the autocorrelation function of the lag $k$, defined as:

\begin{equation}\label{eq:ACFPACF} \mbox{ACF}(k) = \frac{1}{(n-k)\hat{v}}
\sum_{i=1}^{n-k} (x_i - \hat{m}) (x_{i+k} - \hat{m}), \ \ k=1,2,\ldots
\end{equation}
where $\hat{m}$ and $\hat{v}$ are the sample mean and variance of the series, respectively. Often, besides the ACF, one also considers the Partial Auto-Correlation function (PACF). Without fully defining PACF, roughly speaking one may say that
while ACF$(k)$  gives an estimate of the correlation between $X(t_i)$ and $X(t_{i+k})$, PACF$(k)$ informs on the correlation between $X(t_i)$ and $X(t_{i+k})$ \emph{that is not accounted for by the shorter lags from $1$ to $k-1$}, or more precisely it is a sort of sample estimate of:
\[ \mbox{Corr}(X(t_i)-\bar{X}^{i+1,...,i+k-1}_i,\
X(t_{i+k})-\bar{X}^{i+1,...,i+k-1}_{i+k})\]
where $\bar{X}^{i+1,...,i+k-1}_i$ and $\bar{X}^{i+1,...,i+k-1}_{i+k}$ are the best estimates (regressions in a linear context) of $X(t_i)$ and $X(t_{i+k})$ given $X(t_{i+1}),\ldots,X(t_{i+k-1})$.
PACF also gives an estimate of the lag in an autoregressive process.

ACF and PACF for the FTSE 100 equity index returns are shown in the charts in Figure~\ref{ACF_ParCorr}.
\begin{figure}[h]\label{ACF_ParCorr}
\begin{center}
\includegraphics[width=0.45\textwidth]{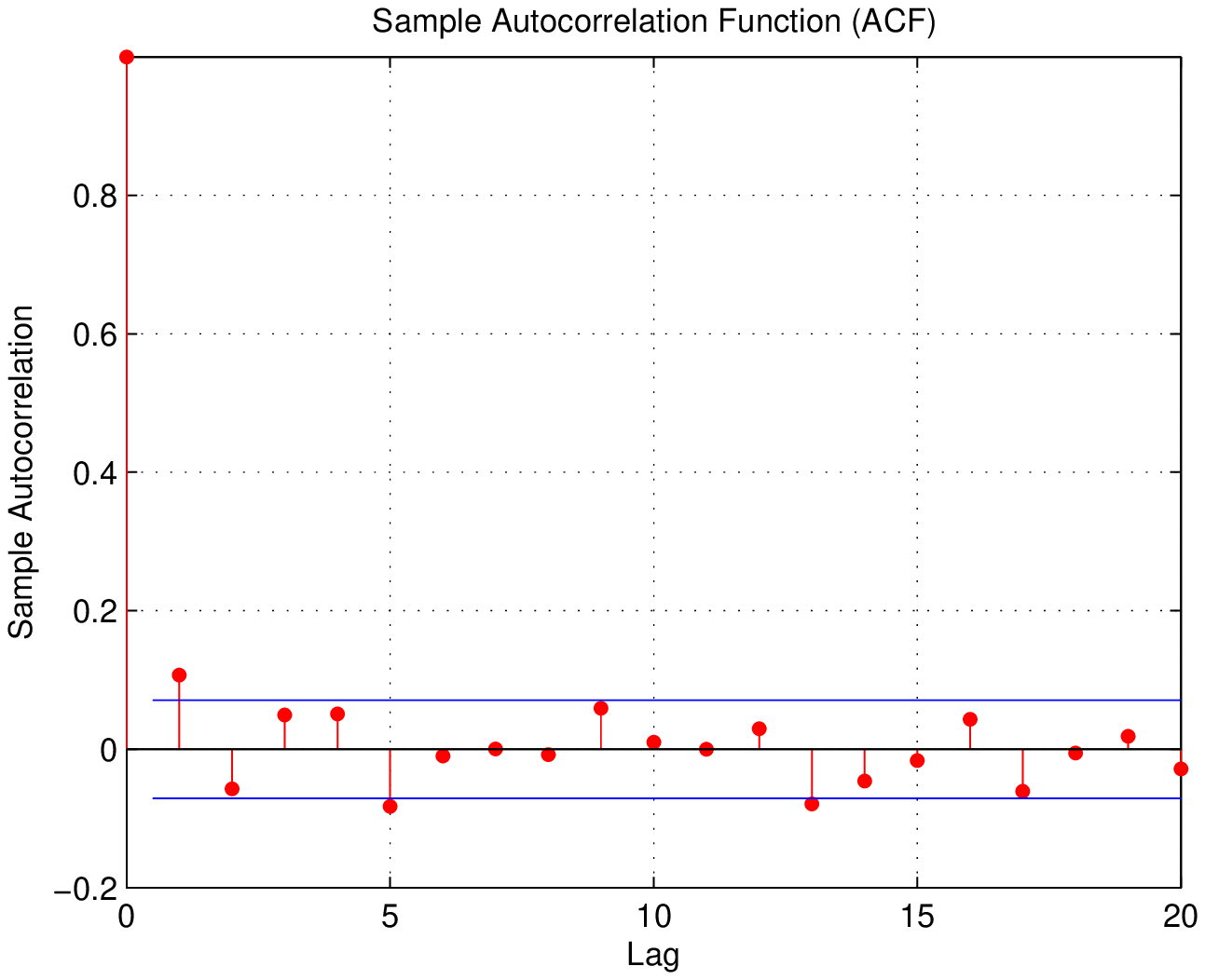}
\includegraphics[width=0.45\textwidth]{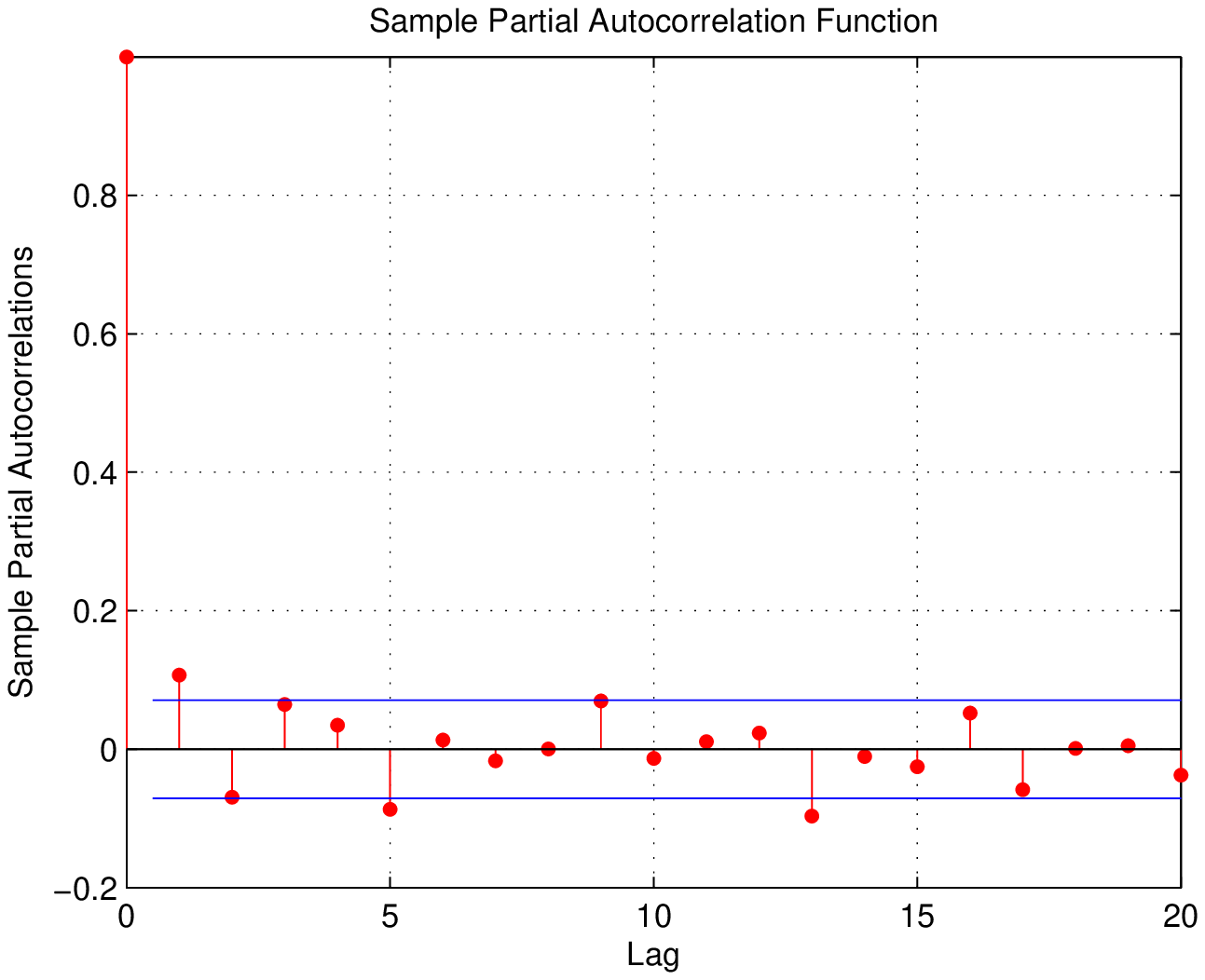}
\caption{Autocorrelation and Partial Autocorrelation Function for FTSE 100}\label{ACF_ParCorr}
\end{center}
\end{figure}
For the FTSE100 one does not observe any significant lags in the historical return time series, which means the independence assumption is acceptable in this example.

\textbf{Maximum Likelihood Estimation}\\

Having tested the properties of independence as well as normality for the underlying historical data one can now proceed to calibrate the parameter set $\Theta=(\mu,\sigma)$ for the GBM based on the historical returns. To find $\Theta$ that yields the best fit to the historical dataset the method of maximum likelihood estimation is used (MLE).\\

MLE can be used for both continuous and discrete random variables. The basic concept of MLE, as suggested by the name, is to find the parameter estimates $\Theta$ for the assumed probability density function $f_\Theta$ (continuous case) or probability mass function (discrete case) that will maximise the likelihood or probability of having observed the given data sample $x_{1},x_{2},x_{3},...,x_{n}$ for the random vector $X_1,...,X_n$. In other terms, the observed sample $x_{1},x_{2},x_{3},...,x_{n}$ is used inside $f_{X_1,X_2,...,X_n;\Theta}$, so that the only variable in $f$ is $\Theta$, and the resulting function is maximised in $\Theta$. The likelihood (or probability) of observing a particular data sample, i.e. the likelihood function, will be denoted by ${\cal L}(\Theta)$.

In general, for stochastic processes, the Markov property is sufficient to be able to write the likelihood along a time series of observations as a product of transition likelihoods on single time steps between two adjacent instants. The key idea there is to notice that for a Markov process $x_t$, one can write, again denoting $f$ the probability density of the vector random sample:
\begin{equation}\label{eq:footnotemarkov}{\cal L}(\Theta) = f_{X(t_0),X(t_1),...,X(t_n);\Theta} = f_{X(t_n)|X(t_{n-1});\Theta}\cdot f_{X(t_{n-1})|X(t_{n-2});\Theta}\cdots f_{X(t_{1})|X(t_{0});\Theta}\cdot f_{X(t_{0});\Theta}
\end{equation}

In the more extreme cases where the observations in the data series are iid, the problem is considerably simplified, since $f_{X(t_i)|X(t_{i-1});\Theta}=f_{X(t_i);\Theta}$ and the likelihood function becomes the product of the probability density of each data point in the sample without any conditioning. In the GBM case, this happens if the maximum likelihood estimation is done on log-returns rather than on the levels. By defining the $X$ as
\begin{equation}\label{eq:XlogreturnS}
X(t_i):= \log S(t_i) - \log S(t_{i-1}),
\end{equation}
 one sees that they are already independent so that one does not need to use explicitly the above decomposition (\ref{eq:footnotemarkov}) through transitions. For general Markov processes different from GBM, or if one had worked at level space $S(t_i)$ rather than in return space $X(t_i)$, this can be necessary though: see, for example, the Vasicek example later on.

The likelihood function for iid variables is in general:
\begin{eqnarray}
	{\cal L}(\Theta)&=&f_{\Theta}\left(x_{1},x_{2},x_{3},...,x_{n} \right)
										=\prod^{n}_{i=1}f_{\Theta} \left(x_{i} \right)
\end{eqnarray}

The MLE estimate $\hat{\Theta}$ is found by maximising the likelihood function. Since the product of density values could become very small, which would cause numerical problems with handling such numbers, the likelihood function is usually converted\footnote{This is allowed, since maxima are not affected by monotone transformations.} to the log likelihood ${\cal L}^{*}=\log {\cal L}$.\footnote{The reader has to be careful not to confuse the log taken to move from levels $S$ to returns $X$ with the log used to move from the likelihood to the log-likelihood. The two are not directly related, in that one takes the log-likelihood in general even in situations when there are no log-returns but just general processes.}
For the iid case the log-likelihood reads:
\begin{eqnarray}
	{\cal L}^{*}(\Theta)=\sum^{n}_{i=1}\log f_{\Theta} \left(x_{i} \right)
\end{eqnarray}

The maximum of the log likelihood usually has to be found numerically using an optimisation algorithm\footnote{For example, Excel Solver or fminsearch in MatLab, although for potentially multimodal likelihoods, such as possibly the jump diffusion cases further on, one may have to preprocess the optimisation through a coarse grid global method such as genetic algorithms or simulated annealing, so as to avoid getting stuck in a local minima given by a lower mode.}. In the case of the GBM, the log-returns increments form normal iid random variables, each with a well known density $f_{\Theta}(x)=f(x;m,v)$, determined by mean and variance. Based on Equation \eqref{eq:logSdistrib} (with $u=t_{i+1}$ and $t=t_i$) the parameters are computed as follows:

\begin{equation}
	m=\left[ \hat{\mu}-\frac{1}{2} \widehat{\sigma^{2}} \right] \Delta t  \qquad\qquad  v=\widehat{\sigma^{2}} \Delta t
\end{equation}

The estimates for the GBM parameters are deduced from the estimates of $m$ and $v$. In the case of the GBM, the MLE method provides closed form expressions for $m$ and $v$ through MLE of Gaussian iid samples\footnote{Hogg and Craig, {\em Introduction to Mathematical Statistics.},
fourth edition, p. 205, example 4.}. This is done by differentiating the Gaussian density function with respect to each parameter and setting the resulting derivative equal to zero. This leads to following well known closed form expressions for the sample mean and variance\footnote{Note that the estimator for the variance is biased, since $\mathbb{E}( \hat{v})=v (n-1)/n$. The bias can be corrected by multiplying the estimator by $n/(n-1)$. However, for large data samples the difference is very small.} of the log returns samples $x_i$ for $X_i=\log S(t_i) - \log S(t_{i-1})$

\begin{equation}
	\hat{m}=\sum^{n}_{i=1}x_{i}/n  \qquad\qquad  \hat{v}=\sum^{n}_{i=1}(x_{i}-\hat{m})^{2}/n .
\end{equation}

Similarly one can find the maximum likelihood estimates for both parameters simultaneously using the following numerical search algorithm. \\

\begin{algorithm}[!h]
\begin{lstlisting}
function [mu sigma] = GBM_calibration(S,dt,params)
    Ret=price2ret(S);
    n=length(Ret);
    options = optimset('MaxFunEvals', 100000, 'MaxIter', 100000);
    fminsearch(@normalLL, params, options);
    function mll = normalLL(params)
        mu=params(1);sigma=abs(params(2));
        ll=n*log(1/sqrt(2*pi*dt)/sigma)+sum(-(Ret-mu*dt).^2/2/(dt*sigma^2));
        mll=-ll;
    end
end
\end{lstlisting}
\caption{$MATLAB^{\circledR}$ MLE Function for iid Normal Distribution.}
\end{algorithm}

An important remark is in order for this important example: given that $x_i = \log s(t_i) - \log s(t_{i-1})$, where $s$ is the observed sample time series for geometric brownian motion $S$, $\hat{m}$ is expressed as:

\begin{equation}
	\hat{m}=\sum^{n}_{i=1}x_{i}/n = \frac{\log(s(t_n))-\log(s(t_0))}{n}
\end{equation}

\emph{The only relevant information used on the whole $S$ sample are its initial and final points. All of the remaining sample, possibly consisting of hundreds of observations, is useless.}\footnote{Jacod, in
{\em Statistics of Diffusion Processes: Some Elements}, CNR-IAMI 94.17, p. 2,
in case $v$ is known, shows that the drift estimation can be statistically consistent, converging in probability to the true drift value, only if the number of points times the time step tends to infinity. This confirms that increasing the observations between two given fixed instants is useless; the only helpful limit would be letting the final time of the sample go to infinity, or let the number of observations increase while keeping a fixed time step}\\

This is linked to the fact that drift estimation for random processes like GBM is extremely difficult. In a way, if one could really estimate the drift one would know locally the direction of the market in the future, which is effectively one of the most difficult problems. The impossibility of estimating this quantity with precision is what keeps the market alive. This is also one of the reasons for the success of risk neutral valuation: in risk neutral pricing one is allowed to replace this very difficult drift with the risk free rate.

The drift estimation will improve considerably for mean reverting processes, as shown in the next sections.

\bigskip

\textbf{Confidence Levels for Parameter Estimates}

For most applications it is important to know the uncertainty around the parameter estimates obtained from historical data, which is statistically expressed by confidence levels. Generally, the shorter the historical data sample, the larger the uncertainty as expressed by wider confidence levels.

For the GBM one can derive closed form expressions for the Confidence Interval (CI) of the mean and standard deviation of returns, using the Gaussian law of independent return increments and the fact that the sum of independent Gaussians is still Gaussian, with the mean given by the sum of the means and the variance given by the sum of variances.  Let $C_{n}=X_{1}+X_{2}+X_{3}+...+X_{n}$, then one knows that $C_{n}\sim N(n m,n v)$ \footnote{ Even without the assumption of \emph{Gaussian}  log returns increments, but knowing only that log returns increments are iid, one can reach a similar Gaussian limit for $C_n$, but only asymptotically in $n$, thanks to the central limit theorem.}. One then has:

\begin{equation}
	P\left(\frac{C_{n}-n m}{\sqrt{v} \sqrt{n}}\leq z \right)=
P\left(\frac{C_{n}/n-m}{\sqrt{v} / \sqrt{n}}\leq z \right)	=\Phi(z)
\end{equation}

where $\Phi()$ denotes the cumulative standard normal distribution. $C_{n}/n$, however, is the MLE estimate for the sample mean $m$. Therefore the 95th confidence level for the sample mean is given by:

\begin{equation}
	\frac{C_{n}}{n}-1.96\frac{\sqrt{v}}{\sqrt{n}} \leq m \leq \frac{C_{n}}{n}+1.96\frac{\sqrt{v}}{\sqrt{n}}
\end{equation}

This shows that as $n$ increases the CI tightens, which means the uncertainty around the parameter estimates declines.\\

Since a standard Gaussian squared is a chi-squared, and the sum of $n$ independent chi-squared is a chi-squared with $n$ degrees of freedom, this gives:


\begin{equation}
	P\left(\sum^{n}_{i=1}\left( \frac{(x_{i}-\hat{m})}{\sqrt{v}} \right)^{2}\leq z \right)
	=P\left(\frac{n}{v} \hat{v} \leq z \right)
		=\chi_{n}^{2}(z),
\end{equation}

hence under the assumption that the returns are normal one finds the following confidence interval for the sample variance:

\begin{equation}
	\frac{n}{q^{\chi}_{U}}\hat{v}  \leq  v \leq  \frac{n}{q^{\chi}_{L}}\hat{v}
\end{equation}

where $q^{\chi}_{L}, q^{\chi}_{U}$ are the quantiles of the chi-squared distribution with $n$ degrees of freedom corresponding to the desired confidence level.

In the absence of these explicit relationships one can also derive the distribution of each of the parameters numerically, by simulating the sample parameters as follows. After the parameters have been estimated (either through MLE or with the best fit to the chosen model, call these the ''first parameters"), simulate a sample path of length equal to the historical sample with those parameters. For this sample path one then estimates the best fit parameters as if the historical sample were the simulated one, and obtains a sample of the parameters. Then again, simulating from the first parameters, one goes through this procedure and obtains another sample of the parameters. By iterating one builds a random distribution of each parameter, all starting from the first parameters. An implementation of the simulated parameter distributions is provided in the next Matlab function. \\

\begin{algorithm}[!h]\label{Bootstrap}
\begin{lstlisting}
%load 'C:\Research\QS CVO Technical Paper\Report\Chapter1\ftse100monthly1940.mat'
load 'ftse100monthly1940.mat'
S=ftse100monthly1940data;
ret=log(S(2:end,2)./S(1:end-1,2));
n=length(ret)

m=sum(ret)/n
v=sqrt(sum((ret-m).^2)/n)

SimR=normrnd(m,v,10000,n);

m_dist=sum(SimR,2)/n;
m_tmp=repmat(m_dist,1,799);
v_dist=sum((SimR-m_tmp).^2,2)/n;

dt=1/12;
sigma_dist=sqrt(v_dist/dt);
mu_dist=m_dist/dt+0.5*sigma_dist.^2;

S0=100;
pct_dist=S0*logninv(0.99,sigma_dist,mu_dist);
\end{lstlisting}
\caption{$MATLAB^{\circledR}$ Simulating distribution for Sample mean and variance.}
\end{algorithm}

\begin{figure}[h!]
\begin{center}
\includegraphics[width=0.45\textwidth]{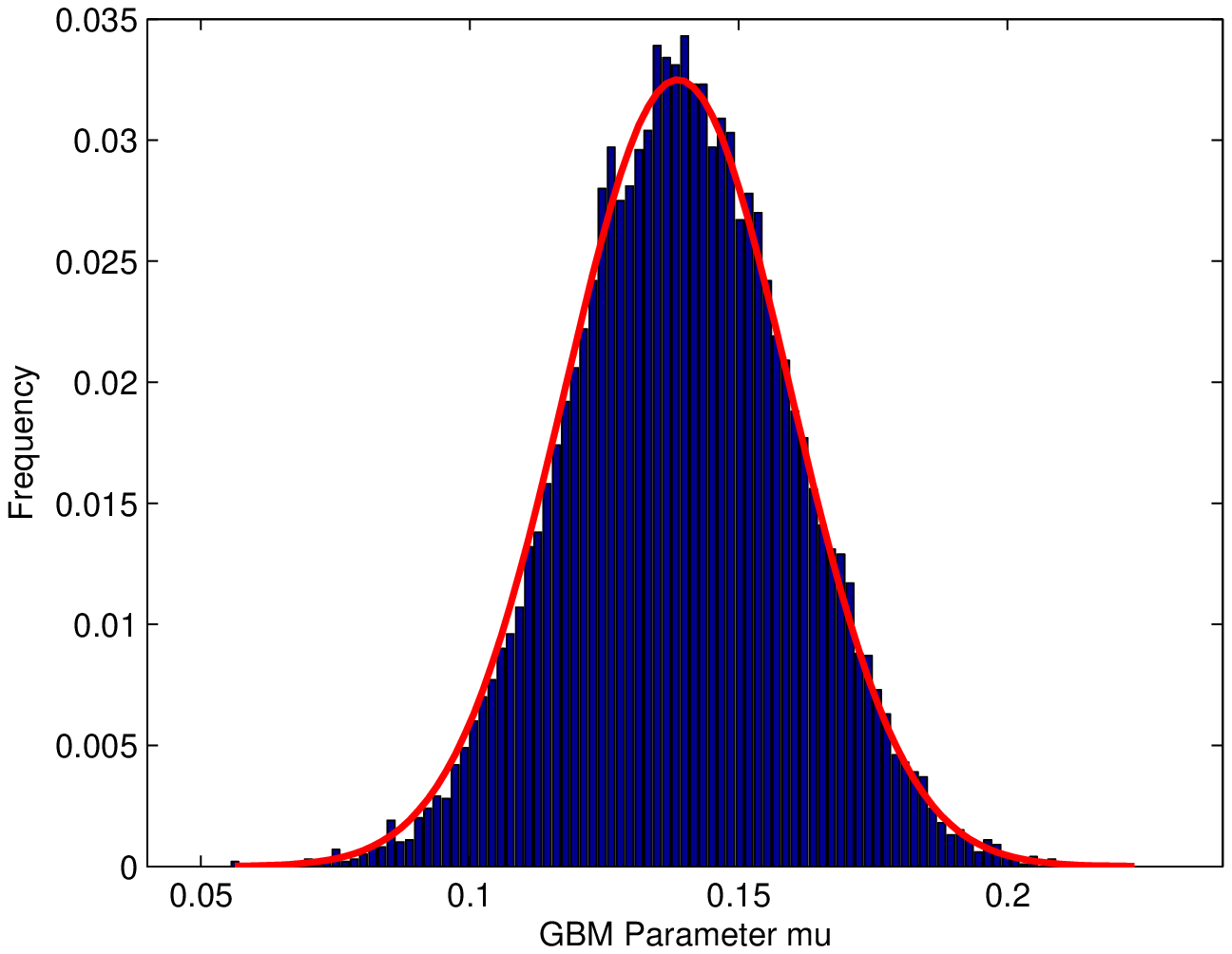}
\includegraphics[width=0.45\textwidth]{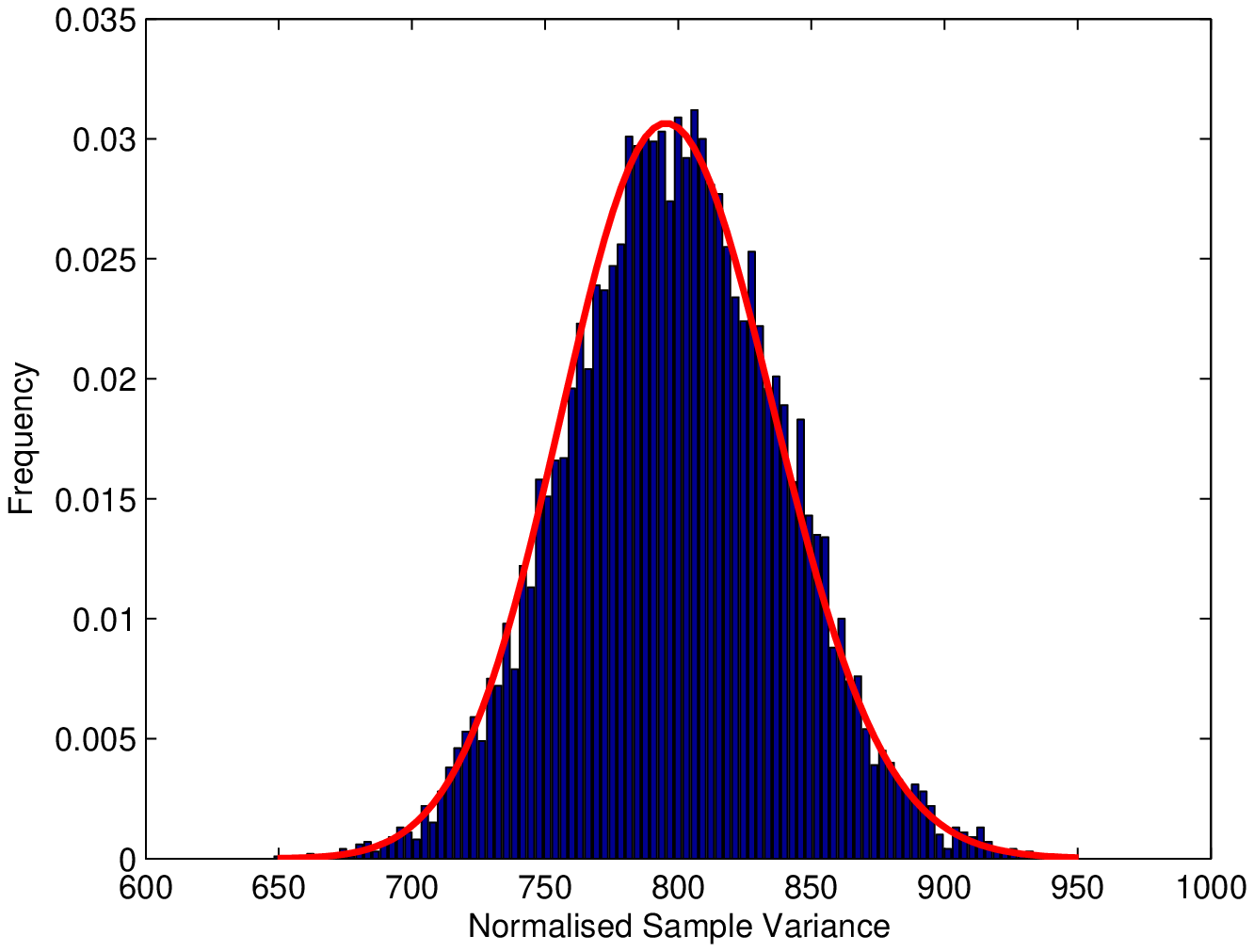}
\caption{Distribution of Sample Mean and Variance}\label{ParaDist}
\end{center}
\end{figure}

\begin{figure}[h!]
\centering
\includegraphics[width=0.45\textwidth]{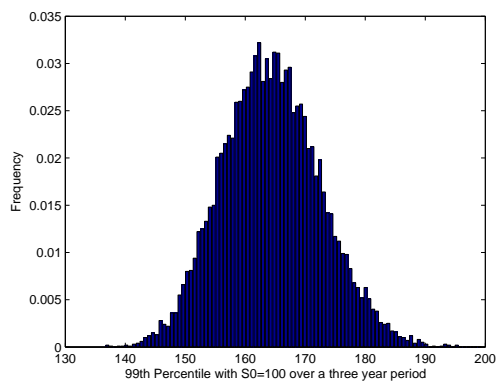}
\caption{Distribution for 99th Percentile of FTSE100 over a three year Period}\label{PctDist}
\end{figure}

The charts in Figure \ref{ParaDist} plot the simulated distribution for the sample mean and sample variance against the normal and Chi-squared distribution given by the theory and the Gaussian distribution. In both cases the simulated distributions are very close to their theoretical limits.
The confidence levels and distributions of individual estimated parameters only tell part of the story. For example, in value-at-risk calculations, one would be interested in finding a particular percentile of the asset distribution. In expected shortfall calculations, one would be interested in computing the expectation of the asset conditional on exceeding such percentile.
In the case of the GBM, one knows the marginal distribution of the asset price to be log normal and one can compute percentiles analytically for a given parameter set. The MATLAB code  \ref{Bootstrap}, which computes the parameter distributions, also simulates the distribution for the 99th percentile of the asset price after three years when re-sampling from simulation according to the first estimated parameters. The histogram is plotted in Figure~\ref{PctDist}.

\subsection{Characteristic and Moment Generating Functions}


Let $X$ be a random variable, either discrete with probability $p_X$ (first case) or continuous with density $f_X$ (second case). The $k-th$ moment $m_k(X)$ is defined as:
\[ m_k(X) = \mathbb{E}(X^k) = \left\{ \begin{array}{l}
         \sum_x x^k p_X(x)\\
        \int_{-\infty}^{\infty} x^k f_X(x)dx.\end{array} \right. \]
Moments are not guaranteed to exist, for example the Cauchy distribution does not even admit the first moment (mean).
The moment generating function $M_X(t)$ of a random variable $X$ is defined, under some technical conditions for existence, as:
\[ M(t)   = \mathbb{E}(e^{tX}) = \left\{ \begin{array}{l}
         \sum_x e^{tx} p_X(x)\\
        \int_{-\infty}^{\infty} e^{tx} f_X(x)dx\end{array} \right. \]
and can be seen to be the exponential generating function of its sequence of moments:
\begin{equation}
M_X(t)  = \sum_{k = 0}^{\infty} \frac{m_k(X)}{k!}t^k.
\end{equation}
The moment generating function (MGF) owes its name to the fact that from its knowledge one may back out all moments of the distribution. Indeed, one sees easily that by differentiating $j$ times the MGF and computing it in $0$ one obtains the $j$-th moment:
\begin{equation}
\frac{d^j}{dt^j} M_X(t)|_{t=0}  = m_j(X) .
\end{equation}

A better and more general definition is that of characteristic function, defined as:
\[\phi_X(t)  = \mathbb{E}(e^{i  t X}),\]
where $i$ is the unit imaginary number from complex numbers theory. The characteristic function exists for every possible density, even for the Cauchy distribution, whereas the moment generating function does not exist for some random variables, notably some with infinite moments. If moments exist, one can derive them also from the characteristic function through differentiation. Basically, the characteristic function forces one to bring complex numbers into the picture but brings in considerable advantages. Finally, it is worth mentioning that the moment generating function can be seen as the Laplace transform of the probability density, whereas the characteristic function is the Fourier transform.

\subsection{Properties of Moment Generating and Characteristic Functions}
If two random variables, $X$ and $Y$, have the same moment generating function $M(t)$, then $X$ and $Y$ have the same probability distribution. The same holds, more generally, for the characteristic function.

If $X$ and $Y$ are independent random variables and $W = X +Y$, then the moment generating function of $W$ is the product of the moment generating functions of $X$ and $Y$:
\begin{equation*}
M_{X+Y}(t)  = \mathbb{E}(e^{t(X+Y)}) = \mathbb{E}(e^{tX}e^{tY})=\mathbb{E}(e^{tX})\mathbb{E}(e^{tY})= M_{X}(t)M_{Y}(t).
\end{equation*}
The same property holds for the characteristic function. This is a very powerful property since the sum of independent returns or shocks is common in modelling. Finding the density of a sum of independent random variables can be difficult, resulting in a convolution, whereas in moment or characteristic function space it is simply a multiplication. This is why typically one moves in moment generating space, computes a simple multiplication and then goes back to density space rather than working directly in density space through convolution.

As an example, if $X$ and $Y$ are two independent random variables and $X$ is normal with mean $\mu_1$ and variance $\sigma_{1}^2$ and $Y$ is normal with mean $\mu_2$ and variance $\sigma_{2}^2$, then one knows that $W = X+Y$ is normal with mean $\mu_1 + \mu_2$ and variance $\sigma_{1}^2 + \sigma_{2}^2$. In fact, the moment generating function of a normal random variable $Z \sim {\cal N}(\mu, \sigma^2)$ is given by:
\begin{equation*}
M_{Z}(t)  = e^{\mu t + \frac{\sigma^2}{2}t^2}
\end{equation*}
Hence
\begin{equation*}
M_{X}(t)  = e^{\mu_1 t + \frac{\sigma_{1}^2}{2}t^2} \quad and \quad  M_{Y}(t)  = e^{\mu_2 t + \frac{\sigma_{2}^2}{2}t^2}
\end{equation*}
From this
\begin{equation*}
M_{W}(t)  = M_{X}(t)M_{Y}(t) = e^{\mu_1 t + \frac{\sigma_{1}^2}{2}t^2}e^{\mu_2 t + \frac{\sigma_{2}^2}{2}t^2} = e^{(\mu_1 +\mu_2) t + \frac{(\sigma_{1}^2 + \sigma_{2}^2)}{2}t^2}
\end{equation*}
one obtains the moment generating function of a normal random variable with mean $(\mu_1 +\mu_2)$ and variance $(\sigma_{1}^2 + \sigma_{2}^2)$, as expected.

\label{CH1}
\section{Fat Tails: GARCH Models}

GARCH\footnote{Generalised Autoregressive Conditional Heteroscedasticity} models were first introduced by \cite{BOLLERSLEV:86}. The assumption underlying a GARCH model is that volatility changes with time and with the past information. In contrast, the GBM model introduced in the previous section assumes that volatility ($\sigma$) is constant. GBM could have been easily generalised to a time dependent but fully deterministic volatility, not depending on the state of the process at any time. GARCH instead allows the volatility to depend on the evolution of the process. In particular the GARCH(1,1) model, introduced by Bollerslev, assumes that the conditional variance (i.e. conditional on information available up to time $t_i$) is given by a linear combination of the past variance $\sigma({t_{i-1}})^{2}$ and squared values of past returns.

\begin{eqnarray}
	\frac{\Delta S(t_i)}{ S(t_i)}&=&  \mu  \Delta t_i +\sigma({t_i}) \Delta W(t_i)\nonumber\\
	\sigma({t_i})^{2}&=&\omega \bar{\sigma}^2+\alpha \sigma({t_{i-1}})^{2}+\beta \epsilon({t_{i-1}})^{2}\nonumber\\
	\epsilon(t_i)^{2}&=& \left(\sigma({t_i}) \Delta W(t_i) \right)^{2}
\end{eqnarray}
where $\Delta S(t_i) = S(t_{i+1}) - S(t_{i})$ and the first equation is just the discrete time counterpart of Equation (\ref{eq:gbm1}).

If one had just written the discrete time version of GBM, $\sigma({t_i})$ would just be a known function of time and one would stop with the first equation above. But in this case, in the above GARCH model $\sigma({t})^{2}$ is assumed itself to be a stochastic process, although one depending only on past information. Indeed, for example
\[ \sigma({t_2})^{2} = \omega \bar{\sigma}^2+ \beta \epsilon({t_{1}})^{2} + \alpha \omega \bar{\sigma}^2+\alpha^2 \sigma({t_{0}})^{2}\]
so that $\sigma({t_2})$ depends on the random variable $\epsilon({t_{1}})^{2}$. Clearly conditional on the information at the previous time $t_1$ the volatility is no longer random, but unconditionally it is, contrary to the GBM volatility.

\begin{figure}[h]
\begin{center}
\includegraphics[width=0.45\textwidth]{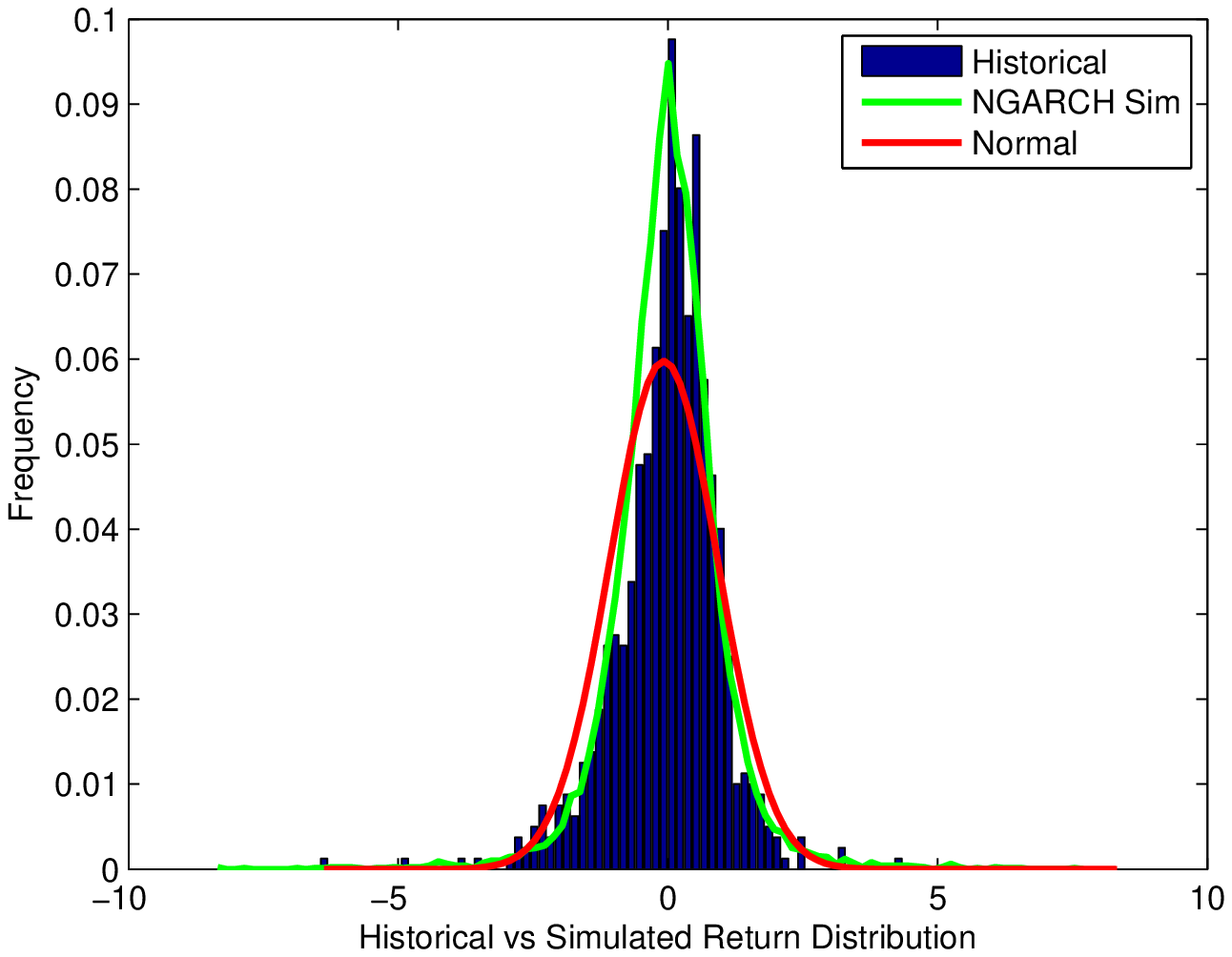}
\includegraphics[width=0.45\textwidth]{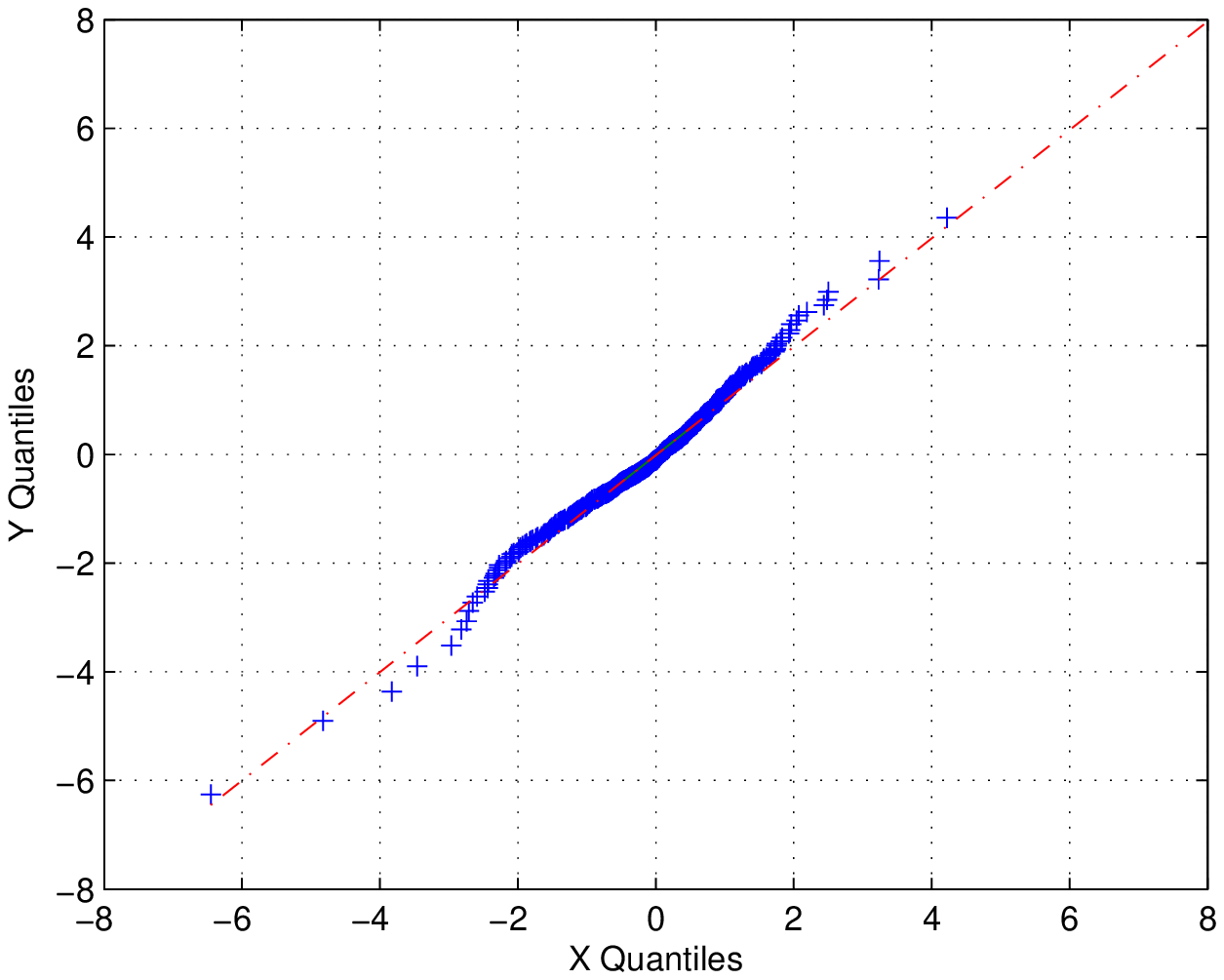}
\caption{Historical FTSE100 Return Distribution vs. NGARCH Return Distribution}\label{QQNGARCH}
\end{center}
\end{figure}

In GARCH the variance is an autoregressive process around a long-term average variance rate $\bar{\sigma}^2$. The GARCH(1,1) model assumes one lag period in the variance. Interestingly, with $\omega=0$ and $\beta=1-\alpha$ one recovers the exponentially weighted        moving average model, which unlike the equally weighted MLE estimator in the previous section, puts more weight on the recent observations. The GARCH(1,1) model can be generalised to a GARCH$(p,q)$ model with $p$ lag terms in the variance and $q$ terms in the squared returns. The GARCH parameters can be estimated by ordinary least squares regression or by maximum likelihood estimation, subject to the constraint $\omega+\alpha+\beta=1$, a constraint summarising of the fact that the variance at each instant is a weighted sum where the weights add up to one.

\begin{figure}[h!t]
\begin{center}
\includegraphics[width=.9\textwidth]{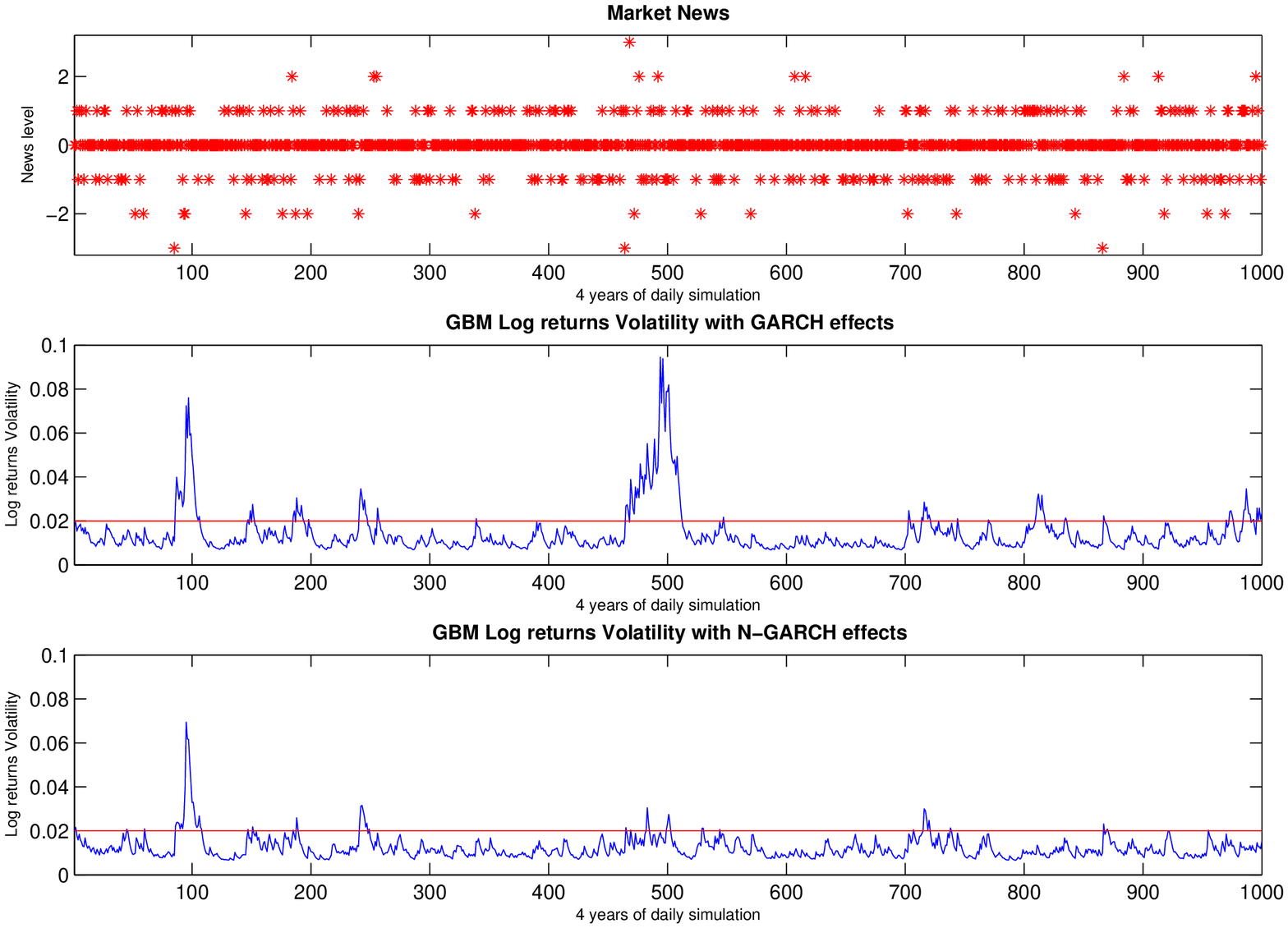}
\includegraphics[width=.9\textwidth]{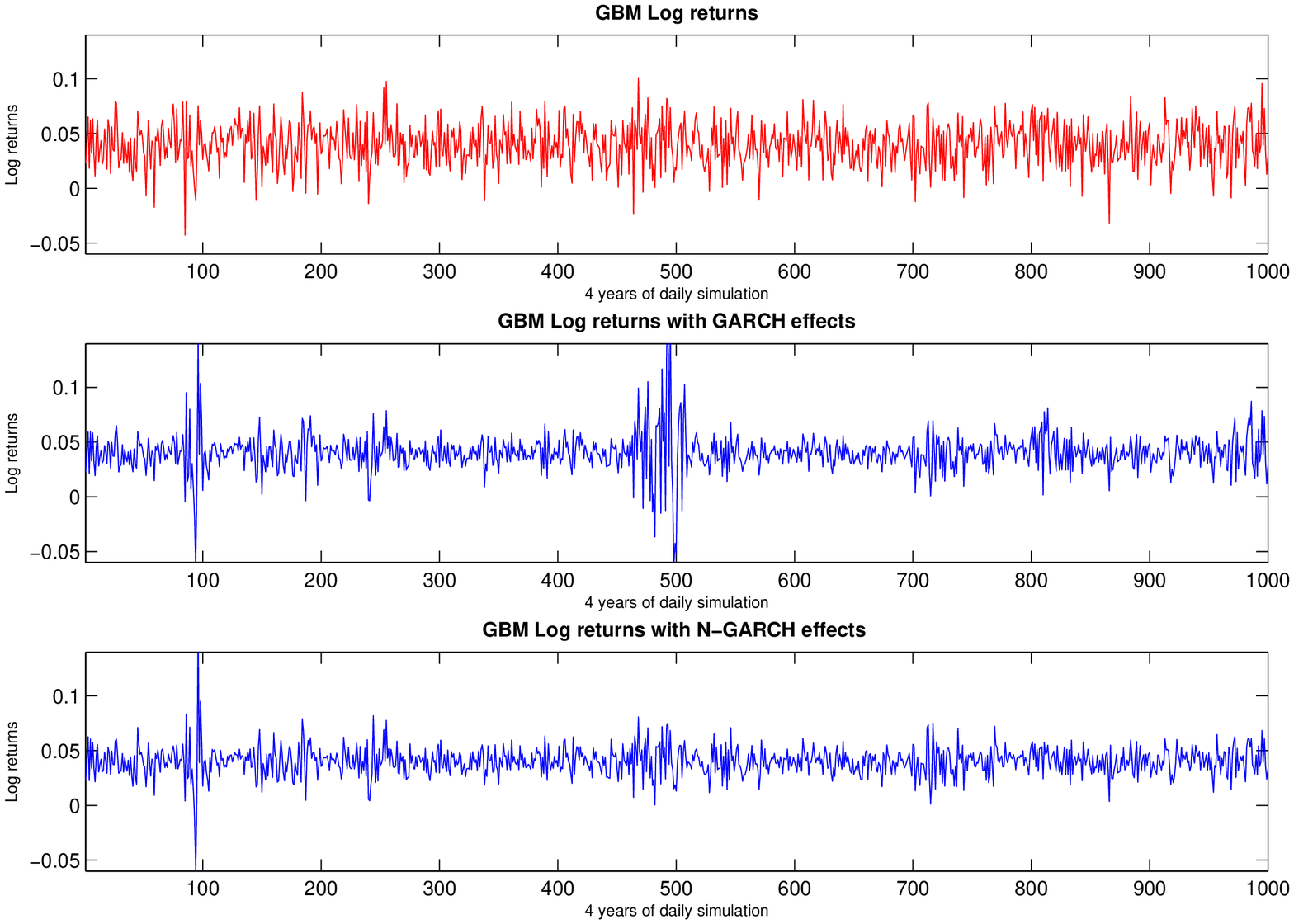}
\caption{NGARCH News Effect vs GARCH News Effect}\label{NGARCHvsGARCH}
\end{center}
\end{figure}

As a result of the time varying state-dependent volatility, the unconditional distribution of returns is non-Gaussian and exhibits so called ``fat" tails. Fat tail distributions  allow the realizations of the random variables having those distributions to assume values that are more extreme than in the normal/Gaussian case, and are therefore more suited to model risks of large movements than the Gaussian. Therefore, returns in GARCH will tend to be ``riskier" than the Gaussian returns of GBM.

GARCH is not the only approach to account for fat tails, and this report will introduce other approaches to fat tails later on.

\begin{figure}[h]
\begin{center}
\includegraphics[width=0.45\textwidth]{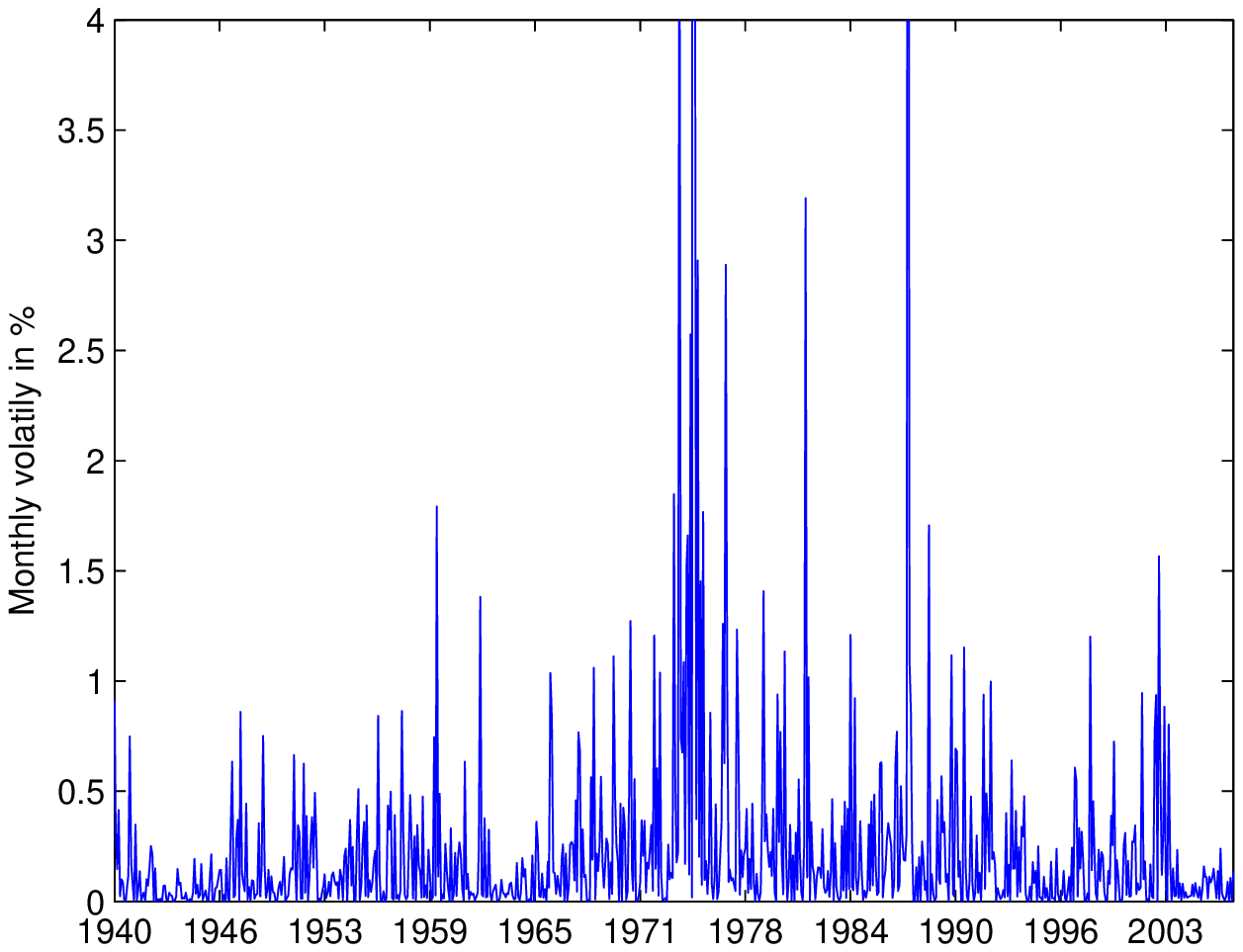}
\includegraphics[width=0.45\textwidth]{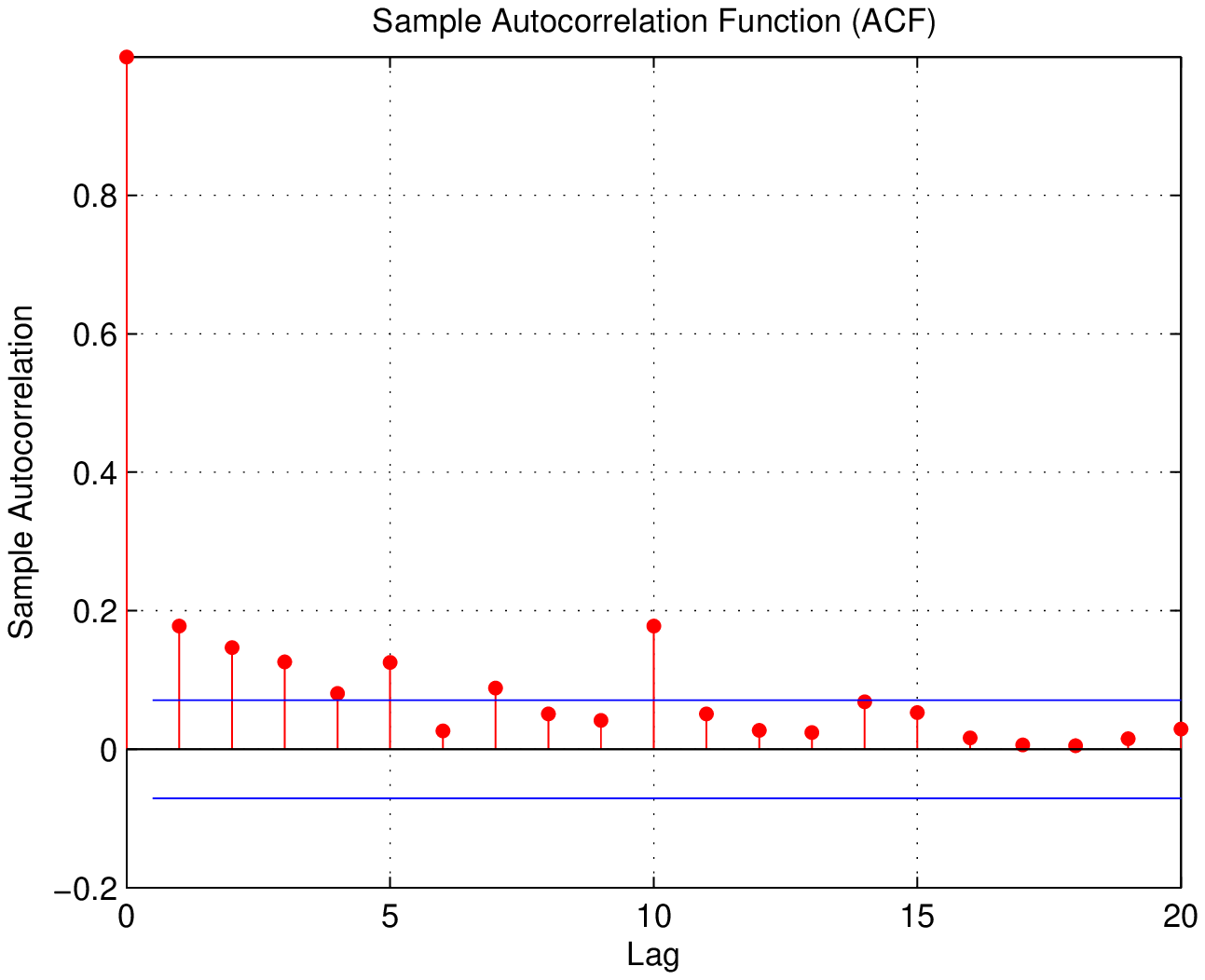}
\caption{Monthly Volatility of the FTSE100 and Autocorrelation Function }\label{ftse100vol}
\end{center}
\end{figure}

The particular structure of the GARCH model allows for volatility clustering (autocorrelation in the volatility), which means a period of high volatility will be followed by high volatility and conversely a period of low volatility will be followed by low volatility. This is an often observed characteristic of financial data. For example, Figure \ref{ftse100vol} shows the monthly volatility of the FTSE100 index and the corresponding autocorrelation function, which confirms the presence of autocorrelation in the volatility.

A long line of research followed the seminal work of Bollerslev leading to customisations and generalisations including exponential GARCH (EGARCH) models, threshold GARCH (TGARCH) models and non-linear GARCH (NGARCH) models among others. In particular, the NGARCH model introduced by \cite{engleng:93} is an interesting extension of the GARCH(1,1) model, since it allows for asymmetric behaviour in the volatility such that ``good news" or positive returns yield subsequently lower volatility, while ``bad news" or negative returns yields a subsequent increase in volatility. The NGARCH is specified as follows:

\begin{eqnarray}
	\frac{\Delta S(t_i)}{ S(t_i)}&=&  \mu  \Delta t_i +\sigma({t_i}) \Delta W(t_i)\nonumber\\
	\sigma({t_i})^{2} &=& \omega +\alpha \sigma({t}_{i-1})^{2}+\beta \left( \epsilon({t_{i-1}})-\gamma \sigma({t_{i-1}})\right)^{2} \nonumber\\
	\epsilon(t_i)^{2}&=& \left(\sigma({t_i}) \Delta W(t_i) \right)^{2}
\end{eqnarray}

NGARCH parameters $\omega$, $\alpha$, and $\beta$ are positive and $\alpha$, $\beta$, and $\gamma$ are subject to the stationarity constraint $\alpha+\beta(1+\gamma^{2})<1$.

Compared to the GARCH(1,1), this model contains an additional parameter $\gamma$, which is an adjustment to the return innovations. Gamma equal to zero leads to the symmetric GARCH(1,1) model, in which positive and negative $\epsilon({t_{i-1}})$ have the same effect on the conditional variance. In the NGARCH $\gamma$ is positive, thus reducing the impact of good news ($\epsilon({t_{i-1}})>0$) and increasing the impact of bad news
($\epsilon({t_{i-1}})<0$) on the variance.
\begin{algorithm}[!h]
\begin{lstlisting}
function S=NGARCH_simulation(N_Sim,T,params,S0)
    mu_=params(1);omega_=params(2);alpha_=params(3);beta_=params(4);gamma_=params(5);
    S=S0*ones(N_Sim,T);
    v0=omega_/(1-alpha_-beta_);
    v=v0*ones(N_Sim,1);
    BM=normrnd(0,1,N_Sim,T);
    for i=2:T
        sigma_=sqrt(v);
        mean=(mu_-0.5*sigma_.^2);
        S(:,i)=S(:,i-1).*exp(mean+sigma_.*BM(:,i));
        v=omega_+alpha_*v+beta_*(sqrt(v).*BM(:,i)-gamma_*sqrt(v)).^2;
    end
end
\end{lstlisting}
\caption{$MATLAB^{\circledR}$ Code to Simulate GBM with NGARCH vol.}\label{GBMNGARCHSim}
\end{algorithm}

MATLAB Routine \ref{GBMNGARCHSim} extends Routine \ref{GBMsim} by simulating the GBM with the NGARCH volatility structure. Notice that there is only one loop in the simulation routine, which is the time loop. Since the GARCH model is path dependent it is no longer possible to vectorise the simulation over the time step. For illustration purpose, this model is calibrated to the historic asset returns of the FTSE100 index. In order to estimate the parameters for the NGARCH geometric Brownian motion the maximum-likelihood method introduced in the previous section is used.\\

Although the variance is changing over time and is state dependent and hence stochastic conditional on information at the initial time, it is locally deterministic, which means that conditional on the information at ${t}_{i-1}$ the variance is known and constant in the next step between time ${t}_{i-1}$ and ${t}_{i}$. In other words, the variance in the GARCH models is locally constant. Therefore, conditional on the variance at ${t}_{i-1}$ the density of the process at the next instant is still normal\footnote{Note that the unconditional density of the NGARCH GBM is no longer normal, but an unknown heavy tailed distribution.}. Moreover, the conditional returns are still independent. The log-likelihood function for the sample $x_1,...,x_i,...,x_n$ of:
\[ X_i = \frac{\Delta S(t_i)}{ S(t_i)} \]
is given by:
\begin{eqnarray}
	{\cal L}^{*}(\Theta)&	=	&\sum^{n}_{i=1}\log f_{\Theta} \left(x_{i};x_{i-1} \right)\\
		f_{\Theta} \left(x_{i};x_{i-1} \right)&	=	& f_{N}(x_i; \mu,\sigma_{i}^{2})=
	\frac{1}{\sqrt{2\pi \sigma_{i}^{2}}}\exp\left(-\frac{(x_{i}-\mu )^{2}}{2\sigma_{i}^{2}}	 \right)\\
	{\cal L}^{*}(\Theta)& = &constant+\frac{1}{2} \left( -\log(\sigma_{i}^{2})-(x_{i}-\mu )^{2}/\sigma_{i}^{2}   \right) \label{loglikfun}
\end{eqnarray}

where $f_{\Theta}$ is the density function of the normal distribution. Note the appearance of $x_{i-1}$ in the function, which indicates that normal density is only locally applicable between successive returns, as noticed previously. The parameter set for the NGARCH GBM includes five parameters, $\Theta=(\mu,\omega,\alpha,\beta,\gamma)$. Due to the time changing state dependent variance one no longer finds explicit solutions for the individual parameters and has to use a numerical scheme to maximise the likelihood function through a solver. To do so one only needs to maximise the term in brackets in (\ref{loglikfun}), since the constant and the factor do not depend on the parameters.

\begin{algorithm}[h!]
\begin{lstlisting}
function [mu_ omega_ alpha_ beta_ gamma_]=NG_calibration(data,params)
    returns = price2ret(data);    
    returnsLength=length(returns);
    options = optimset('MaxFunEvals', 100000, 'MaxIter', 100000);
    fminsearch(@NG_JGBM_LL, params, options);
    function mll = NG_JGBM_LL(params)
        mu_=params(1);omega_=abs(params(2));alpha_=abs(params(3));
        beta_=abs(params(4));gamma_=params(5);
        denum = 1-alpha_-beta_*(1+gamma_^2);
        if denum < 0; mll = intmax; return; end %Variance stationarity test
        var_t = omega_/denum;
        innov = returns(1)-mu_;
        LogLikelihood = log(exp(-innov^2/var_t/2)/sqrt(pi*2*var_t));
        for time=2:returnsLength
            var_t = omega_+alpha_*var_t+beta_*(innov-gamma_*sqrt(var_t))^2;
            if var_t<0; mll = intmax; return; end;
            innov = returns(time)-mu_;
            LogLikelihood = LogLikelihood + log(exp(-innov^2/var_t/2)/sqrt(pi*2*var_t));
        end
        mll = -LogLikelihood;
    end
end
\end{lstlisting}
\caption{$MATLAB^{\circledR}$ Code to Estimate Parameters for GBM NGARCH.}\label{GBMNGARCHMLE}
\end{algorithm}

Having estimated the best fit parameters one can verify the fit of the NGARCH model for the FTSE100 monthly returns using the QQ-plot. A large number of monthly returns using Routine \ref{GBMNGARCHMLE} is first simulated. The QQ plot of the simulated returns against the historical returns shows that the GARCH model does a significantly better job when compared to the constant volatility model of the previous section. The quantiles of the simulated NGARCH distribution are much more aligned with the quantiles of the historical return distribution than was the case for the plain normal distribution (compare QQ-plots in Figures \ref{ftse100} and \ref{QQNGARCH}).

\section{Fat Tails: Jump Diffusion Models}

Compared with the normal distribution of the returns in the GBM model, the log-returns of a GBM model with jumps are often leptokurtotic\footnote{Fat-tailed distribution.}. This section discusses the implementation of a jump-diffusion model with compound Poisson jumps. In this model, a standard time homogeneous Poisson process dictates the arrival of jumps, and the jump sizes are random variables with some preferred distribution (typically Gaussian, exponential or multinomial). \cite{MERTON:85b} applied this model to option pricing for stocks that can be subject to idiosyncratic shocks, modelled through jumps. The model SDE can be written as:

\begin{equation}
    dS(t)=\mu S(t) dt +\sigma S(t) dW(t)+ S(t) dJ_{t}\label{eq:jumpsSDE}\\
\end{equation}

where again $W_{t}$ is a univariate Wiener process and $J_{t}$ is the univariate jump process defined by:
\begin{eqnarray}
    J_{T} = \sum^{N_{T}}_{j=1}(Y_{j} - 1), \ \ \mbox{or}\ \  d J(t) = (Y_{N(t)} -1) dN(t),
\end{eqnarray}
where $(N_{T})_{T\geq0}$ follows a homogeneous Poisson process with intensity $\lambda$, and is thus distributed like a Poisson distribution with parameter $\lambda T$. The Poisson distribution is a discrete probability distribution that expresses the probability of a number of events occurring in a fixed period of time. Since the Poisson distribution is the limiting distribution of a binomial distribution where the number of repetitions of the Bernoulli experiment $n$ tends to infinity, each experiment with probability of success $\lambda/n$, it is usually adopted to model the occurrence of rare events. For a Poisson distribution with parameter $\lambda T$, the Poisson probability mass function is:

\begin{equation*}
		f_P(x,\lambda T)=\frac{\exp(-\lambda T)(\lambda T)^{x}}{x!}, \ \  x=0,1,2,\ldots \end{equation*}
(recall that by convention $0!=1$). The Poisson process $(N_{T})_{T\geq0}$ is counting the number of arrivals in $[1, T]$, and $Y_{j}$ is the size of the $j$-th jump.  The $Y$ are i.i.d log-normal variables ($Y_{j} \sim \exp\left(\mathcal{N}\left(\mu_{Y},\sigma_{Y}^{2}\right)\right)$) that are also independent from the Brownian motion $W$ and the basic Poisson process $N$.\\

Also in the jump diffusion case, and for comparison with (\ref{eq:logS}), it can be useful to write the equation for the logarithm of $S$, which in this case reads
\begin{eqnarray}\label{eq:logSJ}
d \log S(t)&=&\left(\mu-\frac{1}{2}\sigma^{2} \right) dt +\sigma dW(t) + \log(Y_{N(t)}) d N(t)  \ \ \mbox{or, \ equivalently}
\\	\nonumber   d \log S(t)&=&\left(\mu+ \lambda \mu_Y-\frac{1}{2}\sigma^{2} \right) dt +\sigma dW(t) + [\log(Y_{N(t)}) d N(t) - \mu_Y \lambda dt]
\end{eqnarray}
where now both the jump (between square brackets) and diffusion shocks have zero mean, as is shown right below. Here too it is convenient to do MLE estimation in log-returns space, so that here too one defines $X(t_i)$'s according to (\ref{eq:XlogreturnS}).

The solution of the SDE for the levels $S$ is given easily by integrating the log equation:
\begin{equation}
    S(T)=S(0)\exp\left(\left(\mu-\frac{\sigma^2}{2}\right) T +\sigma W(T)\right)\prod_{j=1}^{N(T)}Y_{j}\label{eq:jumpsSDE2}\\
\end{equation}

\begin{algorithm}[!h]
\begin{lstlisting}
function S = JGBM_simulation(N_Sim,T,dt,params,S0)
    mu_star=params(1);sigma_=params(2);lambda_=params(3);
    mu_y_=params(4);sigma_y_=params(5);
    M_simul = zeros(N_Sim,T);
    for t=1:T
        jumpnb = poissrnd(lambda_*dt,N_Sim,1);
        jump = normrnd(mu_y_*(jumpnb-lambda_*dt),sqrt(jumpnb)*sigma_y_);
        M_simul(:,t) = mu_star*dt + sigma_*sqrt(dt)*randn(N_Sim,1) + jump;
    end
    S=ret2price(M_simul',S0)';
end
    
\end{lstlisting}
\caption{$MATLAB^{\circledR}$ Code to Simulate GBM with Jumps.}
\end{algorithm}

The discretisation of the Equation (\ref{eq:jumpsSDE2}) for the given time step $\Delta t$ is:
\begin{equation}
    S(t)=S(t-\Delta t)\exp\left(\left(\mu-\frac{\sigma^2}{2}\right)\Delta t +\sigma\sqrt{\Delta t} \varepsilon_{t}\right)\prod_{j=1}^{n_{t}}Y_{j}
\end{equation}

where $\varepsilon \sim\mathcal{N}\left(0,1\right)$ and $n_{t} = N_{t}-N_{t-\Delta t}$ counts the jumps between time $t-\Delta t$ and $t$. When rewriting this equation for the log returns

$X(t):=\Delta \log(S(t)) =\log(S(t))-\log(S(t-\Delta t))$ we find:
\begin{equation}
    X(t) = \Delta \log(S(t)) =\mu^{*}\Delta t +\sigma\sqrt{\Delta t}\ \varepsilon_{t}+\Delta J^{*}_{t} \label{eq:jumpsEQ}\\
\end{equation}

\begin{figure}[h]
\begin{center}
\includegraphics[width=0.9\textwidth]{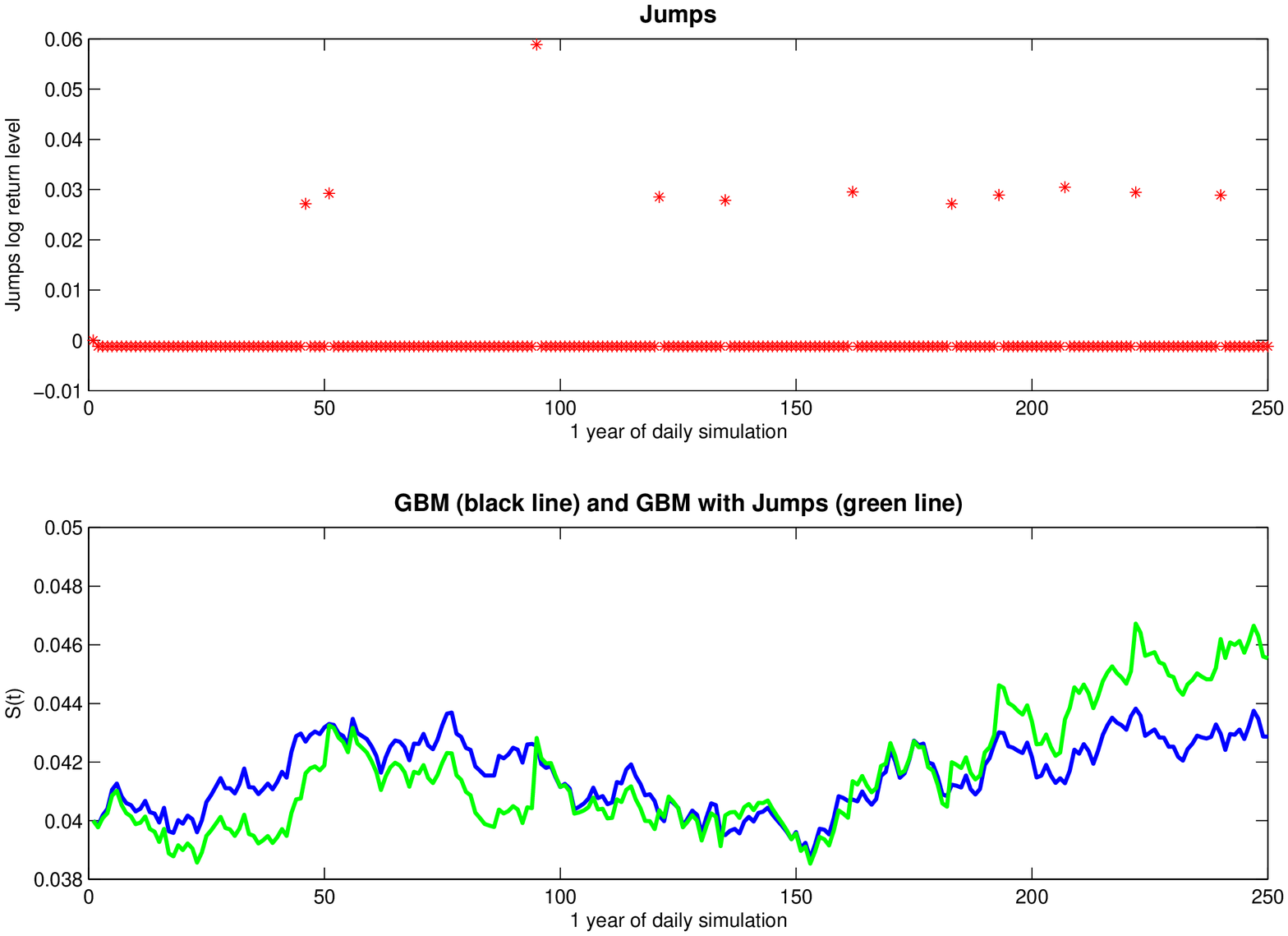}
\includegraphics[width=0.9\textwidth]{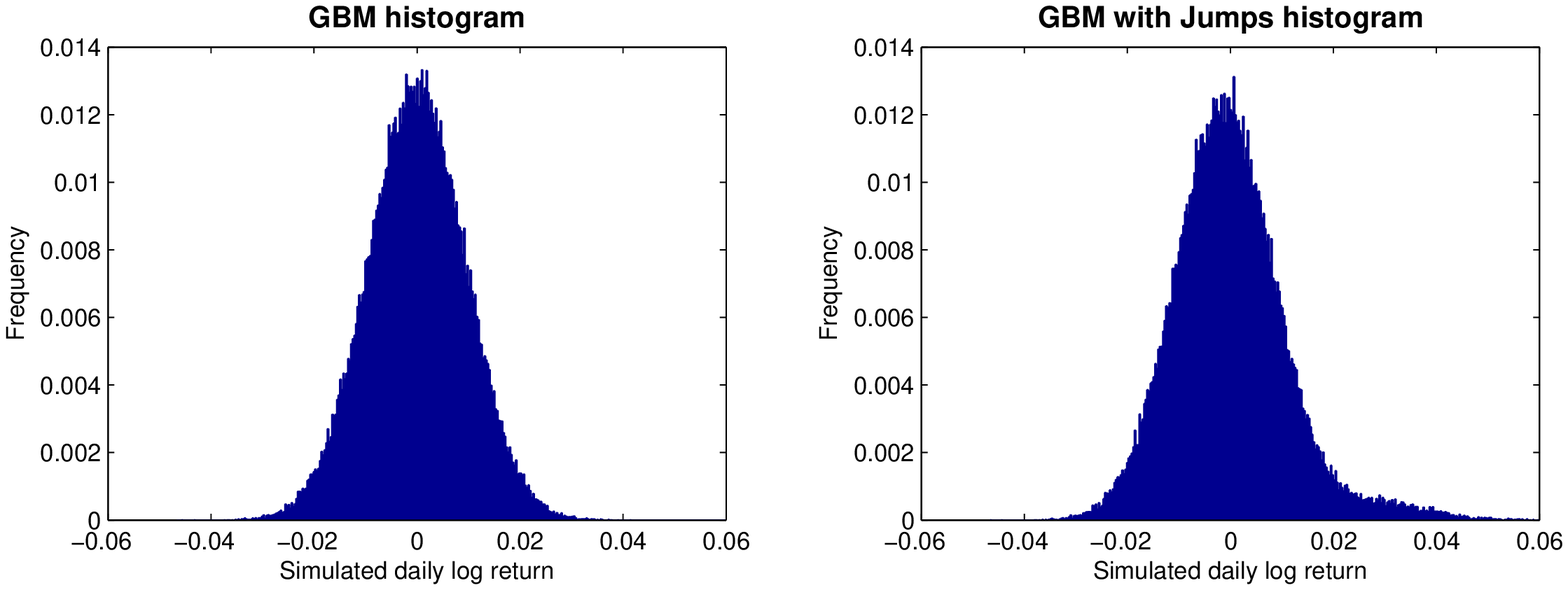}
\caption{The Effect of the Jumps on the Simulated GBM path and on the Density Probability of the Log Returns. $\mu^{*}=0$, $\sigma=0.15$, $\lambda=10$, $\mu_{Y}=0.03$, $\sigma_{Y}=0.001$}
\end{center}
\end{figure}

where the jumps $\Delta J^{*}_{t}$  in the time interval $\Delta t$ and the drift $\mu^{*}$ are defined as:
\begin{eqnarray}
     \Delta J^{*}_{t}= \sum^{n_{t}}_{j=1} \log(Y_{j}) - \lambda\ \Delta t\ \mu_{Y}, \ \ \  \mu^{*} = \left(\mu+ \lambda \mu_Y-\frac{1}{2}\sigma^{2} \right)  \label{eq:jumps}
\end{eqnarray}
so that the jumps $\Delta J^{*}_{t}$ have zero mean.
Since the calibration and simulation are done in a discrete time framework, we use the fact that for the Poisson process with rate $\lambda$, the number of jump events occurring in the fixed small time interval $\Delta t$ has the expectation of $\lambda \Delta t$. Compute
 \begin{eqnarray}
    \mathbb{E}(\Delta J^{*}_{t}) &=& E\left(\sum^{n_{t}}_{j=1} \log Y_{j}\right) - \lambda\ \Delta t\  \mu_{Y} =E\left(E\left(\sum^{n_{t}}_{j=1} \log Y_{j}|n_{t}\right)\right)- \lambda\ \Delta t\  \mu_{Y}\nonumber\\
    &=&E\left(n_{t}\mu_{Y}\right)-\lambda\ \Delta t\  \mu_{Y}=0\nonumber
\end{eqnarray}

The simulation of this model is best done by using Equation (\ref{eq:jumpsEQ}).

\begin{algorithm}[!h]
\begin{lstlisting}
function [mu_star sigma_ lambda_ mu_y_ sigma_y_] = JGBM_calibration(data,dt,params)
    returns = price2ret(data);
    dataLength=length(data);
    options = optimset('MaxFunEvals', 100000, 'MaxIter', 100000);
    params = fminsearch(@FT_JGBM_LL, params, options);
    mu_star=params(1);sigma_=abs(params(2));lambda_=abs(params(3));
    mu_y_=params(4);sigma_y_=abs(params(5));
    
    function mll = FT_JGBM_LL(params)
	    mu_=params(1);sigma_=abs(params(2));lambda_=abs(params(3));
	    mu_y_=params(4);sigma_y_=abs(params(5));
        Max_jumps = 5;
        factoriel = factorial(0:Max_jumps);
        LogLikelihood = 0;
        for time=1:dataLength
        ProbDensity = 0;
            for jumps=0:Max_jumps-1
                jumpProb = exp(-lambda_*dt)*(lambda_*dt)^jumps/factoriel(jumps+1);
                condVol = dt*sigma_^2+jumps*sigma_y_^2;
                condMean= mu_*dt+mu_y_*(jumps-lambda_*dt);
                condToJumps = exp(-(data(time)-condMean)^2/condVol/2)/sqrt(pi*2*condVol);
                ProbDensity = ProbDensity + jumpProb*condToJumps;
            end
            LogLikelihood = LogLikelihood + log(ProbDensity);
        end
        mll = -LogLikelihood;
        
    end
end
\end{lstlisting}
\caption{$MATLAB^{\circledR}$ Code to Calibrate GBM with Jumps.}
\end{algorithm}

Conditional on the jumps occurrence $n_{t}$, the jump disturbance $\Delta J^{*}_{t}$ is normally distributed \[\Delta J^{*}_{t}|_{n_t} \sim\mathcal{N}\left((n_{t}-\lambda \Delta t) \mu_{Y},n_{t}\sigma_{Y}^{2}\right).\] Thus the conditional distribution of $X(t)=\Delta \log(S(t))$ is also normal and the first two conditional moments are:

 \begin{eqnarray}
    \mathbb{E}(\Delta \log(S(t))|n_{t}) &=& \mu^{*}\Delta t + (n_{t}-\lambda\ \Delta t)\mu_{Y} = \left( \mu - \sigma^2/2 \right) \Delta t + n_t \mu_Y, \label{eq:cm}\\
    {\mbox{Var}}(\Delta \log(S(t))|n_{t})&=& \sigma^{2}\Delta t + n_{t}\sigma_{Y}^{2}\label{eq:cv}
\end{eqnarray}

The probability density function is the sum of the conditional probabilities density weighted by the probability of the conditioning variable: the number of jumps. Then the calibration of the model parameters can be done using the MLE on log-returns $X$ conditioning on the jumps.

\begin{eqnarray}
	&&\argmax_{\mu^{*},\mu_{Y},\lambda>0,\sigma>0,\sigma_{Y}>0} {\cal{L}}^{*}=\log({\cal L})
\end{eqnarray}

The log-likelihood function for the returns $X(t)$ at observed times $t=t_1,\ldots,t_n$ with values $x_1,\ldots,x_n$, with $\Delta t = t_i - t_{i-1}$,  is then given by:

\begin{figure}[t]
\begin{center}
\includegraphics[width=0.475\textwidth]{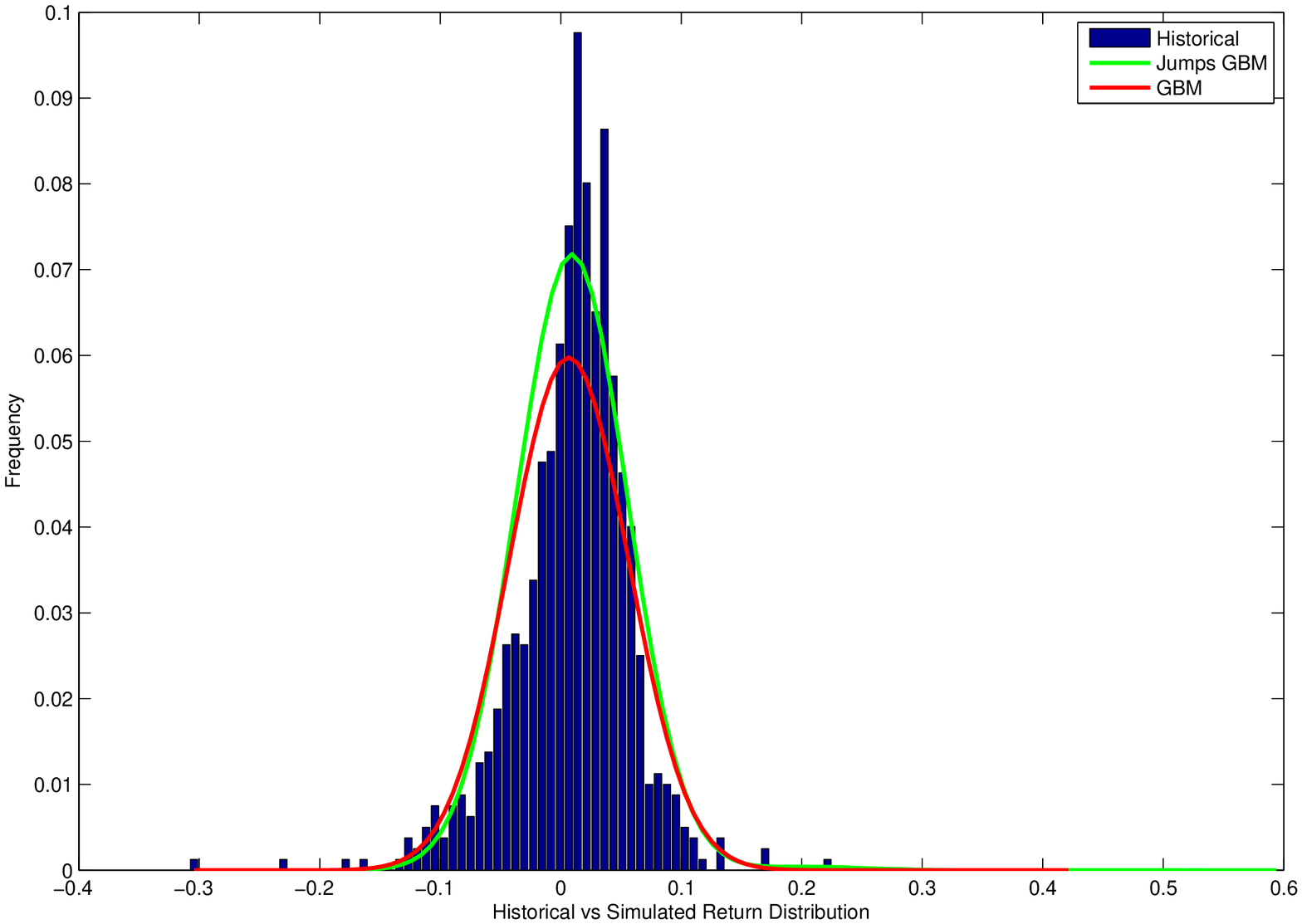}
\includegraphics[width=0.475\textwidth]{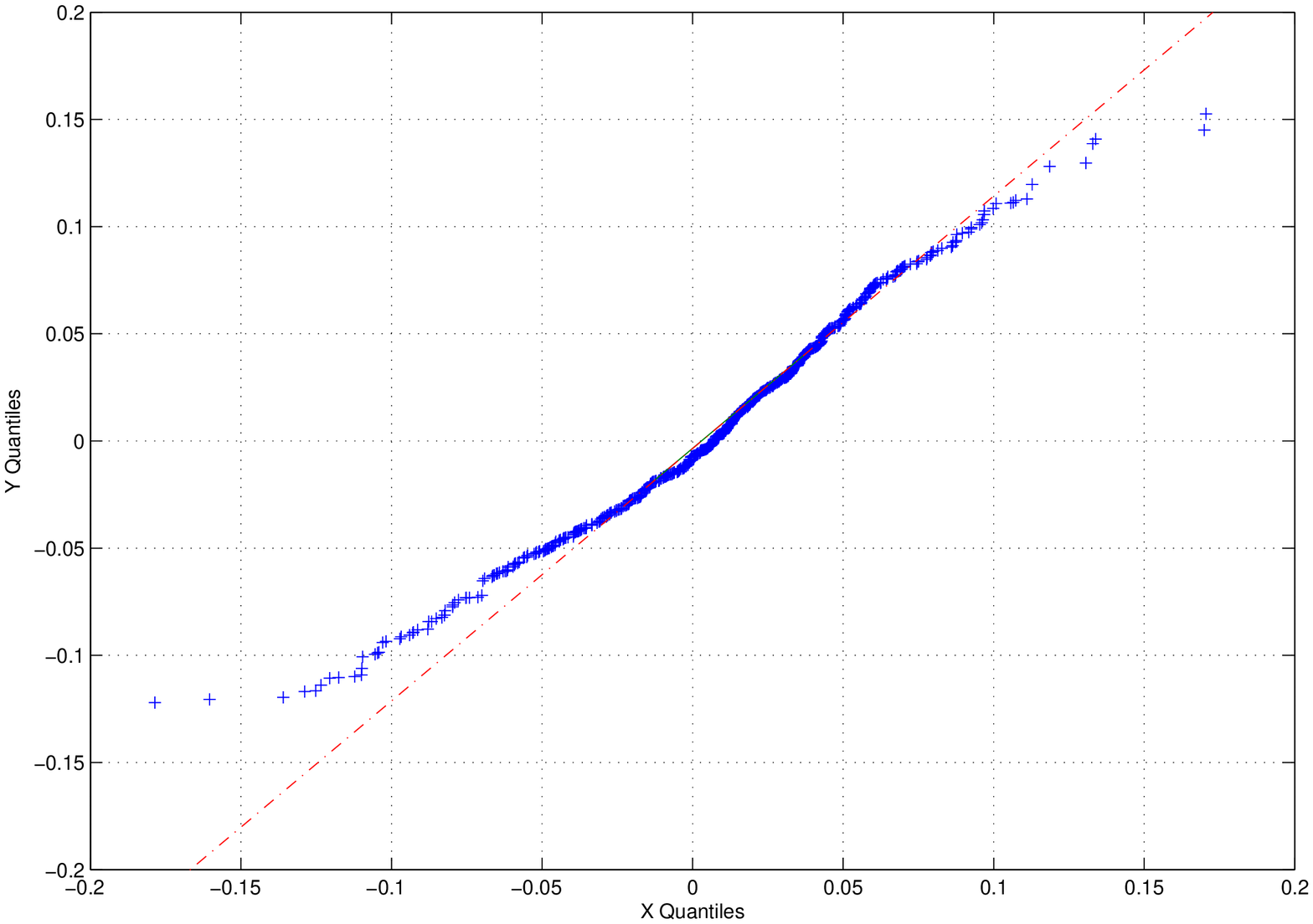}
\caption{Historical FTSE100 Return Distribution vs. Jump GBM Return Distribution}\label{fig:JGBM}
\end{center}
\end{figure}

\begin{eqnarray}
	{\cal{L}}^{*}(\Theta)&	=&\sum^{n}_{i=1}\log f \left(x_{i};\mu,\mu_{Y},\lambda,\sigma,\sigma_{Y}\right)\\
		f\left(x_{i};\mu,\mu_{Y},\lambda,\sigma,\sigma_{Y}\right)&=& \sum^{+\infty}_{j=0}P(n_{t}=j)f_{\cal N}\left(x_{i};\left( \mu - \sigma^2/2 \right) \Delta t + j \mu_Y, \sigma^{2}\Delta t + j\sigma_{Y}^{2}\right)
\end{eqnarray}

This is an infinite mixture of Gaussian random variables, each weighted by a Poisson probability $P(n_{t}=j) = f_P(j;\lambda \Delta t)$.
Notice that if $\Delta t$ is small, typically the Poisson process jumps at most once. In that case, the above expression simplifies in
\begin{eqnarray*}
		f\left(x_{i};\mu,\mu_{Y},\lambda,\sigma,\sigma_{Y}\right) &=& (1 - \lambda \Delta t) f_{\cal N}\left(x_{i};\left( \mu - \sigma^2/2 \right) \Delta t, \sigma^{2}\Delta t\right)\\ &+& \lambda \ \Delta t \ f_{\cal N}\left(x_{i};\left( \mu - \sigma^2/2 \right) \Delta t +  \mu_Y, \sigma^{2}\Delta t + \sigma_{Y}^{2}\right)
\end{eqnarray*}
i.e. a mixture of two Gaussian random variables weighted by the probability of zero or one jump in $\Delta t$.

 The model parameters are estimated by MLE for FTSE100 on monthly returns and then different paths are simulated. The goodness to fit is seen by showing the QQ-plot of the simulated monthly log returns against the historical FTSE100 monthly returns in Fig. \ref{fig:JGBM}. This plot shows that the Jumps model fits the data better then the simple GBM model.

Finally, one can add markedly negative jumps to the arithmetic Brownian motion return process (\ref{eq:logSJ}), to then exponentiate it to obtain a new level process for $S$ featuring a more complex jump behaviour. Indeed, if $\mu_Y$ is larger than one and $\sigma^2_Y$ is not large,  jumps in the log-returns will tend to be markedly positive only, which may not be realistic in some cases.  For details on how to incorporate negative jumps and on the more complex mixture features this induces, the reader can refer to Section~\ref{sec:meanrevjumps}, where the more general mean reverting case is described.
\label{CH2}
\newcommand{\mubar}{\bar{\mu}}
\newcommand{\sigbar}{\bar{\sigma}}

\section{Fat Tails Variance Gamma (VG) process}

 The two previous sections have addressed the modelling of phenomena where numerically large changes in value (\lq\lq fat tails") are more likely than in the case of the normal distribution. This section presents yet another method to accomplish fat tails, and this is through a process obtained by \emph{time changing} a Brownian motion (with drift $\mubar$ and volatility $\sigbar$) with an independent subordinator, that is an increasing process with independent and stationary increments. In a way, instead of making the volatility state dependent like in GARCH or random like in stochastic volatility models, and instead of adding a new source of randomness like jumps, one makes the time of the process random through a new process, the subordinator. A possible related interpretative feature is that financial time, or market activity time, related to trading, is a random transform of calendar time.

Two models based on Brownian subordination are the Variance Gamma (VG) process and Normal Inverse Gaussian (NIG) process. They are a subclass of the generalised hyperbolic (GH) distributions. They are characterised by drift parameters, volatility parameter of the Brownian motion, and a variance parameter of the subordinator.

Both VG and NIG have exponential tails and the tail decay rate is the same for both VG and NIG distributions and it is smaller than for the Normal distribution.

Therefore these models allow for more flexibility in capturing fat tail features and high kurtosis of some historical financial distributions with respect to GBM.

Among fat tail distributions obtained through subordination, this section will focus on the VG distribution, that can also be thought of as a mixture of normal distributions, where the mixing weights density is given by the Gamma distribution of the subordinator. The philosophy is not so different from the GBM with jumps seen before, where the return process is also a mixture. The VG distribution was introduced in the financial literature by\footnote{For a more advanced introduction and background on the above processes
refer to \cite{Seneta2004,ContTan04,TM2006} and \cite{vgMCC1998}.} \cite{MadSen90}.

The model considered for the log-returns of an asset price is of this type:
\begin{equation}\label{VGassetEq}
d \log S(t) =  \mubar\ dt + {\bar{\theta}} \ d g(t) + \sigbar\ d W(g(t))
\hspace{1cm} S(0)=S_0
\end{equation}
where $\mubar$, ${\bar{\theta}} $ and $\sigbar$ are real constants and $\sigbar \geq 0$\footnote{In addition one may say that (\ref{VGassetEq}) is a $L\grave{e}vy$ process of finite variation but of relatively low activity with small jumps.}.
This model differs from the \lq \lq usual" notation of Brownian motion mainly in the term $g_t$. In fact one introduces $g_t$ to characterise the market activity time.
One can define the \lq market time' as a positive increasing random process, $g(t)$,  with stationary increments $g(u) - g(t)$ for $u\ge  t \ge 0$.  In probabilistic jargon $g(t)$ is a subordinator. An important assumption is that $\mathbb{E}[g(u)-g(t) ] = u-t$. This means that the market time has to reconcile with the calendar time between $t$ and $u$ on average. Conditional on the subordinator $g$, between two instants $t$ and $u$:
\begin{equation}
\sigbar (\ W(g(u))- W(g(t))\ )|_g \sim \sigbar \sqrt{(g(u)- g(t))} \epsilon,
\end{equation}
where $\epsilon$ is a standard Gaussian, and hence:
\begin{equation}\label{eq:densityVGcond}
(\log (S(u)/S(t)) - \mubar (u- t))|g \sim N({\bar{\theta}} (g(u)-g(t)), \sigbar^2(g(u)- g(t)))
\end{equation}

VG assumes $\{g(t)\} \sim \Gamma(\frac{t}{\nu},{\nu})$, a Gamma process with parameter $\nu$ that is independent from the standard Brownian motion $\{W_t\}_{t\geq 0}$.
Notice also that the gamma process definition implies
$\{g(u)-g(t)\} \sim \Gamma(\frac{u-t}{\nu},{\nu})$ for any $u$ and $t$. Increments are independent and stationary Gamma random variables. Consistently with the above condition on the expected subordinator, this Gamma assumption implies
$\mathbb{E}[g(u)-g(t)]=u-t$ and Var$(g(u)-g(t)) = \nu (u-t)$.

Through iterated expectations, one has easily that:
\[ \mathbb{E}[\log(S(u)) - \log(S(t))] = (\mubar  + {\bar{\theta}} )(u-t), \ \ \
\mbox{Var}(\log S(u)- \log S(t))  = (\sigbar^2 + {\bar{\theta}} ^2 \nu ) (u-t).\]
Mean and variance of these log-returns can be compared to the log-returns of GBM in Equation (\ref{eq:logSdistrib}), and found to be the same if
$(\mubar  + {\bar{\theta}} ) = \mu - \sigma^2/2$ and $\sigbar^2 + {\bar{\theta}} ^2 \nu = \sigma^2$. In this case, the differences will be visible with higher order moments.

\begin{algorithm}[!h]\label{VG_simulation}
\begin{lstlisting}
function S = VG_simulation(N_Sim,T,dt,params,S0)
theta = params(1); nu = params(2); sigma = params(3); mu = params(4);
g = gamrnd(dt/nu,nu,N_Sim,T);
S = S0*cumprod([ones(1,T); exp(mu*dt + theta*g + sigma*sqrt(g).*randn(N_Sim,T))]); 

\end{lstlisting}
\caption{$MATLAB^{\circledR}$ code that simulate the asset model in eq. (\ref{VGassetEq}). }\label{VG}
\end{algorithm}
\noindent

The process in equation (\ref{VGassetEq}) is called a Variance-Gamma process.

\subsection{Probability Density Function of the VG Process}
The unconditional density may then be obtained by integrating out $g$ in Eq~(\ref{eq:densityVGcond}) through its Gamma density. This corresponds to the following density function, where one sets $u-t =: \Delta t$ and $X_{\Delta t} := \log (S(t+\Delta t)/S(t))$:
\begin{equation}\label{densVG}
f_{(X_{\Delta t})}(x)=\int_0^{\infty} f_N(x;{\bar{\theta}}  g,\sigbar^2 g) f_\Gamma\left(g;\frac{\Delta t}{\nu},{\nu}\right) dg
\end{equation}
As mentioned above, this is again a (infinite) mixture of Gaussian densities, where the weights are given by the Gamma densities rather than Poisson probabilities.
The above integral converges and the PDF of the VG process $X_{\Delta t}$ defined in Eq. (\ref{VGassetEq}) is
\begin{equation}\label{pdfVG}
f_{(X_{\Delta t})}(x)= \frac{2e^{\frac{{\bar{\theta}} (x-\mubar)}{\sigbar^2}}}{\sigbar \sqrt{2 \pi} \nu^{\frac{\Delta t}{\nu}}\Gamma(1/\nu)}\Bigg(\frac{|x-\mubar|}{\sqrt{\frac{2\sigbar^2}{\nu} + {\bar{\theta}} ^2}} \Bigg)^{\Delta t/\nu -1/2}K_{\Delta t/\nu -1/2}\Bigg(\frac{|s-\mubar| \sqrt{\frac{2\sigbar^2}{\nu} + {\bar{\theta}} ^2}}{\sigbar^2} \Bigg)
\end{equation}
Here $K_{\eta}(\cdot)$ is a modified Bessel function of the third kind with index $\eta$, given for $\omega>0$ by
\begin{equation*}
K_{\eta}(x)= \frac{1}{2} \int^{\infty}_{0} y^{\eta -1} \exp\left\{- \frac{x}{2}(y^{-1} + y)\right\}dy
\end{equation*}

\begin{algorithm}[!h]
\begin{lstlisting}
function fx = VGdensity(x, theta, nu, sigma, mu, T )
    
    v1 = 2*exp((theta*(x-mu)) / sigma^2) / ( (nu^(T/nu)) * sqrt(2*pi) * sigma * gamma(T/nu) );
    M2 = (2*sigma^2)/nu + theta^2;
    v3 = abs(x-mu)./sqrt(M2);    
    v4 = v3.^(T/nu - 0.5) ;  
    v6 = (abs(x-mu).*sqrt(M2))./sigma^2;
    K  = besselk(T/nu - 0.5, v6);
    fx = v1.*v4.*K;   
    


\end{lstlisting}
\caption{$MATLAB^{\circledR}$ code of the variance gamma density function as in equation (\ref{pdfVG}). }
\end{algorithm}

\begin{figure}[ht]
    \begin{minipage}[b]{0.5\linewidth}
        \centering
        \includegraphics[scale=.5]{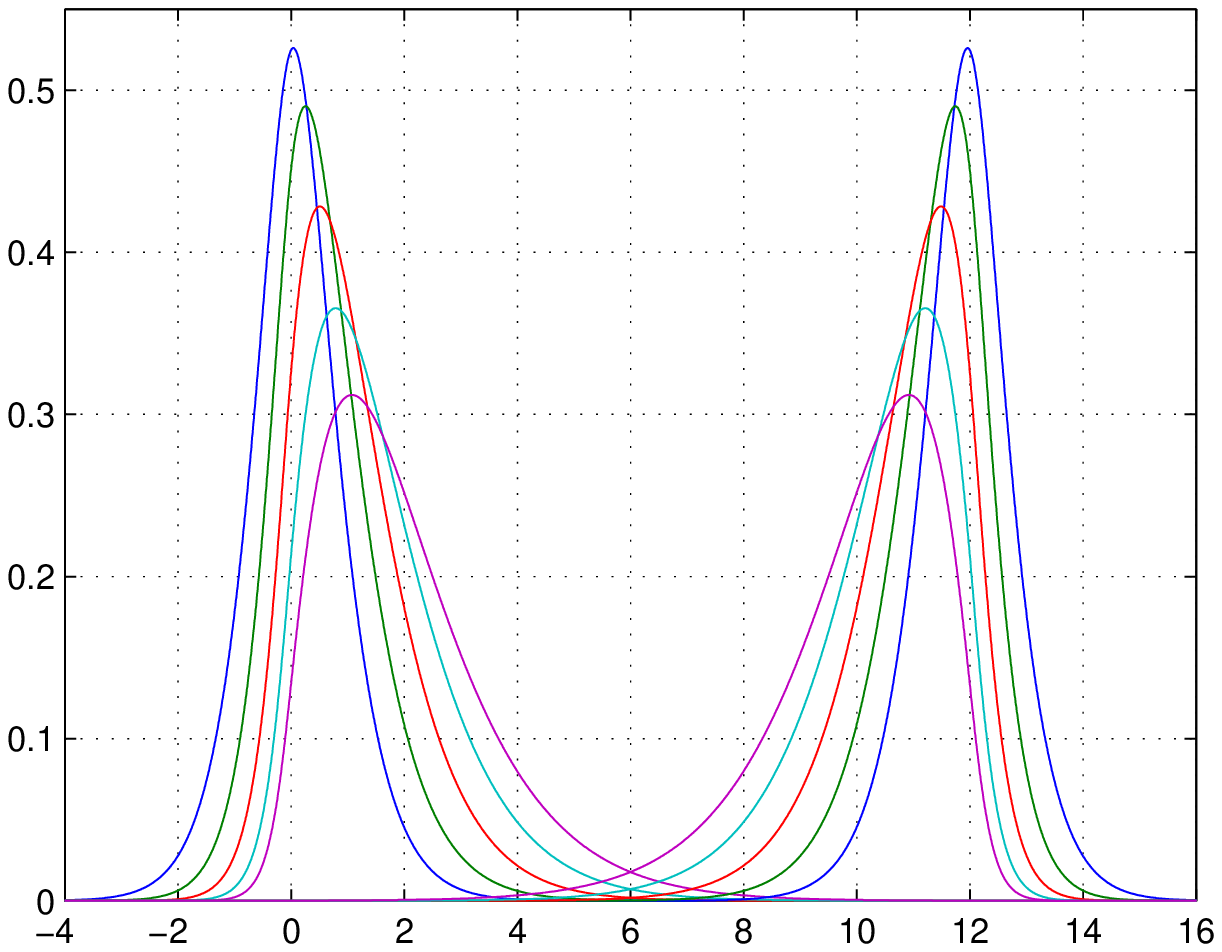}
        \label{VGd}
    \end{minipage}
    \hspace{0.5cm}
    \begin{minipage}[b]{0.5\linewidth}
        \centering
        \includegraphics[scale=.5]{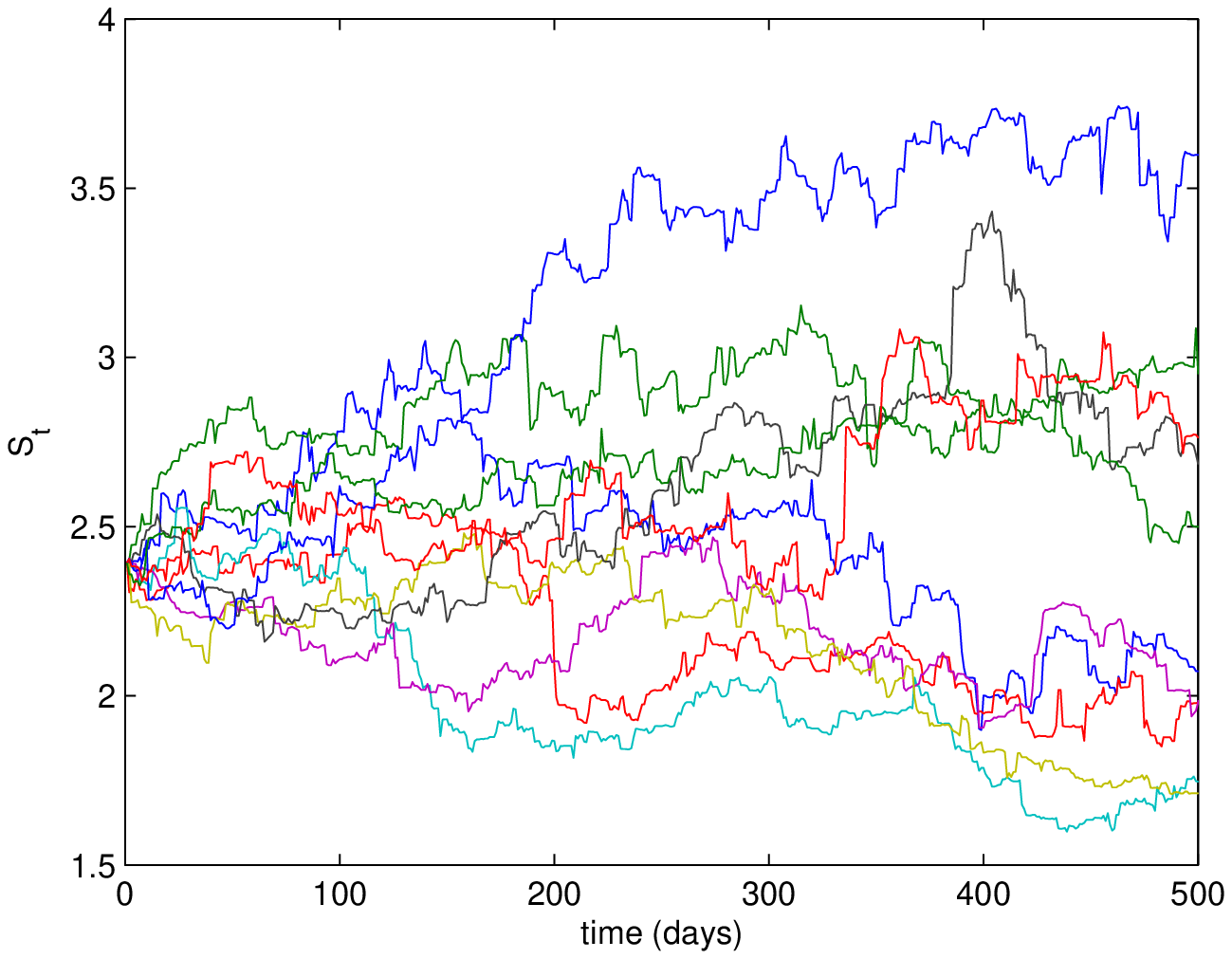}
        \label{VGpaths}
    \end{minipage}
    \caption{On the left side, VG densities for $\nu=.4$, $\sigbar=.9$, $\mubar=0$ and various values of of the parameter ${\bar{\theta}} $. On the right, simulated VG paths samples. To replicate the simulation, use code (\ref{VG_simulation}) with the parameters $\bar{\mu} = \bar{\theta} = 0$, $\nu =6$ and $\bar{\sigma} = .03$.}
\end{figure}

\subsection{Calibration of the VG Process}
Given a time series of independent log-returns $\log(S(t_{i+1})/S(t_i))$ (with $t_{i+1} - t_i= \Delta t$), say  $(x_1, x_2, \ldots, x_n)$, and denoting by $\pi = \{\mubar, {\bar{\theta}} ,  \sigbar, \nu \}$ the variance-gamma probability density function parameters that one would like to estimate, the maximum likelihood estimation (MLE) of $\pi$ consists in finding $\bar{\pi}$ that maximises the logarithm of the likelihood function ${\cal L}(\pi) = \Pi_{i=1}^{x}f(x_i;\pi)$.
The moment generating function of $X_{\Delta t}:= \log(S(t+\Delta t)/S(t))$, $\mathbb{E}[e^{zX}]$ is given by:
\begin{equation}
M_X(z) = e^{\mubar z} \Big( 1-{\bar{\theta}}  \nu z - \frac{1}{2}\nu \sigbar^2 z^2 \Big)^{-\frac{\Delta t}{\nu}}
\end{equation}
and one obtains the four central moments of the return distribution over an interval of length $\Delta t$:
\begin{eqnarray*}
E[X] &=& \bar{X} = (\mubar + {\bar{\theta}} ) \Delta t\\
E[(X - \bar{X})^2] &=& (\nu {\bar{\theta}} ^2 + \sigbar^2)\Delta t\\
E[(X - \bar{X})^3] &=& (2{\bar{\theta}} ^3 \nu^2 +3 \sigbar^2 \nu {\bar{\theta}} )\Delta t\\
E[(X - \bar{X})^4] &=& (3\nu \sigbar^4 + 12{\bar{\theta}} ^2\sigbar^2\nu^2 +6 {\bar{\theta}} ^4\nu^3)\Delta t + (3\sigbar^4 + 6{\bar{\theta}} ^2\sigbar^2\nu  + 3 {\bar{\theta}} ^4\nu^2)(\Delta t)^2.
\end{eqnarray*}
Now considering the skewness and the kurtosis:
\begin{eqnarray*}
S &=& \frac{3 \nu {\bar{\theta}} \Delta t}{((\nu {\bar{\theta}} ^2 + \sigbar^2)\Delta t)^{3/2}}\\
K &=& \frac{(3\nu \sigbar^4 + 12{\bar{\theta}} ^2\sigbar^2\nu^2 +6 {\bar{\theta}} ^4\nu^3)\Delta t + (3\sigbar^4 + 6{\bar{\theta}} ^2\sigbar^2\nu  + 3 {\bar{\theta}} ^4\nu^2)(\Delta t)^2}{((\nu {\bar{\theta}} ^2 + \sigbar^2)\Delta t)^{2}}.
\end{eqnarray*}
and assuming ${\bar{\theta}}$ to be small, thus ignoring ${\bar{\theta}} ^2$, ${\bar{\theta}} ^3$, ${\bar{\theta}} ^4$, obtain:
\begin{eqnarray*}
S &=& \frac{3 \nu {\bar{\theta}} }{\sigbar \sqrt{\Delta t}}\\
K &=& 3 \big(1 + \frac{\nu}{\Delta t}\big).
\end{eqnarray*}
So the initial guess for the parameters $\pi_0$ will be:
\begin{eqnarray*}
\sigbar &=&  \sqrt{\frac{V}{\Delta t}}\\
\nu &=& \Big(\frac{K}{3}-1\Big)\Delta t\\
{\bar{\theta}}  &=& \frac{S \sigbar \sqrt{\Delta t} }{3 \nu}\\
\mubar &=& \frac{\bar{X}}{\Delta t} - {\bar{\theta}}
\end{eqnarray*}

\begin{algorithm}[!h]
\begin{lstlisting}
%function params = VG_calibration(data,dt)
% computation of the initial parameters

data = (EURUSDsv(:,1));
hist(data)
dt   = 1;

data = price2ret(data);
M = mean(data);
V = std(data);
S = skewness(data);
K = kurtosis(data);                               
sigma  = sqrt(V/dt);
nu     = (K/3-1)*dt;
theta  = (S*sigma*sqrt(dt))/(3*nu);
mu     = (M/dt)-theta;
% VG MLE
pdf_VG = @(data,theta, nu, sigma, mu) VGdensity(data, theta, nu, sigma, mu,dt); 
start  = [theta, nu, sigma, mu];
lb              = [-intmax 0 0 -intmax];
ub              = [intmax intmax intmax intmax];
options         = statset('MaxIter',1000, 'MaxFunEvals',10000);
params = mle(data, 'pdf',pdf_VG, 'start',start, 'lower',lb, 'upper',ub, 'options',options);

\end{lstlisting}
\caption{$MATLAB^{\circledR}$ code of the MLE for variance gamma density function in (\ref{pdfVG}).}
\end{algorithm}

\begin{figure}[h!]
        \centering
        \includegraphics[scale=.7]{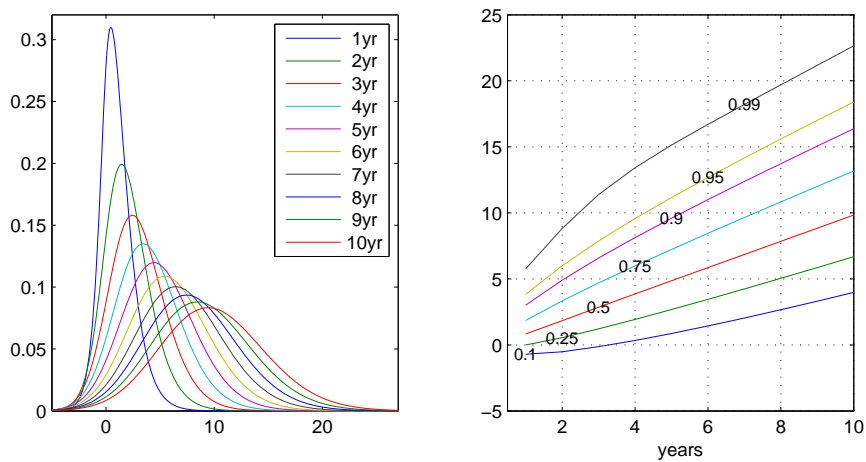}
    \caption{On the left side, VG densities for ${\bar{\theta}}  = 1$, $\nu=.4$, $\sigbar = 1.4$, $\mubar=0$ and time from 1 to 10 years. On the right side, on the $\mbox{y-axis}$, percentiles of these densities as time varies, $\mbox{x-axis}$.}\label{VGperc}
\end{figure}

\begin{figure}[h!]
        \centering
        \includegraphics[scale=.5]{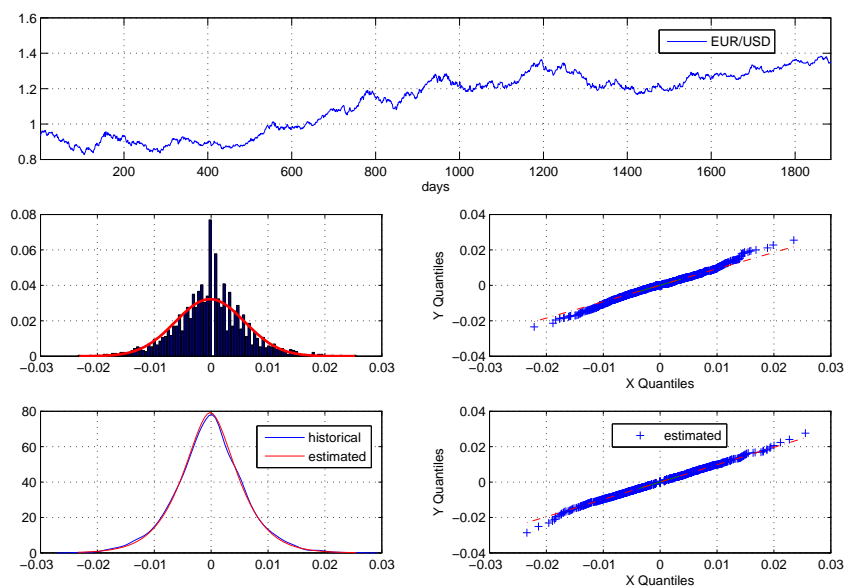}
        \label{VGmoments}
    \caption{On the top chart the historical trend of the EUR/USD is reported. In the middle chart the daily return distribution presents fat tails with respect to a normal distribution. The MLE parameters for the normal R.V. are $\mu=-.0002$ and $\sigma = 0.0061$. The QQ-plot in the middle shows clearly the presence of tails more pronounced than in the normal case. The bottom chart is the plot of the daily return density against the estimated VG one, and the QQ-plot of the fitted daily historical FX return time series EUR/USD using a VG process with parameters ${\bar{\theta}}  = -.0001$, $\nu = 0.4$, $\sigbar= 0.0061$, $\mubar=-.0001$. }\label{fig:VGfit}
\end{figure}

\begin{algorithm}[h!]
\begin{lstlisting}
clear all; close all

lb      = 0; ub      = 20;
options = optimset('Display','off','TolFun',1e-5,'MaxFunEvals',2500);

X     = linspace(-15,30,500);
sigma = 1.4;

f  = @(x,t) VGdensity(x, 1, .4, sigma, 0, t);
nT = 10;
t  = linspace(1,10,nT);

pc = [.1 .25 .5 .75 .9 .95 .99];

for k = 1:nT
    Y(:,k) = VGdensity(X, 1, .4, sigma, 0, t(k) );
    f      = @(x) VGdensity(x, 1, .4, sigma, 0, t(k));
    VG_PDF = quadl(f,-3,30);
    
for zz = 1:length(pc)
    percF       = @(a) quadl(f,-3,a) - pc(zz);
    PercA(k,zz) = fzero(percF,1,options);
end
end

subplot(1,2,1)
plot(X,Y); 
legend(' 1yr',' 2yr',' 3yr',' 4yr',' 5yr',' 6yr',' 7yr',' 8yr',' 9yr','10yr')
xlim([-5 27]);ylim([0 .32]); subplot(1,2,2); plot(t,PercA); 
for zz = 1:length(pc)
    text(t(zz),PercA(zz,zz),[{'\fontsize{9}\color{black}',num2str(pc(zz))}],'HorizontalAlignment',...
    'center','BackgroundColor',[1 1 1]); hold on
end 
xlim([.5 10]); xlabel('years'); grid on
\end{lstlisting}
\caption{$MATLAB^{\circledR}$ code to calculate the percentiles of the VG distribution as in fig (\ref{VGperc}).}
\end{algorithm}
An example can be the calibration of the historical time series of the EUR/USD exchange rate, where the VG process is a more suitable choice than the geometric Brownian motion. The fit is performed on daily returns and the plot in Fig.~\ref{fig:VGfit} illustrates the results. Notice in particular that the difference between the Arithmetic Brownian motion for the returns and the VG process is expressed by the time change parameter $\nu$, all other parameters being roughly the same in the ABM and VG cases.
 \label{CH4}
\section{The Mean Reverting Behaviour}\label{Sc:mr}

One can define mean reversion as the property to always revert to a certain constant or time varying level with limited variance around it. This property is true for an AR(1) process if the absolute value of the autoregression coefficient is less than one  ($|\alpha|<1$). To be precise, the formulation of the first order autoregressive process AR(1) is:
\begin{equation}\label{eq:MR0}
x_{t+1} =\mu + \alpha x_t + \sigma \epsilon_{t+1} \Rightarrow \DD x_{t+1} = (1-\alpha) \left( \frac{\mu}{1-\alpha} - x_t \right) + \sigma \epsilon_{t+1}
\end{equation}

All the mean reverting behaviour in the processes that are introduced in this section is due to an AR(1) feature in the discretised version of the relevant SDE.

In general, before thinking about fitting a specific mean reverting process to available data, it is important to check the validity of the mean reversion assumption. A simple way to do this is to test for stationarity. Since for the AR(1) process $|\alpha|<1$ is also a necessary and sufficient condition for stationarity, testing for mean reversion is equivalent to testing for stationarity. Note that in the special case where $\alpha=1$, the process behaves like a pure random walk with constant drift, $\DD x_{t+1} = \mu + \sigma \epsilon_{t+1}$.\\

There is a growing number of stationarity statistical tests available and ready for use in many econometric packages.\footnote{For example Stata, SAS, R, E-Views, etc.} The most popular are:

\begin{itemize}

\item the \cite[][DF]{DF:79} test;
\item the Augmented DF (ADF) test of \cite{ADF:84} test;
\item the \cite[][PP]{PhillipsPerron:88};
\item the Variance Ratio (VR) test of \cite{PoterbaSummers:86} and \cite{Cochrane:01}; and
\item the \cite[][KPSS]{kpss:92} test.
\end{itemize}

All tests give results that are asymptotically valid, and are constructed under specific assumptions.

\begin{figure}[h!b]
\begin{center}
\includegraphics[width=0.7\textwidth]{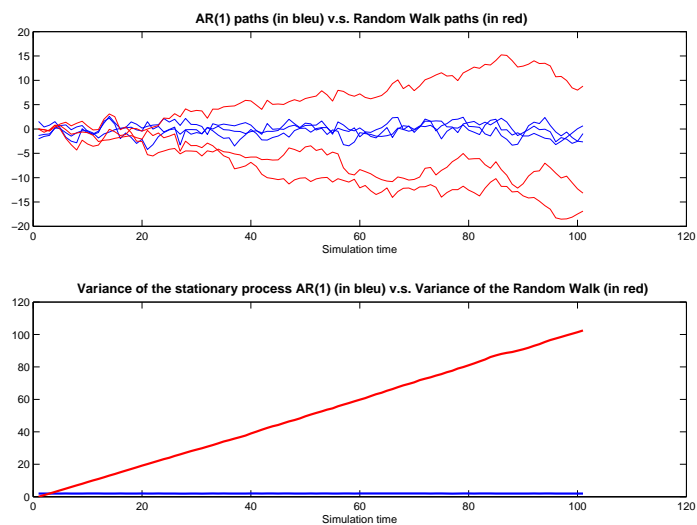}
\caption{The Random walk vs the AR(1) stationary process. (AR(1): $\mu=0,\alpha=0.7,\sigma=1$ and Random Walk: $\mu=0,\sigma=1$)}
\end{center}
\end{figure}

Once the stationarity has been established for the time series data, the only remaining test is for the statistical significance of the autoregressive lag coefficient. A simple Ordinary Least Squares of the time series on its one time lagged version (Equation (\ref{eq:MR0})) allows to assess whether $\alpha$ is effectively significant.\\

\begin{figure}
\includegraphics[width=1.0\textwidth]{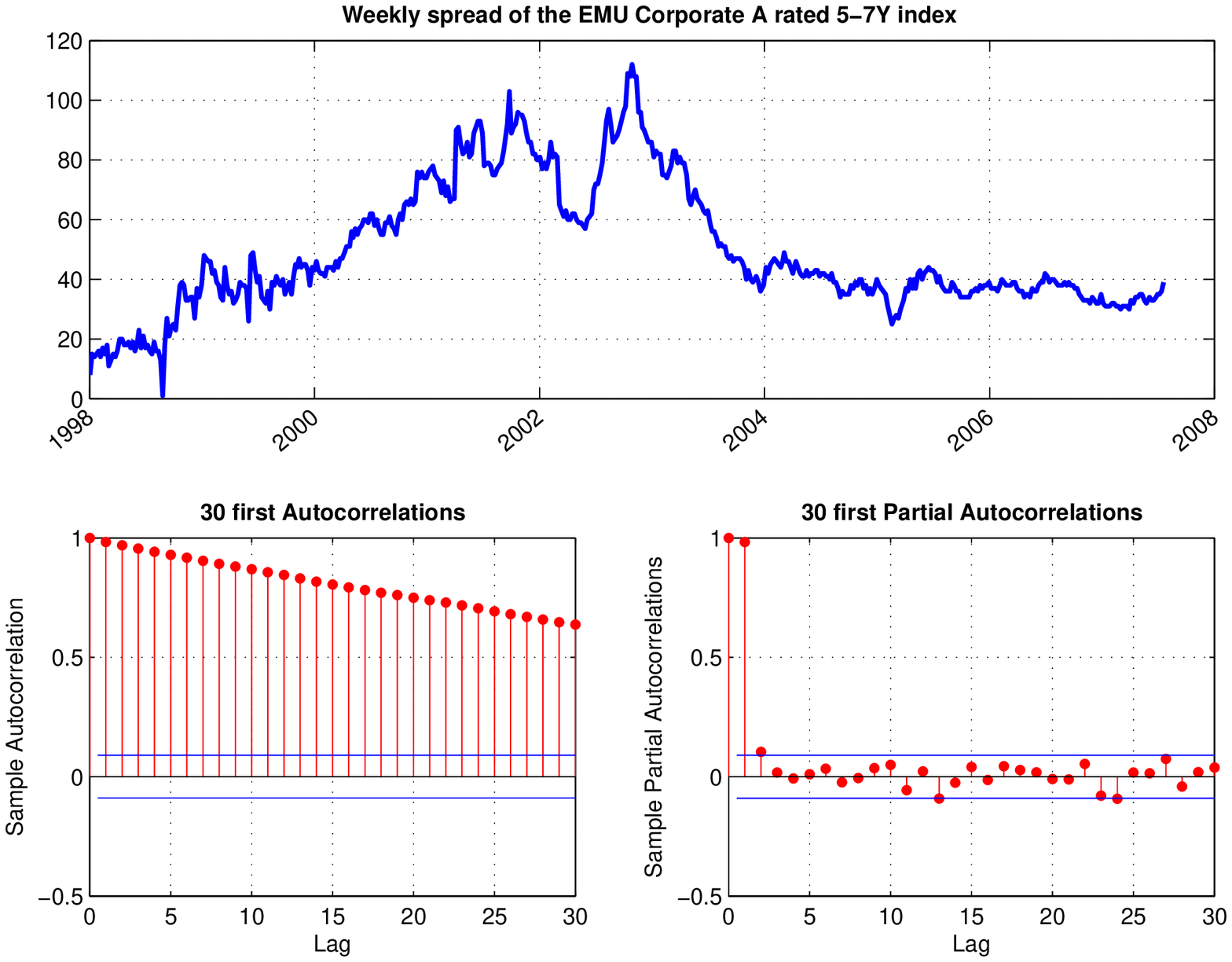}
\caption{On the top: the EMU Corporate A rated 5-7Y weekly index from 01/98 to 08/07. On the bottom: the ACF and PACF plot of this time series.} \label{logweeklyindex}
\end{figure}

Suppose that, as an example, one is interested in analysing the GLOBOXX index history.
Due to the lack of Credit Default Swap (CDS) data for the time prior to 2004, the underlying GLOBOXX spread history has to be proxied by an Index based on the Merrill Lynch corporate indices\footnote{Asset Swap Spreads data of a portfolio of equally weighted Merrill Lynch Indices that match both the average credit quality of GLOBOXX, its duration, and its geographical composition.}. This proxy index has appeared to be mean reverting. Then the risk of having the CDS spreads rising in the mean term to revert to the index historical "long-term mean" may impact the risk analysis of a product linked to the GLOBOXX index.

The mean reversion of the spread indices is supported by the intuition that the credit spread is linked to the economic cycle. This assumption also has some support from the econometric analysis. To illustrate this, the mean reversion of a Merrill lynch index, namely the EMU Corporate A rated 5-7Y index\footnote{Bloomberg Tickers ER33 - Asset Swap Spreads data from 01/98 to 08/07.}, is tested using the Augmented Dickey-Fuller test (\cite{ADF:84}).\\

The ADF test for mean reversion is sensitive to outliers\footnote{For an outlier due to a bad quote, the up and down movement can be interpreted as a strong mean reversion in that particular data point and is able to influence the final result.}. Before testing for mean reversion, the index data have been cleaned by removing the innovations that exceed three times the volatility, as they are considered outliers. Figure \ref{logweeklyindex} shows the EMU Corporate A rated 5-7Y index cleaned from outliers and shows also the autocorrelations and partial autocorrelation plots of its log series. From this chart one can deduce that the weekly log spread of the EMU Corporate A rated 5-7Y index shows a clear AR(1) pattern. The AR(1) process is characterised by  exponentially decreasing autocorrelations and by a partial autocorrelation that reaches zero after the first lag (see discussions around Equation (\ref{eq:ACFPACF}) for the related definitions):

\begin{eqnarray}
ACF(k)=\alpha^{k},\ \ k=0,1,2,...; \ \ \ PACF(k)=\alpha, k=1; \ \ PACF(k)=0, k=2,3,4,...
\end{eqnarray}

The mean reversion of the time series has been tested using the Augmented Dickey-Fuller test (\cite{ADF:84}). The ADF statistics obtained by this time series is -3.7373. The more negative the ADF statistics, the stronger the rejection of the unit root hypothesis\footnote{Before cleaning the data from the outliers the ADF statistic was -5.0226. This confirms the outliers effect on amplifying the mean reversion detected level.}. The first and fifth percentiles are -3.44 and -2.87. This means that the test rejects the null hypothesis of unit root at the 1\% significance level.

\begin{figure}
\centering
\includegraphics[width=0.6\textwidth]{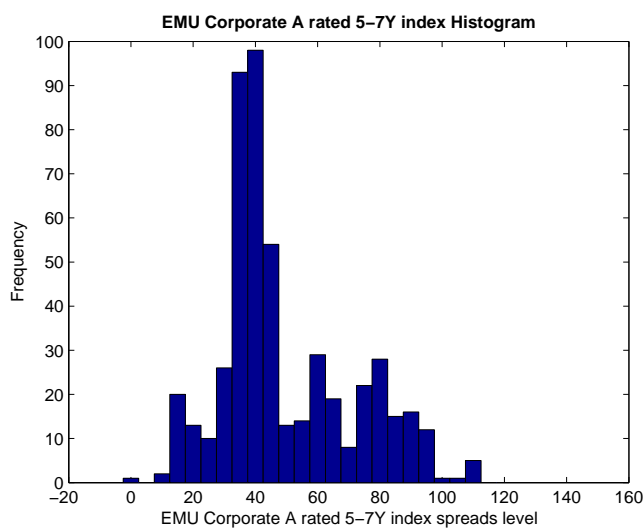}
\caption{The histogram of EMU Corporate A rated 5-7Y index.} \label{indexhist}
\end{figure}

Having discussed AR(1) stylised features in general, this tutorial now embarks on describing specific models featuring this behaviour in continuous time.

\section{Mean Reversion: The Vasicek Model}\label{sc:cal}

The Vasicek model, owing its name to \cite{vasicek:77}, is one of the earliest stochastic models of the short-term interest rate. It assumes that the instantaneous spot rate (or "short rate") follows an Ornstein-–Uhlenbeck process with constant coefficients under the statistical or objective measure used for historical estimation:
\begin{equation}
    dx_{t} = \alpha(\theta - x_{t})dt + \sigma dW_{t} \label{eq:vasi1}
\end{equation}
with $\alpha$,$\theta$ and $\sigma$ positive and $dW_{t}$ a standard Wiener process.
Under the risk neutral measure used for valuation and pricing one may assume a similar parametrisation for the same process but with one more parameter in the drift modelling the market price of risk; see for example \cite{brigo:06}.

This model assumes a mean-reverting stochastic behaviour of interest rates. The Vasicek model has a major shortcoming that is the non null probability of negative rates. This is an unrealistic property for the modelling of positive entities like interest rates or credit spreads. The explicit solution to the SDE (\ref{eq:vasi1}) between any two instants $s$ and $t$, with $0\le s< t$, can be easily obtained from the solution to the Ornstein-–Uhlenbeck SDE, namely:

\begin{equation}
x_{t} =  \theta\left(1-e^{-\alpha(t-s)}\right) + x_{s}e^{-\alpha(t-s)} + \sigma e^{-\alpha t}\int^t_{s}e^{\alpha u}dW_{u} \label{eq:eq1}
\end{equation}

The discrete time version of this equation, on a time grid $0=t_0,t_1,t_2,\ldots$ with (assume constant for simplicity) time step $\DD t = t_{i}-t_{i-1}$ is:

\begin{equation}
x(t_{i}) =  c + b x(t_{i-1})+\delta\epsilon({t_i}) \label{eq:eq2}
\end{equation}

where the coefficients are:
\begin{eqnarray}
c &=& \theta \left(1-e^{-\alpha \Delta t}\right) \\
b &=& e^{-\alpha \Delta t}
\end{eqnarray}

and $\epsilon(t)$ is a Gaussian white noise ($ \epsilon \sim {N}\left(0,1\right)$). The volatility of the innovations  can be deduced by the \Ito isometry

\begin{equation}
\delta = \sigma\sqrt{\left(1-e^{-2\alpha \Delta t}\right)/2\alpha}
\end{equation}

whereas if one used the Euler scheme to discretise the equation this would simply be $\sigma {\sqrt{\DD}} t$, which is the same (first order in $\DD t$) as the exact one above.

Equation (\ref{eq:eq2}) is exactly the formulation of an AR(1) process (see section \ref{Sc:mr}). Having $0<b<1$ when $0<\alpha$ implies that this AR(1) process is stationary and mean reverting to a long-term mean given by $\theta$. One can check this also directly by computing mean and variance of the process. As the distribution of $x(t)$ is Gaussian, it is characterised by its first two moments. The conditional mean and the conditional variance of $x(t)$ given $x(s)$ can be derived from Equation (\ref{eq:eq1}):

\begin{equation}
\mathbb{E}_{s} [x_{t}] = \theta + (x_{s}-\theta)e^{-\alpha (t-s)} \label{eq:expVacisekMean}
\end{equation}
\begin{equation}
\mbox{Var}_{s}[x_{t}] = \frac{\sigma^{2}}{2\alpha} \left(1-e^{-2\alpha (t-s)}\right) \label{eq:expVacisekVar}
\end{equation}

One can easily check that as time increases the mean tends to the long-term value $\theta$ and the variance remains bounded, implying mean reversion. More precisely, the long-term distribution of the Ornstein–-Uhlenbeck process is stationary and is Gaussian with $\theta$ as mean and $\sqrt{\sigma^{2}/2\alpha}$ as standard deviation.\\

\begin{figure}
\includegraphics[width=1.0\textwidth]{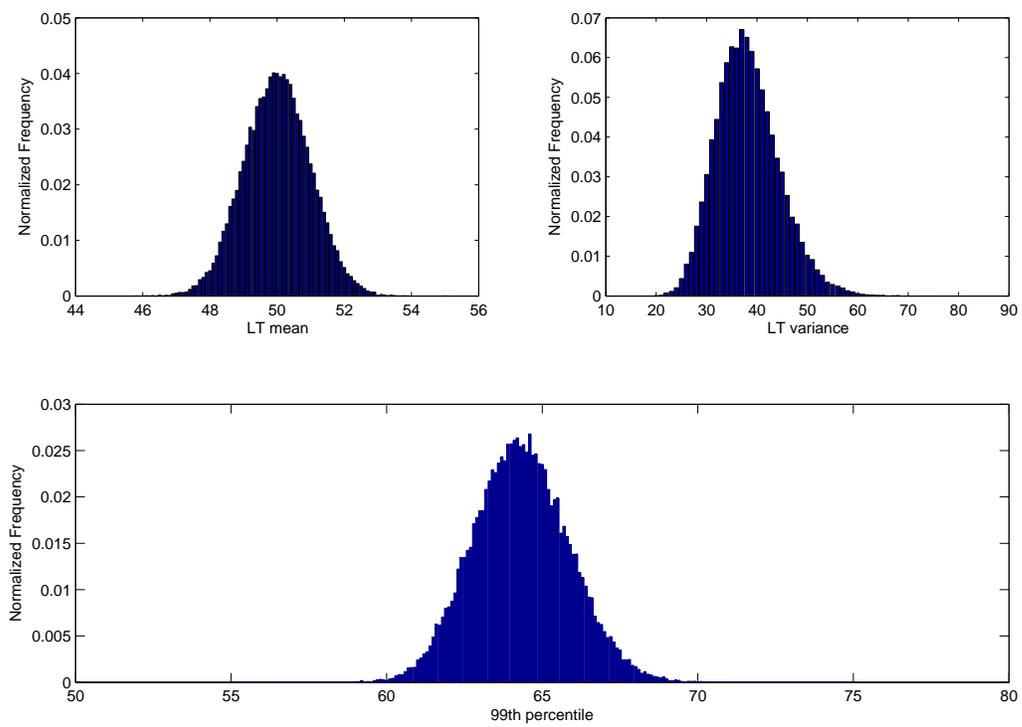}
\caption{The distribution of the long-term mean and long-term variance and the 99th percentile estimated by several simulations of a Vasicek process with $\alpha=8$, $\theta=50$ and $\sigma=25$} \label{vasicekpercdist}
\end{figure}

To calibrate the Vasicek model the discrete form of the process is used, Equation (\ref{eq:eq2}). The coefficients $c$, $b$ and $\delta$ are calibrated using the equation (\ref{eq:eq2}). The calibration process is simply an OLS regression of the time series $x(t_{i})$ on its lagged form $x(t_{i-1})$. The OLS regression provides the maximum likelihood estimator for the parameters $c$, $b$ and $\delta$. By resolving the three equations system one gets the following $\alpha$, $\theta$ and $\sigma$ parameters:
\begin{eqnarray}
\alpha &=& -\ln(b)/\Delta t \label{Sc:calib1} \\
\theta &=& c/(1-b) \label{Sc:calib2} \\
\sigma &=& \delta / \sqrt{(b^{2}-1)\Delta t/2\ln(b)}\label{Sc:calib3}
\end{eqnarray}

One may compare the above formulas with the estimators derived directly via maximum likelihood in \cite{brigo:06}, where for brevity $x(t_{i})$ is denoted by $x_i$ and $n$ is assumed to be the number of observations.

\begin{equation}
\widehat{b}=\frac{n\sum_{i=1}^n x_i x_{i-1}-\sum_{i=1}^n x_i
\sum_{i=1}^n x_{i-1}} {n\sum_{i=1}^n x_{i-1}^2 -\left(\sum_{i=1}^n
x_{i-1}\right)^2},
\end{equation}

\begin{equation}
\widehat{\theta}=\frac{\sum_{i=1}^n [x_i-\widehat{b} x_{i-1}]}{n(1-\widehat{b})},
\end{equation}

\begin{equation}
\widehat{\delta^2}=\frac{1}{n}\sum_{i=1}^n\left[x_i-\widehat{b}x_{i-1}
-\widehat{\theta}(1-\widehat{b})\right]^2.
\end{equation}

From these estimators, using (\ref{Sc:calib1}) and (\ref{Sc:calib3}) one obtains immediately the estimators for the parameters $\alpha$ and $\sigma$.

\begin{algorithm}[!h]
\begin{lstlisting}
function ML_VasicekParams = Vasicek_calibration(V_data,dt)

    N=length(V_data);
    x = [ones(N-1,1) V_data(1:N-1)];
    ols = (x'*x)^(-1)*(x'*V_data(2:N));
    resid = V_data(2:N) - x*ols;
    
    c=ols(1);
    b=ols(2);
    delta=std(resid);
    
    alpha=-log(b)/dt;
    theta=c/(1-b);
    sigma=delta/sqrt((b^2-1)*dt/(2*log(b)));
    
    ML_VasicekParams = [alpha theta sigma];

    

    
\end{lstlisting}
\caption{$MATLAB^{\circledR}$ Calibration of the Vasicek process.}
\end{algorithm}

For the Monte Carlo simulation purpose the discrete time expression of the Vasicek model is used. The simulation is straightforward using the calibrated coefficients $\alpha$, $\theta$ and $\sigma$. One only needs to draw a random normal process $\epsilon_{t}$ and then generate recursively the new $x_{t_i}$ process, where now $t_i$ are future instants, using equation (\ref{eq:eq2}), starting the recursion with the current value $x_{0}$.\\

\section{Mean Reversion: The Exponential Vasicek Model}

To address the positivity of interest rates and ensuring their distribution to have fatter tails, it is possible to transform a basic Vasicek process $y(t)$ in a new positive process $x(t)$.

The most common transformations to obtain a positive number $x$ from a possibly negative number $y$ are the square and the exponential functions.

By squaring or exponentiating a normal random variable $y$ one obtains a random variable $x$ with fat tails. Since the exponential function increases faster than the square and is larger ($\exp(y) >> y^2$ for positive and large $y$) the tails in the exponential Vasicek model below will be fatter than the tails in the squared Vasicek (and CIR as well). In other terms, a lognormal distribution (exponential Vasicek) has fatter tails than a chi-squared (squared Vasicek) or even noncentral chi-squared (CIR) distribution.

Therefore, if one assumes $y$ to follow a Vasicek process,
\begin{equation}
    dy_{t} = \alpha(\theta - y_{t})dt + \sigma dW_{t}
\end{equation}
then one may transform $y$. If one takes the square to ensure positivity and define $x(t) = y(t)^2$, then by \Ito's formula $x$ satisfies:
\[ d x(t) = (B + A \sqrt{x(t)} - C x(t))dt + \nu \sqrt{x(t)} dW_t \]
for suitable constants $A,B,C$ and $\nu$. This ``squared Vasicek" model is similar to the CIR model that the subsequent section will address, so it will not be addressed further here.

If, on the other hand, to ensure positivity one takes the exponential, $x(t) = \exp(y(t))$, then one obtains a model that at times has been termed Exponential Vasicek (for example in \cite{brigo:06}).

\Ito's formula this time reads, for $x$ under the objective measure:

\begin{equation}
\boxed{    dx_{t} = \alpha x_{t}(m  - \log x_{t})dt + \sigma x_{t} dW_{t} } \label{eq:expvas0}
\end{equation}

\vspace{0.3cm}
where $\alpha$ and $\sigma$ are positive, and a similar parametrisation can be used under the risk neutral measure. Also, $m= \theta  + \frac{\sigma^{2}}{2\alpha}$.

In other words, this model for $x$ assumes that the \emph{logarithm} of the short rate (rather than the short rate itself) follows an Ornstein-–Uhlenbeck process.

The explicit solution of the Exponential Vasicek equation is then, between two any instants $0<s<t$:
\begin{equation}
x_{t} =  \exp\left\{\theta\left(1-e^{-\alpha(t-s)}\right) + \log(x_{s}) e^{-\alpha(t-s)} + \sigma e^{-\alpha t}\int^t_{s}e^{\alpha u}dW_{u}\right\} \label{eq:expvas}
\end{equation}

\begin{figure}
\includegraphics[width=1.0\textwidth]{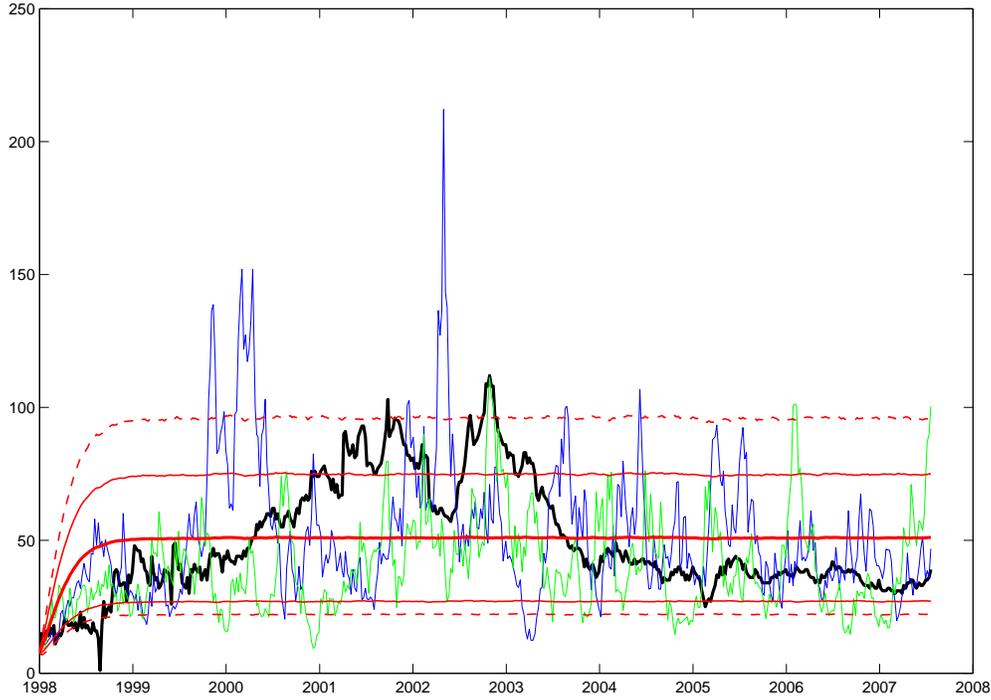}
\caption{EMU Corporate A rated 5-7Y index (in dark black) vs two simulated Exponential Vasicek paths (in blue and green). The dashed red lines describe the 95th percentile and the 5th percentile of the exponential Vasicek distribution. The dark red line is the distribution mean. The two remaining red lines describe one standard deviation above and below the mean.} \label{indexsimul}
\end{figure}

The advantage of this model is that the underlying process is mean reverting, stationary in the long run, strictly positive and having a distribution skewed to the right and characterised by a fat tail for large positive values. This makes the Exponential Vasicek model an excellent candidate for the modelling of the credit spreads when one needs to do historical estimation. As shortcomings for the Exponential Vasicek model one can identify the explosion problem that appears when this model is used to describe the interest rate dynamic in a pricing model involving the bank account evaluation (like for the Eurodollar future pricing, \cite{brigo:06}). One should also point out that if this model is assumed for the short rate or stochastic default intensity, there is no closed form bond pricing or survival probability formula and all curve construction and valuation according to this model has to be done numerically through finite difference schemes, trinomial trees or Monte Carlo simulation.

The historical calibration and simulation of the Exponential Vasicek model can be done by calibrating and simulating its log-process $y$ as described in the previous Section. To illustrate the calibration, the EMU Corporate A rated 5-7Y index described in Section~\ref{Sc:mr} is used. The histogram of the index shows a clear fat tail pattern. In addition, the log spread process is tested to be mean reverting (see again Section~\ref{Sc:mr}). This leads to think about using the Exponential Vasicek model that has a log normal distribution and an AR(1) disturbance in the log prices.

The OLS regression of the log-spread time series on its first lags (Equation (\ref{eq:eq2})) gives the coefficients $c$, $b$ and $\delta$. Then the $\alpha$, $\theta$ and $\sigma$ parameters in Equations (\ref{eq:expvas0}) are deduced using equations (\ref{Sc:calib1}), (\ref{Sc:calib2}) and (\ref{Sc:calib3}):
\begin{eqnarray*}
 c = 0.3625	&&\alpha = 4.9701 \\
 b = 0.9054	&&\theta =  3.8307\\
 \delta = 0.1894 &&\sigma = 1.4061
\end{eqnarray*}

Figure \ref{indexsimul} shows how the EMU Corporate A rated 5-7Y index time series fits into the simulation of 50,000 paths using the Equation for the exponential of (\ref{eq:eq2}) starting with the first observation of the index (first week of 1998). The simulated path (in blue) illustrates that the Exponential Vasicek process can occasionally reach values that are much higher than its 95--percentile level.

\section{Mean Reversion: The CIR Model}

\begin{figure}
\includegraphics[width=1.0\textwidth]{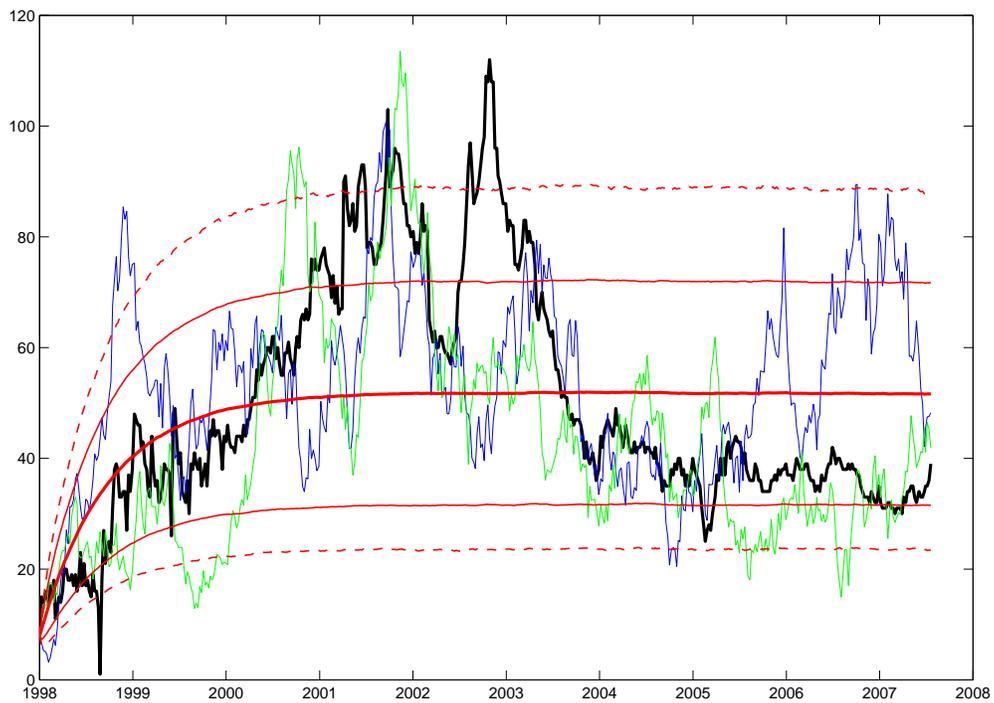}
\caption{EMU Corporate A rated 5-7Y index (in dark black) vs two simulated CIR paths (in blue and green). The dashed red lines describe the 95th percentile and the 5th percentile of the distribution. The dark red line is the distribution mean. The two remaining red lines describe one standard deviation above and below the mean} \label{cir_indexsimul}
\end{figure}

Originally the model proposed by \cite{CIR:85a} was introduced in the context of the general equilibrium model of the same authors \cite{CIR:85b}. By choosing an appropriate market price of risk, the $x_{t}$ process dynamics can be written with the same structure under both the objective measure and risk neutral measure. Under the objective measure one can write

\begin{equation}
    dx_{t} = \alpha(\theta - x_{t})dt + \sigma \sqrt{x_{t}} dW_{t}\label{eq:cir}
\end{equation}

with $\alpha,\theta,\sigma>0$, and $dW_{t}$ a standard Wiener process. To ensure that $x_{t}$ is always strictly positive one needs to impose  $\sigma^{2}\leq 2\alpha\theta$. By doing so the upward drift is sufficiently large with respect to the volatility to make the origin inaccessible to $x_{t}$.

The simulation of the CIR process can be done using a recursive discrete version of Equation (\ref{eq:cir}) at discretisation times $t_i$:
\begin{equation}
  x(t_{i})= \alpha\theta\Delta t + (1-\alpha\Delta t) x(t_{i-1}) + \sigma \sqrt{x(t_{i-1}) \ \Delta t}\ \epsilon_{t_i}\label{eq:cir2}
\end{equation}

with $ \epsilon \sim\mathcal{N}\left(0,1\right)$. If the time step is too large, however, there is risk that the process approaches zero in some scenarios, which could lead to numerical problems and complex numbers surfacing in the scheme. Alternative positivity preserving schemes are discussed, for example in \cite{brigo:06}. For simulation purposes it is more precise but also more computationally expensive to draw the simulation directly from the following exact conditional distribution of the CIR process:

\begin{equation}
 f_{CIR}\left(x_{t_{i}}|x_{t_{i-1}}\right) = ce^{-u-v}\left(\frac{v}{u}\right)^{q/2}I_{q}\left(2\sqrt{uv}\right)
\end{equation}
\begin{eqnarray}
&& c=\frac{2\alpha}{\sigma^{2}(1-e^{-\alpha\Delta t})} \\
&& u=cx_{t_{i}}e^{-\alpha\Delta t}\\
&& v=cx_{t_{i-1}}\\
&& q=\frac{2\alpha\theta}{\sigma^{2}} - 1
\end{eqnarray}

with $I_{q}$ the modified Bessel function of the first kind of order $q$ and $\Delta t=t_{i}-t_{i-1}$. The process $x$ follows a noncentral chi-squared distribution $x(t_i)|x(t_{i-1}) \sim \chi^2(2cx_{t_{i-1}},2q+2,2u)$.\\

To calibrate this model the Maximum Likelihood Method is applied:

\begin{eqnarray}
	&&\argmax_{\alpha>0,\theta>0,\sigma>0} \log({\cal{L}})
\end{eqnarray}

The Likelihood can be decomposed, as for general Markov processes\footnote{Equation (\ref{eq:cir2}) shows that $x(t_i)$ depends only on $x(t_{i-1})$ and not on previous values $x(t_{i-2}),x(t_{i-3})...$, so CIR is Markov.}, according to Equation~(\ref{eq:footnotemarkov}), as a product of transition likelihoods:
\begin{eqnarray}
	&&{\cal{L}}=\prod^{n}_{i=1} f_{\mbox{CIR}}\left(x(t_i)|x(t_{i-1});\alpha,\theta,\sigma\right)
\end{eqnarray}

Since this model is also a fat tail mean reverting one, it can be calibrated to the EMU Corporate A rated 5-7Y index described in Section \ref{Sc:mr}. The calibration is done by Maximum Likelihood. To speed up the calibration one uses as initial guess for $\alpha$ the AR(1) coefficient of the Vasicek regression (Equation (\ref{eq:eq2})) and then infer  $\theta$ and $\sigma$ using the first two moments of the process. The long-term mean of the CIR process is $\theta$ and its long-term variance is $\theta\sigma^{2}/(2\alpha)$. Then:
\begin{eqnarray}
 \alpha_{0} &=& -\log(b)/\Delta t\\
 \theta_{0} &=& \mathbb{E}(x_{t})\\
 \sigma_{0} &=& \sqrt{2\alpha_{0} Var(x_{t})/\theta_{0}}
\end{eqnarray}

\begin{algorithm}[!h]
\begin{lstlisting}
function ML_CIRparams = CIR_calibration(V_data,dt,params)
    % ML_CIRparams = [alpha theta sigma]
    
    N=length(V_data);
    if nargin <3
        x = [ones(N-1,1) V_data(1:N-1)];
        ols = (x'*x)^(-1)*(x'*V_data(2:N));
        m=mean(V_data); v=var(V_data);
        params = [-log(ols(2))/dt,m,sqrt(2*ols(2)*v/m)];
    end
    options = optimset('MaxFunEvals', 100000, 'MaxIter', 100000);
    ML_CIRparams = fminsearch(@FT_CIR_LL_ExactFull, params, options);
    
    function mll = FT_CIR_LL_ExactFull(params)
        alpha = params(1);teta = params(2);sigma = params(3);
        c = (2*alpha)/((sigma^2)*(1-exp(-alpha*dt)));
        q = ((2*alpha*teta)/(sigma^2))-1;
        u = c*exp(-alpha*dt)*V_data(1:N-1);
        v = c*V_data(2:N);
        mll = -(N-1)*log(c)+sum(u+v-log(v./u)*q/2-...
            log(besseli(q,2*sqrt(u.*v),1))-abs(real(2*sqrt(u.*v))));
    end
end
\end{lstlisting}
\caption{$MATLAB^{\circledR}$ MLE calibration of the CIR process.}
\end{algorithm}

The initial guess and fitted parameters for the EMU Corporate A rated 5-7Y index are the following:
\begin{eqnarray*}
 \alpha_{0} = 0.8662	&&\alpha = 1.2902 \\
 \theta_{0} = 49.330	&&\theta = 51.7894\\
 \sigma_{0} = 4.3027	&&\sigma = 4.4966
\end{eqnarray*}

\begin{figure}[h!t]
\includegraphics[width=1.0\textwidth]{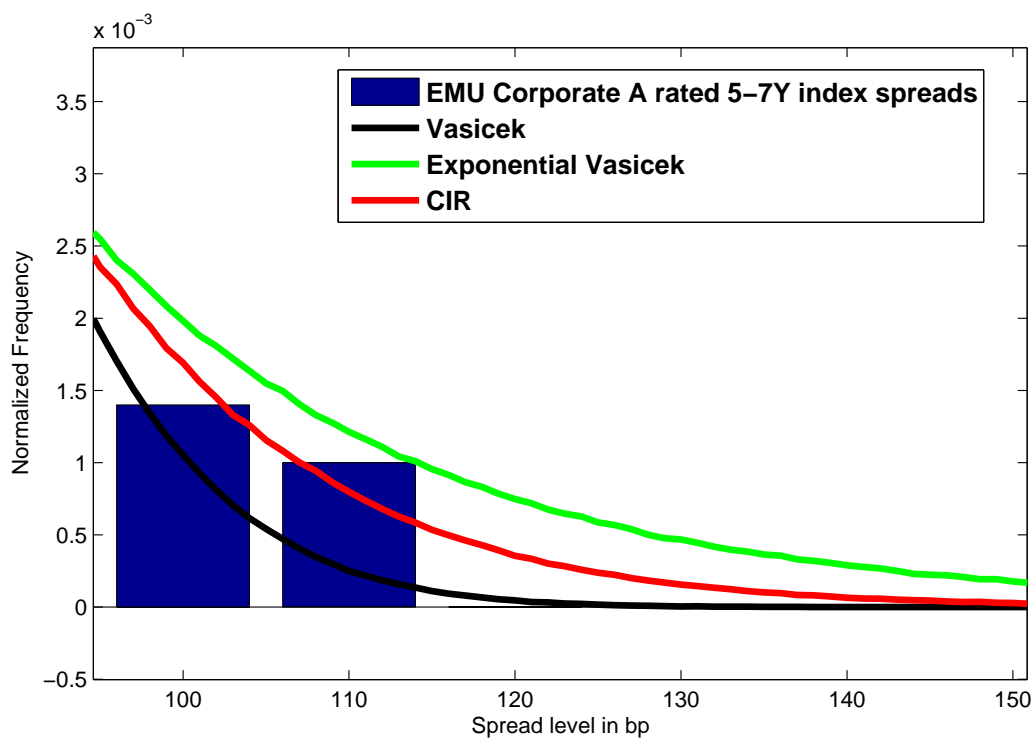}
\caption{The tail of the EMU Corporate A rated 5-7Y index histogram (in blue) compared to the fitted Vasicek process probability density (in black), to the fitted Exponential Vasicek process probability density (in green) and to the fitted CIR process probability density (in red)} \label{cir_vas_expvas}
\end{figure}

Figure~\ref{cir_indexsimul} shows how the EMU Corporate A rated 5-7Y index time series fits into the simulation of 40,000 paths using Equation (\ref{eq:cir2}) starting with the first observation of the index (first week of 1998).

As observed before, the CIR model has less fat tails than the Exponential Vasicek, but avoids the above-mentioned explosion problem of the log-normal processes bank account. Furthermore, when used as an instantaneous short rate or default intensity model under the pricing measure, it comes with closed form solutions for bonds and survival probabilities. Figure~(\ref{cir_vas_expvas}) compares the tail behaviour of the calibrated CIR process with the calibrated Vasicek and Exponential Vasicek processes.
Overall, the exponential Vasicek process is easier to estimate and to simulate, and features fatter tails, so it is preferable to CIR. However, if used in conjunction with pricing applications, the tractability of CIR and non-explosion of its associated bank account can make CIR preferable in some cases.
\label{CH3}

\section{Mean Reversion and Fat Tails together}\label{sec:meanrevjumps}

In the previous sections, two possible features of financial time series were studied: fat tails and mean reversion. The first feature accounts for extreme movements in short time, whereas the second one accounts for a process that in long times approaches some reference mean value with a bounded variance. The question in this section is: can one combine the two features and have a process that in short time instants can feature extreme movements, beyond the Gaussian statistics, and in the long run features mean reversion?

In this section one example of one such model is formulated: a mean reverting diffusion process with jumps. In other terms, rather than adding jumps to a GBM, one may decide to add jumps to a mean reverting process such as Vasicek. Consider:

\begin{equation}
    d x(t)=\alpha(\theta -  x(t))  dt +\sigma dW(t)+ dJ_{t}\label{eq:jumpsvacSDE}
\end{equation}
where the jumps $J$ are defined as:
\[   J(t) = \sum_{j=1}^{N(t)} Y_j, \ \  dJ(t) = Y_{N(t)} d N(t) \]
where $N(t)$ is a Poisson process with intensity $\lambda$ and $Y_j$ are iid random variables modelling the size of the $j$-th jump, independent of $N$ and $W$.

The way the previous equation is written is slightly misleading, since the mean in the shocks in nonzero. To have zero-mean shocks, so as to be able to read the long-term mean level from the equation, one needs to rewrite it as:
\begin{equation}
    d x(t)=\alpha\left(\theta_Y  -  x(t)\right)  dt +\sigma dW(t)+ [dJ_{t} - \lambda \mathbb{E}[Y] \ dt ] \label{eq:jumpsvacomSDE}
\end{equation}
 where the actual long term mean level is:
 \[\theta_Y := \theta + \lambda \mathbb{E}[Y]/\alpha,\]
 and can be much larger than $\theta$, depending on $Y$ and $\lambda$. For the distribution of the jump sizes, Gaussian jump size are considered, but calculations are possible also with exponential or other jump sizes, particularly with stable or infinitely divisible distributions.

\begin{figure}
\begin{center}
\includegraphics[width=1\textwidth]{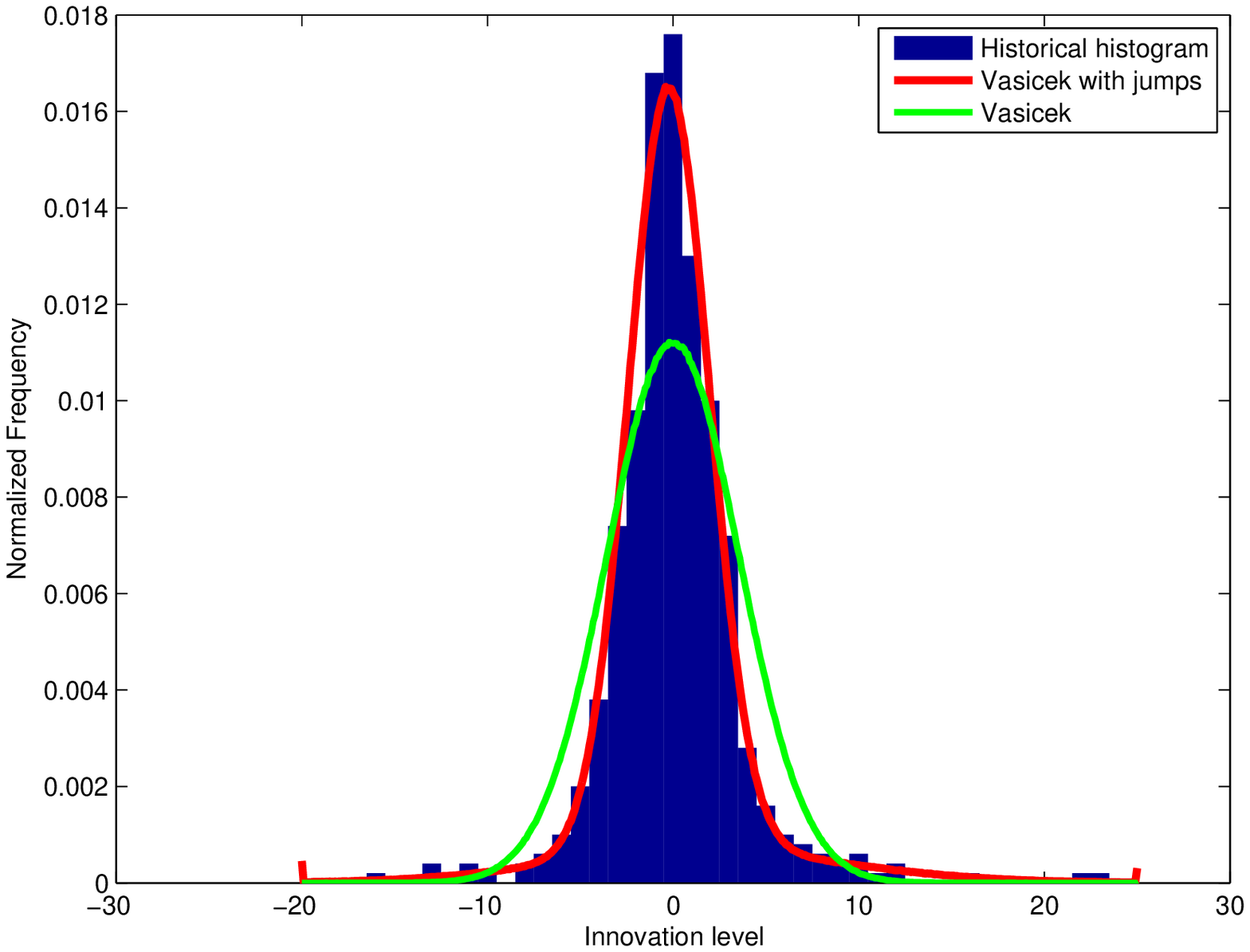}
\includegraphics[width=0.45\textwidth]{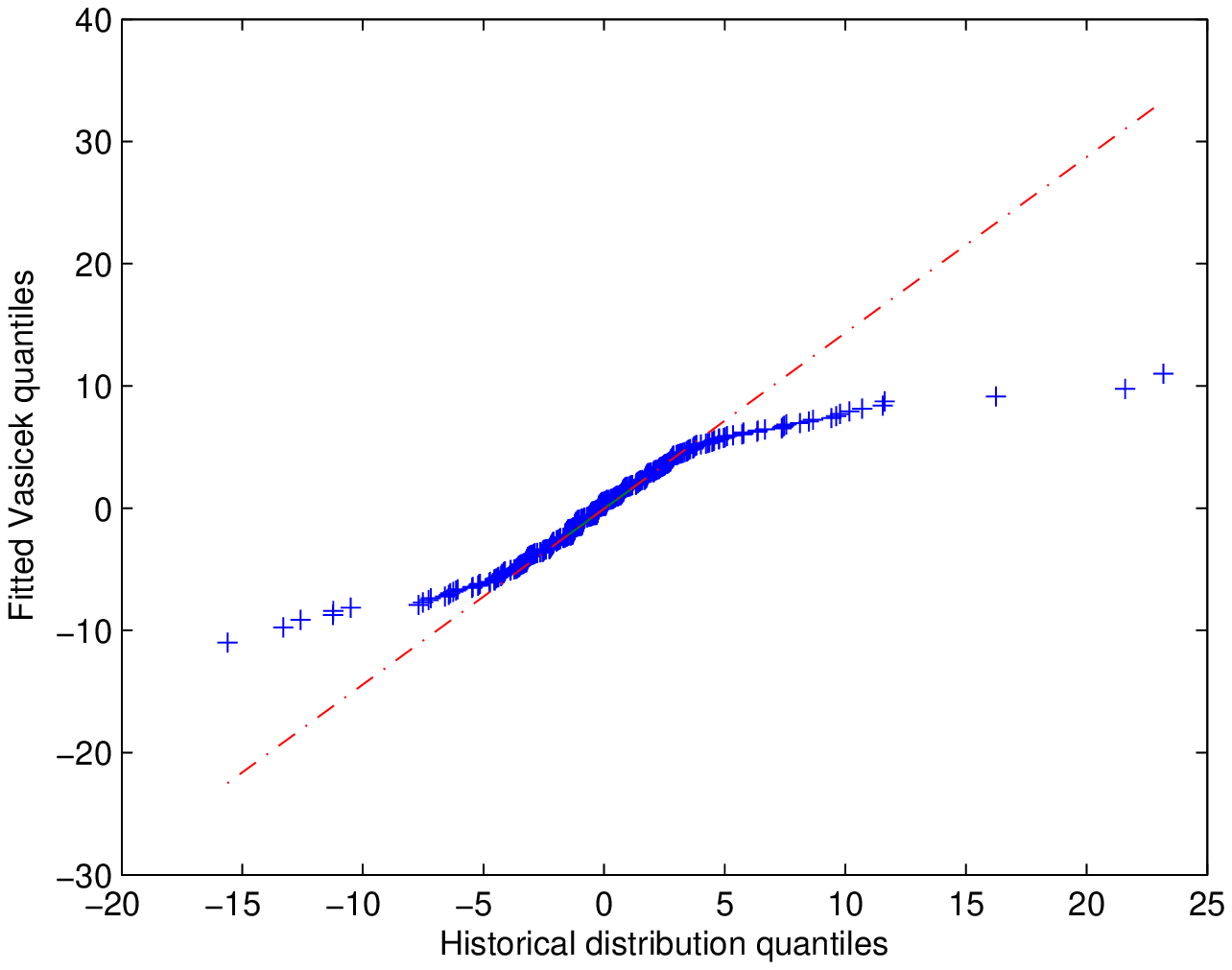}
\includegraphics[width=0.45\textwidth]{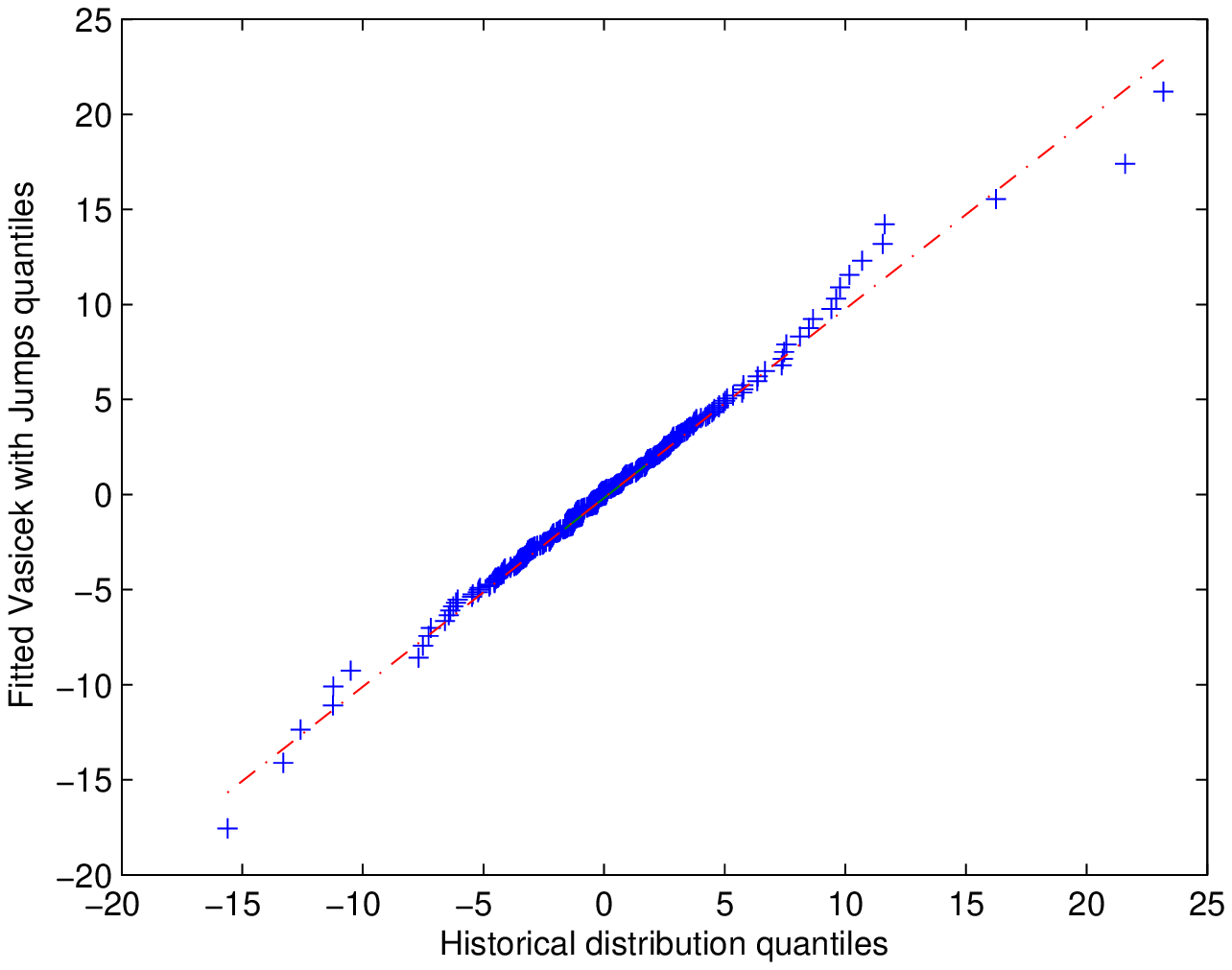}
\caption{EMU Corporate A rated 5-7Y index charts. The top chart compares the innovations fit of the Vasicek process to the fit of the Vasicek process with normal jumps (in red). The bottom left figure is the QQ-plot of the Vasicek process innovations against the historical data innovations. The bottom right figure is the QQ-plot of the Vasicek process with normal jumps innovations against the historical data innovations. }\label{fig:vasjumpemu0}
\end{center}
\end{figure}

Otherwise, for a general jump size distribution, the process integrates into:

\begin{eqnarray}
x_{t} =  m_x(x_s,t-s) + \sigma e^{-\alpha t}\int^t_{s}e^{\alpha z}dW_{z} + e^{-\alpha t} \int^t_{s}e^{\alpha z}dJ_{z} \label{eq:eq1jumps} \\ \nonumber
m_x(x_s,t-s) := \theta\left(1-e^{-\alpha(t-s)}\right) + x_{s}e^{-\alpha(t-s)},\\ \nonumber
v_x(t-s) := \frac{\sigma^2}{2\alpha} \left(1-e^{-2\alpha (t-s)}\right),
\end{eqnarray}
where $m_x$ and $v_x$ are, respectively, mean and variance of the Vasicek model without jumps.

Notice that taking out the last term in (\ref{eq:eq1jumps}) one still gets~(\ref{eq:eq1}), the earlier equation for Vasicek. Furthermore, the last term is independent of all others. Since a sum of independent random variables leads to convolution in the density space and to product in the moment generating function space, the moment generating function of the transition density between any two instants is computed:
\begin{equation}
M_{t|s}(u) = \mathbb{E}_{x(s)} \left[ \exp( u x(t)) \right]
\end{equation}
that in the model is given by
\begin{eqnarray}
M_{t|s}(u) = \exp\left\{ m_x(x_s,t-s) u +
 \frac{v_x(t-s)}{2} u^2+ \lambda \int_s^t \left( M_Y(u e^{-\alpha (t-z)}) -1\right) dz \right\}
\end{eqnarray}
where the first exponential is due to the diffusion part (Vasicek) and the second one comes from the jumps\footnote{As stated previously, the characteristic function is a better notion than the moment generating function. For the characteristic function one would obtain:
\begin{eqnarray}
\phi_{t|s}(u) = \exp\left\{ m_x(x_s,t-s) i u -
 \frac{v_x(t-s)}{2} u^2 + \lambda \int_s^t \left( \phi_Y(u e^{-\alpha (t-z)}) -1\right) dz \right\}.
\end{eqnarray}
As recalled previously, the characteristic function always exists, whereas the moment generating function may not exist in some cases.}.

Now, recalling Equation~(\ref{eq:footnotemarkov}), one knows that to write the likelihood of the Markov process it is enough to know the transition density between subsequent instants. In this case, such density between any two instants $s=t_{i-1}$ and $t=t_i$ can be computed by the above moment generating function through inverse Laplace transform\footnote{Or better, from the above characteristic function through inverse Fourier transform.}.

There are however some special jump size distributions where one does not need to use transform methods. These include the Gaussian jump size case, that is covered in detail now.

Suppose then that the Jump size $Y$ is normally distributed, with mean $\mu_Y$ and variance $\sigma^2_Y$. For small $\Delta t$, one knows that there will be possibly just one jump. More precisely, there will be one jump in$(t-\Delta t, t]$ with probability $\lambda \Delta t$, and no jumps with probability $1- \lambda \Delta t$.

Furthermore, with the approximation:
\begin{equation}\label{eq:jump:approx} \int^{t}_{t-\Delta t}e^{\alpha z}dJ_{z} \approx  e^{\alpha (t-\Delta t)} \Delta J(t), \ \ \Delta J(t) := J(t) - J(t-\Delta t).
\end{equation}

one can write the transition density between $s=t-\Delta t$ and $t$ as:

\begin{eqnarray}
f_{x(t)|x(t-\Delta t)}(x) = \lambda\ \Delta t\  f_{N}(x;m_x(\Delta t,x(t-\Delta t))+\mu_Y, v_x(\Delta t)+\sigma_Y^2 )\\ \nonumber  + (1-\lambda \Delta t)f_{N}(x;m_x(\Delta t,x(t-\Delta t)),v_x(\Delta t))
\end{eqnarray}

This is a mixture of Gaussians, weighted by the arrival intensity $\lambda$. It is known that mixtures of Gaussians feature fat tails, so this model is capable of reproducing more extreme movements than the basic Vasicek model.

To check that the process is indeed mean-reverting, in the Gaussian jump size case, through straightforward calculations, one can prove that the exact solution of Equation (\ref{eq:jumpsvacomSDE}) or equivalently Equation(\ref{eq:jumpsvacSDE})) satisfies:

\begin{equation}
\mathbb{E} [x_{t}] = \theta_Y + (x_{0}-\theta_Y)e^{-\alpha t} \label{eq:jumpVacisekMean}
\end{equation}
\begin{equation}
Var[x_{t}] = \frac{\sigma^{2}+\lambda (\mu_Y^2 + \sigma_Y^2)}{2\alpha} \left(1-e^{-2\alpha t}\right) \label{eq:jumpVacisekVar}
\end{equation}

This way one sees that the process is indeed mean reverting, since as time increases the mean tends to $\theta_Y$ while the variance remains bounded, tending to $(\sigma^{2}+\lambda (\mu_Y^2 + \sigma_Y^2))/(2\alpha)$. Notice that, either letting $\lambda$ or $\mu_Y$ and $\sigma_Y$ going to zero (no jumps) gives back the mean and variance for the basic Vasicek process without jumps.

It is important that the jump mean $\mu_Y$ not be too close to zero, or one gets again almost symmetric Gaussian shocks that are already there in Brownian motion and the model does not differ much from a pure diffusion Vasicek model. On the contrary, it is best to assume $\mu_Y$ to be significantly positive or negative with a relatively small variance, so the jumps will be markedly positive or negative. If one aims at modelling both markedly positive and negative jumps, one may add a second jump process $J_{-}(t)$ defined as:
\[ d J_{-}(t) =  Z_{M(t)} d M(t) \]
where now $M$ is an arrival Poisson process with intensity $\lambda_{-}$ counting the jumps, independent of $J$, and $Z$ are again iid normal random variables with mean $\mu_Z$ and variance $\sigma^2_Z$, independent of $M,J,W$. The overall process reads:
\begin{eqnarray}
    d x(t)&=&\alpha\left(\theta -  x(t)\right)  dt +\sigma dW(t)+ dJ(t) -dJ_{-}(t) \ \ \mbox{or} \label{eq:2jumpsvacSDE} \\
    d x(t)&=&\alpha\left(\theta_J  -  x(t)\right)  dt +\sigma dW(t)+ [dJ(t) - \lambda \mu_Y \ dt ] - [dJ_{-}(t) - \lambda_{-} \mu_Z \ dt ],  \nonumber \\  \theta_J &:=& \theta + \frac{\lambda \mu_Y - \lambda_{-} \mu_{Z}}{\alpha} \nonumber
\end{eqnarray}

This model would feature as transition density, over small intervals $[s,t]$:
\begin{eqnarray}
f_{x(t)|x(t-\Delta t)}(x) = \lambda\ \Delta t\  f_{N}(x;m_x(\Delta t,x(t-\Delta t))+\mu_Y, v_x(\Delta t)+\sigma_Y^2 )\\ \nonumber
+ \lambda_{-} \Delta t\  f_{N}(x;m_x(\Delta t,x(t-\Delta t))-\mu_Z,v_x(\Delta t)+\sigma_Z^2 )\\ \nonumber  + (1-(\lambda+\lambda_{-}) \Delta t)\ f_{N}(x;m_x(\Delta t,x(t-\Delta t)),v_x(\Delta t))
\end{eqnarray}

This is now a mixture of three Gaussian distributions, and allows in principle for more flexibility in the shapes the final distribution can assume.

Again, one can build on the Vasicek model with jumps by squaring or exponentiating it. Indeed, even with positive jumps the basic model still features positive probability for negative values. One can solve this problem by exponentiating the model: define a stochastic process $y$ following the same dynamics as $x$ in (\ref{eq:2jumpsvacSDE}) or (\ref{eq:jumpsvacSDE}), but then define your basic process $x$ as $x(t)=\exp(y(t))$. Then historical estimation can be applied working on the log-series, by estimating $y$ on the log-series rather than on the series itself.

\begin{figure}
\begin{center}
\includegraphics[width=1\textwidth]{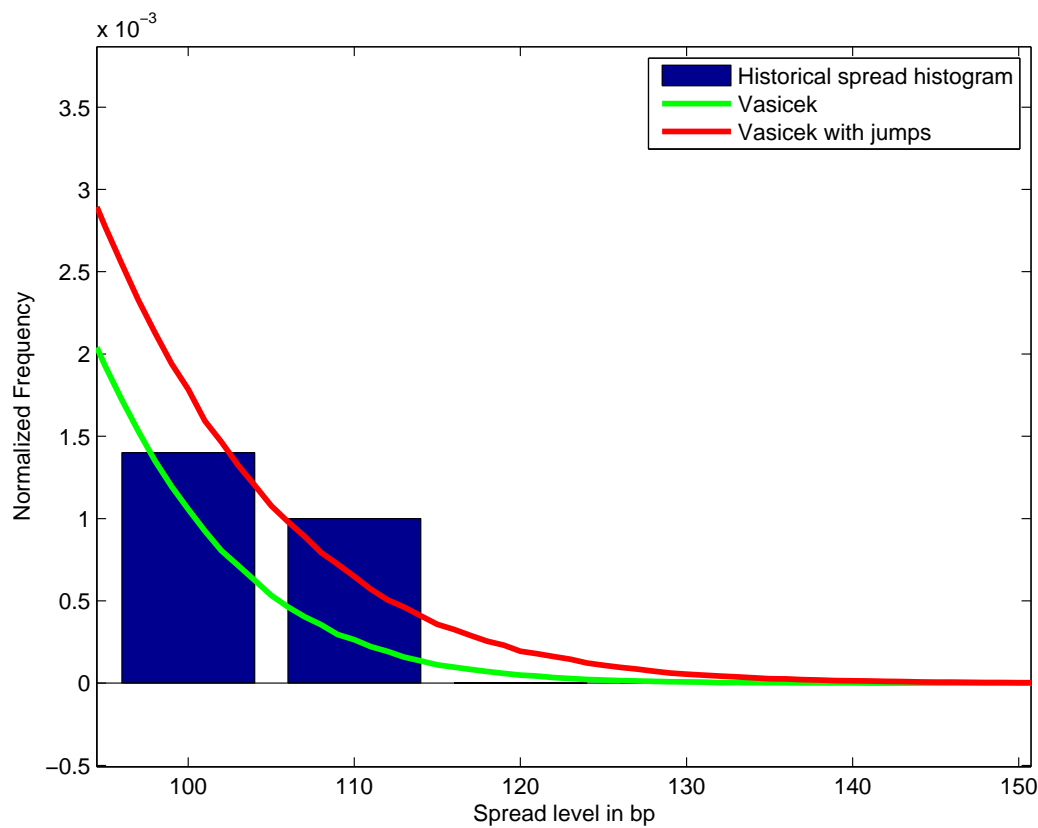}
\caption{The tail of the EMU Corporate A rated 5-7Y index histogram (in blue) compared to the fitted Vasicek process probability density (in green) and to the fitted Vasicek with jumps process probability density (in red)}\label{fig:vasjumpemu}
\end{center}
\end{figure}

From the examples in Figures~\ref{fig:vasjumpemu0} and \ref{fig:vasjumpemu} one sees the goodness of fit of Vasicek with jumps, and note that the Vasicek model with positive-mean Gaussian size jumps can be an alternative to the exponential Vasicek. Indeed, Figure~\ref{cir_vas_expvas} shows that the tail of exponential Vasicek almost exceeds the series, whereas Vasicek with jumps can be a weaker tail alternative to exponential Vasicek, somehow similar to CIR, as one sees from Figure~\ref{fig:vasjumpemu}. However, by playing with the jump parameters or better by exponentiating Vasicek with jumps one can easily obtain a process with tails that are fatter than the exponential Vasicek when needed. While, given Figure~\ref{cir_vas_expvas}, it may not be necessary to do so in this example, in general the need may occur in different applications.

\medskip

More generally, given Vasicek with positive and in case negative jumps, by playing with the jump parameters and in case by squaring or exponentiating, one can obtain a large variety of tail behaviours for the simulated process. This may be sufficient for several applications. Furthermore, using a more powerful numerical apparatus, especially inverse Fourier transform methods, one may generalise the jump distribution beyond the Gaussian case and attain an even larger flexibility.

\label{CH5}

\newpage
\section*{Notation}

\vspace{1cm}

\begin{tabbing}
{\bf General}\\
$\mathbb{E}$  \qquad  \qquad \qquad \qquad \= Expectation\\
Var  \>  Variance\\
SDE \> Stochastic Differential Equation\\
$P(A)$ \> Probability of event $A$\\
$M_X(u)$ \> Moment generating function of the random variable $X$ computed at $u$\\
$\phi_X(u)$ \> Characteristic function of $X$ computed at $u$\\
$q$ \> Quantiles\\
$f_X(x)$ \> Density of the random variable $X$ computed at $x$\\
$F_X(x)$ \> Cumulative distribution function of the random variable $X$ computed at $x$\\
$f_{X|Y=y}(x)$ \> \\
or $f_{X|Y}(x;y)$:  \> Density of the random variable $X$ computed at $x$ \\ \> conditional on the random variable $Y$ being $y$\\
$N(m,V)$ \> Normal (Gaussian) random variable with mean $m$ and Variance $V$  \\
$f_N(x;m,V)$ \> Normal density with mean $m$ and Variance $V$ computed at $x$ \\
\> $=(2\pi V)^{-\frac{1}{2}}\exp\left(-\frac{(x-m )^{2}}{2 V}\right)$\\
$\Phi$ \> Standard Normal Cumulative Distribution Function\\
$f_P(x;L)$ \> Poisson probability mass at $x=0,1,2,\ldots$ with parameter $L$\\
     \> $=\exp(-L) (L)^x/(x!)$\\
$f_\Gamma(x;\kappa,\zeta)$ \> Gamma probability density at $x\ge 0$ with parameters $\kappa$ and $\zeta$:\\
     \> $=   {x^{\kappa-1}
\exp(-{x}/{\zeta})}/({\zeta^{\kappa} \Gamma(\kappa)})$\\
$\sim$ \> Distributed as\\ \\
{\bf Statistical Estimation }\\
CI, ACF, PACF \> Confidence Interval, Autocorrelation Function, Partial ACF\\
MLE \> Maximum Likelihood Estimation\\
OLS \> Ordinary Least Squares \\
$\Theta$ \> Parameters for MLE\\
${\cal L}(\Theta)(x_1,\ldots,x_n) := f_{X_1,X_2,\ldots,X_n|\Theta}(x_1,\ldots,x_n)$ or $f_{X_1,X_2,\ldots,X_n}(x_1,\ldots,x_n;\Theta)$: \\ \> likelihood distribution function with parameter $\Theta$ in the assumed distribution,\\ \> for random variables $X_1,\ldots,X_n$, computed at $x_1,\ldots,x_n$ \\
${\cal L}^\ast(\Theta)(x_1,\ldots,x_n) :=$ \> \ $\log({\cal L}(\Theta)(x_1,\ldots,x_n))$, log-likelihood.\\
$\hat{\Theta}$ \> Estimator for $\Theta$ \\ \\
{\bf Geometric Brownian Motion (GBM)}\\
ABM  \> Arithmetic  Brownian Motion\\
GBM  \> Geometric Brownian Motion\\
$\mu, \ \hat{\mu}$ \> Drift parameter and Sample drift estimator of GBM \\
$\sigma, \ \hat{\sigma}$ \> Volatility and sample volatility \\
$W_t$ or $W(t)$ \>   Standard Wiener process \\ \\
{\bf Fat Tails: GBM with Jumps, Garch Models}\\
$J_t, J(t)$ \> Jump process at time $t$\\
$N_t, N(t)$ \> Poisson process counting the number of jumps,\\
 \> jumping by one whenever a new jump occurs, with intensity $\lambda$ \\
$Y_1,Y_2,\ldots,Y_i,...$ \> Jump amplitudes random variables\\ \\
{\bf Time, discretisation, Maturities, Time Steps}\\
$\Delta t$  \>  Time step in discretization \\
$s,t,u$ \>  Three time instants, $0\le s \le t \le u$.\\
$0=t_0,t_i,\ldots,t_i,\ldots$ \>  A sequence of equally spaced time instants, $t_i - t_{i-1} = \Delta t$.\\
$T$ \> Maturity\\
$n$ \> Sample size\\ \\
{\bf Variance Gamma Models}\\
VG \> Variance Gamma \\
$\bar{\mu}$ \> Log-return drift in calendar-time\\
$g(t)\sim \Gamma(t/\nu,\nu)$ \> Transform of calendar time $t$ into market activity time\\
$\bar{\theta}$ \> Log-return drift in  market-activity-time \\
$\bar{\sigma}$ \> volatility \\ \\
{\bf Vasicek and CIR Mean Reverting Models}\\
$\theta, \alpha, \sigma$ \> Long-term mean reversion level, speed, and volatility\\
\end{tabbing}

\newpage
\bibliographystyle{chicago}
\bibliography{mybib}

\end{document}